%% file: d1.tex
\documentclass[a4paper, 11pt, twopage]{book}
\pdfoutput=1
\usepackage[english,UKenglish]{babel}
\usepackage[utf8]{inputenc}
\usepackage{graphicx} % For graphics
\usepackage{setspace}
\usepackage{amssymb} % Maths symbols
\usepackage{amsmath} % Maths rendering
\usepackage{booktabs} % Ensures correct lines ruled in tables
\usepackage{dcolumn} % provides table column type D
\newcolumntype{d}[1]{D{.}{.}{#1}} % aligns decimal places in tables
\usepackage{microtype}
\usepackage{cite} % Cleaner reference numbering 
\usepackage{bm} % Bold mathematical symbols
\usepackage[bf]{caption} % Fancy captions, 'Figure' label in bold font
\usepackage{subcaption}
\usepackage{rotating} % Allows figure rotation
\newcommand\myeq{\mathrel{\stackrel{\makebox[0pt]{\mbox{\normalfont\tiny def}}}{=}}}

\input{abbrev}

\begin{document}
\onehalfspacing
\bibliographystyle{tim}
\include{title}

\include{author}
\include{quo}

\include{sum}
\tableofcontents
% Chapters
\include{intro}
\include{rev}
\include{stl}
\include{ltl2}
\include{uni}

% \include{nag}
\include{con}
% Appendices
\appendix
\include{css}
\include{rom}
\include{aps}
\include{uniap}
% Permutational invariance, expansion of T_{i0} in imaginary time (amalgamate this with result in uniqueness chapter)
\singlespacing
\addcontentsline{toc}{chapter}{References}
\bibliography{/home/tjhh2/Projects/articles/refbig}
%\bibliography{refbig}
\end{document}

%% file: abbrev.tex
\newcommand{\cfss}{c_{\rm fs}^{\rm sym}(t)}

\newcommand{\Cssnv}[1]{C_{\rm ss}^{[N]}(#1)}
\newcommand{\Cfsnv}[1]{C_{\rm fs}^{[N]}(#1)}

\newcommand{\Cfsov}[1]{C_{\rm fs}^{[1]}(#1)}

\newcommand{\Cfsxv}[1]{C_{\rm fs}^{[\bm{\Xi}]}(#1)}
\newcommand{\Dfsnv}[1]{D_{\rm fs}^{[N]}(#1)}
\newcommand{\Cfsbnv}[1]{\bar C_{\rm fs}^{[N]}(#1)}
\newcommand{\Csfbnv}[1]{\bar C_{\rm sf}^{[N]}(#1)}
\newcommand{\Cfsun}[1]{C_{\rm fs \neq}^{[\bm{\Xi}]}(#1)}
\newcommand{\eb}{e^{-\beta \hat H}}

\newcommand{\ebN}{e^{-\beta_N \hat H}}
\newcommand{\ebt}{e^{-\beta \hat H/2}}

\newcommand{\etf}{e^{-i \hat H t/\hbar}}
\newcommand{\etb}{e^{i \hat H t/\hbar}}
\newcommand{\etfo}{e^{-i \hat H_0 t/\hbar}}
\newcommand{\etbo}{e^{i \hat H_0 t/\hbar}}

\newcommand{\bra}[1]{\langle #1 |}
\newcommand{\ket}[1]{| #1 \rangle}
\newcommand{\kb}[1]{\ket{#1}\bra{#1}}
\newcommand{\bk}[2]{\bra{#1} #2 \rangle}
\newcommand{\dfs}[1]{\delta\!\left[ #1 \right]}
\newcommand{\dfq}{\delta[f(\bq)]}
\newcommand{\dfQ}{\delta[f({\bf Q})]}

\newcommand{\dd}[2]{\frac{d #1}{d #2}}

\newcommand{\ddp}[2]{\frac{\partial #1}{\partial #2}}

\newcommand{\ppi}{\psi_{p_i}}

\newcommand{\ppim}{\psi_{p_{i-1}}}

\newcommand{\piNz}{\prod_{i=0}^{N-1}}

\newcommand{\smiN}{\sum_{i=1}^N}
\newcommand{\smiNz}{\sum_{i=0}^{N-1}}

\newcommand{\smjNz}{\sum_{j=0}^{N-1}}

\newcommand{\lttz}{\lim_{t\to 0_+}}
\newcommand{\ltti}{\lim_{t\to \infty}}

\newcommand{\lNti}{\lim_{N\to \infty}}
\newcommand{\shortt}{$t\to 0_+$}
\newcommand{\longt}{$t\to \infty$}
\newcommand{\largeN}{$N\to\infty$}

\newcommand{\tr}{ {\rm Tr} }
\newcommand{\qdd}{q^\ddag}

\newcommand{\td}{\tilde}
\newcommand{\tdo}{\tilde D_0}
\newcommand{\tpo}{\tilde P_0}
\newcommand{\Qrb}{Q_{\rm r}(\beta)}

\newcommand{\no}{\nonumber}
\newcommand{\ort}{\tfrac{1}{\sqrt{2}}}
\newcommand{\otrt}{\tfrac{1}{2\sqrt{2}}}

\newcommand{\ntn}{\mathcal{N}_{2N}}
\newcommand{\betaN}{\beta_N}

\newcommand{\bp}{ {\bf p} }
\newcommand{\btp}{{\bf \tilde p}}
\newcommand{\bq}{ {\bf q} }
\newcommand{\bs}{ {\bf s} }

\newcommand{\by}{ {\bf y} }
\newcommand{\bz}{ {\bf z} }
\newcommand{\bDelta}{ {\bf \Delta} }
\newcommand{\btDelta}{\bm{\tilde \Delta}}
\newcommand{\bmeta}{ \bm{\eta} }
\newcommand{\bpi}{ \bm{\pi} }
\newcommand{\bQ}{ {\bf Q} }
\newcommand{\bZ}{ {\bf Z} }
\newcommand{\bP}{ {\bf P} }
\newcommand{\bD}{ {\bf D} }

\newcommand{\fq}{ f({\bf q}) }
\newcommand{\fQ}{ f({\bf Q}) }

\newcommand{\fz}{ f({\bf z}) }

\newcommand{\gz}{ g({\bf z}) }
\newcommand{\gq}{ g({\bf q}) }
\newcommand{\gQ}{ g({\bf Q}) }

\newcommand{\ap}{\bm{A}(\bp)}
\newcommand{\api}{\bm{A}(\bpi)}
\newcommand{\tph}{\frac{1}{2\pi\hbar}}

\newcommand{\tphN}{\frac{1}{(2\pi\hbar)^N}}
\newcommand{\tphtN}{\frac{1}{(2\pi\hbar)^{2N}}}
\newcommand{\tphF}{\frac{1}{(2\pi\hbar)^F}}
\newcommand{\inti}{\int_{-\infty}^{\infty}}
\newcommand{\eqr}[1]{Eq.~\eqref{eq:#1}}
\newcommand{\eqsr}[2]{Eqs.~\eqref{eq:#1} and \eqref{eq:#2}}
\newcommand{\eql}[1]{\label{eq:#1}}
\newcommand{\figr}[1]{Fig.~\ref{fig:#1}}
\newcommand{\figl}[1]{\label{fig:#1}}

%% file: title.tex
\thispagestyle{empty}
\pagenumbering{roman}
\begin{center}
\vspace*{30mm}
{\LARGE \bf Quantum Transition-State Theory}
\vspace*{50mm}

\includegraphics[width=0.3\textwidth]{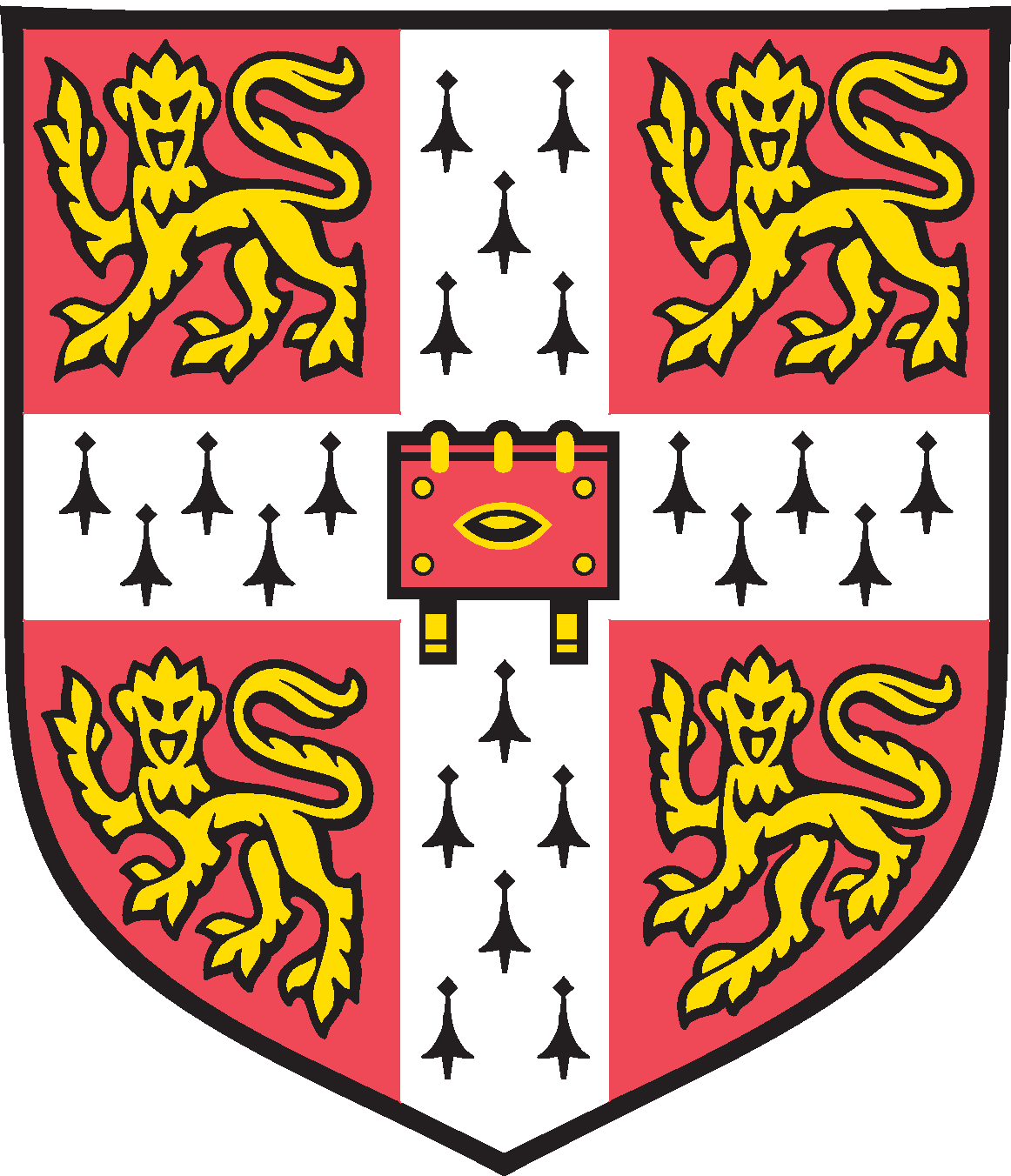}

\vspace*{30mm}

{\large \rm
{Timothy John Harvey Hele}

{Trinity College}

{University of Cambridge}

}
\vspace*{15mm}

{\large \it
A dissertation submitted for the degree of

Doctor of Philosophy

June 2014
}
\vspace*{5mm}

Redacted Version

\end{center}

%% file: author.tex
\section*{Authorship}
This dissertation is the result of my own work and includes nothing which is the outcome of work done in collaboration except where specifically indicated in the text.
In accordance with university regulations, I acknowledge my ownership of the copyright of this dissertation and assert my moral right to be identified as its author.

\section*{Acknowledgements}
I acknowledge guidance and intellectual support from Prof.\ Althorpe, and I am grateful to Michael Willatt for proof-reading this dissertation, and to the other members of the Althorpe group for their advice. I further acknowledge support from my friends and family.

\section*{Length}
This dissertation does not exceed the word limit for the Physics and Chemistry Degree Committee.

\section*{Redaction}
Two chapters of the original dissertation, containing research which is yet to be published, have
been omitted from this version of the dissertation deposited online. The central message and logical argument of the
dissertation is unaffected, and the author intends to make the complete dissertation available as soon as practicable.

% \section*{Copyright}

%I further acknowledge the support provided by the Althorpe group, particularly from Michael Willatt for proof-reading this dissertation remarkably quickly, and from my friends and family.
\clearpage

%% file: quo.tex
\begin{flushright}
 \textrm{\ldots the notion of an activated or transition state is not strictly compatible \\  with the laws of quantum mechanics.}
 \\
 \textit{Hirschfelder and Wigner, 1939}
 
 \vspace{1cm}
 
 \textrm{\ldots the inherent structure of quantum mechanics does not allow one to formulate a  \\ quantum transition-state theory\ldots}
 \\
 \textit{Voth, 1993}
 
 \vspace{1cm}
 
 \textrm{Unfortunately, despite the ongoing research effort on constructing quantum transition state theories for the last few decades, nothing has emerged that one can properly call a rigorous quantum TST.}
 \\
 \textit{Small, Predescu and Miller, 2005}
 \\
%  \vfill
%  \textrm{\ldots there exists a true quantum transition-state theory\ldots}
%  \\
%  \textit{Hele and Althorpe, 2013}
\end{flushright}

%% file: sum.tex
\cleardoublepage
\begin{center}

\vspace*{10mm}

{\Large \bf Quantum Transition-State Theory}
 
\vspace*{10mm}

{Timothy John Harvey Hele}

{Trinity College}

{University of Cambridge}

\vspace{10mm}

{\large \bf Summary}

\vspace{5mm}

\end{center}

The calculation of chemical reaction rates is vital to our understanding of chemical, physical and biological processes. This dissertation unifies one of the central methods of classical rate calculation, `Transition-State Theory' (TST), with quantum mechanics, thereby deriving a rigorous `Quantum Transition-State Theory' (QTST), which since the 1930s had been considered impossible. 
The resulting QTST is identical to ring polymer molecular dynamics transition-state theory (RPMD-TST), which was previously considered a heuristic method, and whose results we thereby validate. Furthermore, strong evidence is presented that this is the only QTST with positive-definite Boltzmann statistics and therefore the pre-eminent method for computation of thermal quantum rates in direct reactions.

The rationale for this development is that many processes, particularly for light atoms at low temperatures, are governed by quantum mechanics, often leading to counter-intuitive results. The equations for exact quantum calculation were derived in a theoretical framework in the 1970s, but due to their high computational cost, scaling exponentially with the dimensionality of the system, are only viable for very small or model systems.

The key step in deriving a QTST is alignment of the flux and side dividing surfaces in path-integral space. This initially leads to a rate theory proposed by Wigner on heuristic grounds, but possesses non positive-definite Boltzmann statistics, producing erroneous results at low temperatures. To circumvent this, we polymerize the quantum flux-side time-correlation function in path-integral space, obtaining as a short-time limit a positive-definite expression for the instantaneous thermal quantum flux through a dividing surface. We then prove that this produces the exact quantum rate in the absence of recrossing by the exact quantum dynamics, fulfilling the requirements of a QTST. Remarkably, the rate expression is identical to RPMD-TST.

% The dissertation also presents numerical results corroborating the algebraic derivation and a preliminary extension of QTST to electronically non-adiabatic systems.

%% file: intro.tex
\chapter{Introduction}
\label{ch:int}
\pagenumbering{arabic}

The calculation of chemical reaction rates is fundamental to our understanding of chemistry, physics and biology \cite{han90}. Many such physical processes are dominated by counter-intuitive quantum effects such as delocalization, tunnelling and electronically non-adiabatic transitions, which are particularly pronounced at low temperatures and for light atoms. The exact theoretical expressions for classical and quantum rate calculation are known, by correlating the thermal flux through a dividing surface with the side of the products at later time, producing a flux-side time-correlation function \cite{yam59,cha78,mil74,mil83}. However, for all except the simplest systems the exact quantum calculation remains computationally unfeasible\cite{sch10,all13}.

There has been much effort in obtaining approximate methods which possess lower computational cost, but result in a minimal loss in accuracy\footnote{There exists an enormous literature, for which the reader is referred to various review articles \cite{han90,tru96,ber88,pol05}.}. In the 1930s `Transition-State Theory' (TST) was proposed as a method of calculating reaction rates for systems obeying classical mechanics, with the central assumption that the reaction possesses a well-defined dividing surface separating products and reactants (the `Transition-State'), and that all systems which pass this point react, such that the rate can be accurately approximated as the classical flux through the dividing surface\cite{eyr35,eyr35rev,tru96,han90}. It was subsequently realised \cite{cha78} that classical TST corresponded to the short-time (\shortt) limit of a classical flux-side time-correlation function, which would be equal to the exact (\longt) rate in the absence of recrossing of the dividing surface by classical dynamics of the system \cite{sun98}. 

Classical TST has been extremely successful for calculating reaction rates for classical systems (those with heavy atoms at high temperatures), but fails, often underestimating the rate by many orders of magnitude, in the quantum regime\cite{ker06}. There has therefore been a scientific need for a quantum analogue of classical TST, a `Quantum Transition-State Theory' (QTST): a rate equation which measures the instantaneous thermal quantum flux through a dividing surface, such that the exact quantum rate is obtained in the absence of recrossing 
by the exact quantum dynamics.

However, since the late 1930s it was believed impossible to form a QTST, due to a number of factors including concerns over the
uncertainty principle \cite{tru96}, the delocalization of the quantum Boltzmann operator \cite{wig38,wig39}, or that the short-time
limit of proposed quantum flux-side time-correlation functions appears to give zero, as illustrated in \figr{clasvqm}
\cite{mil93,vot93}. 

\begin{figure}[tb]
\begin{subfigure}{0.49\linewidth}
% \begin{figure}[h]
  \includegraphics[width=\textwidth]{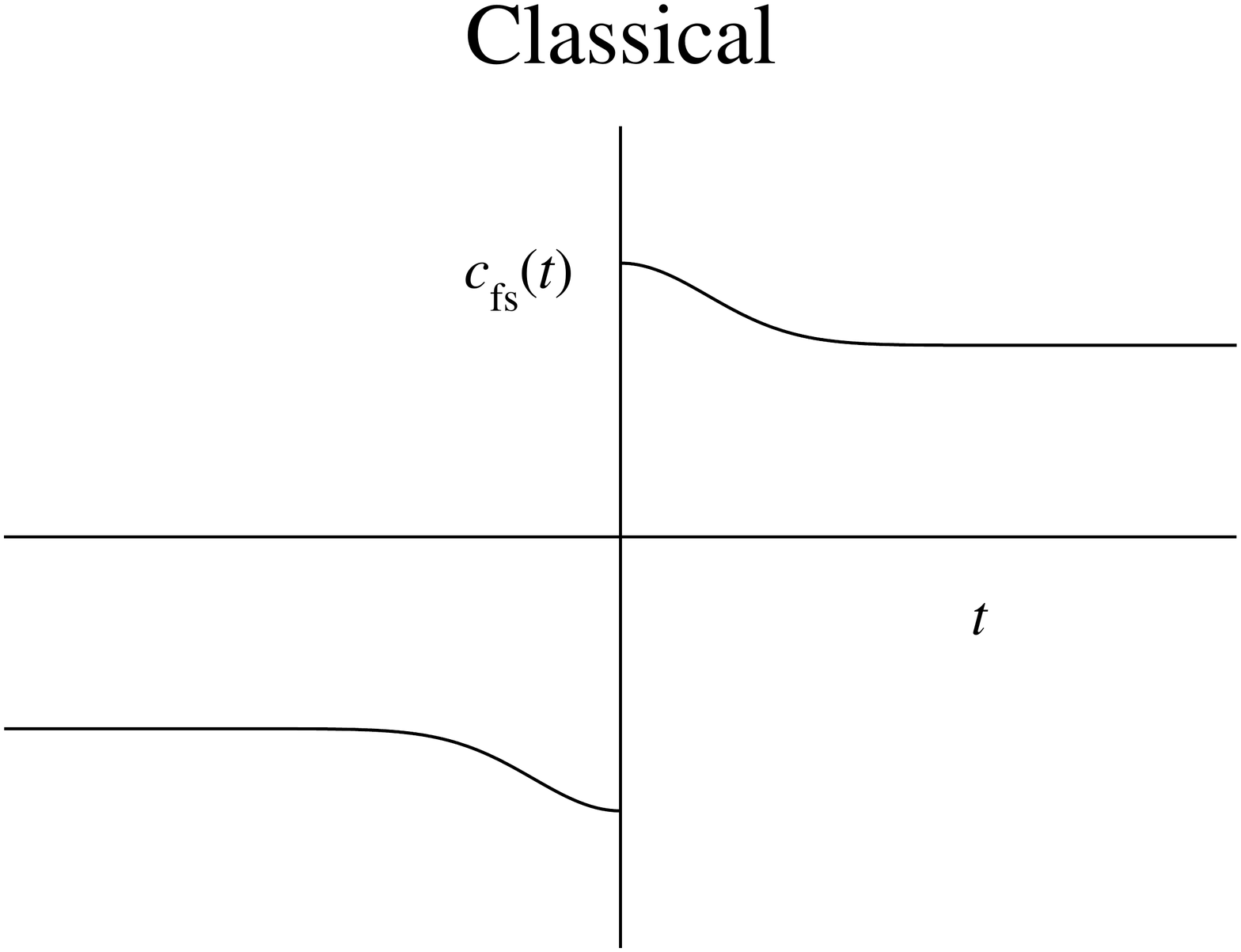}
  \caption{Non-zero \shortt~classical TST}
 \figl{clasrate}
% \end{figure}
\end{subfigure}
\begin{subfigure}{0.49\linewidth}
% \begin{figure}[h]
  \includegraphics[width=\textwidth]{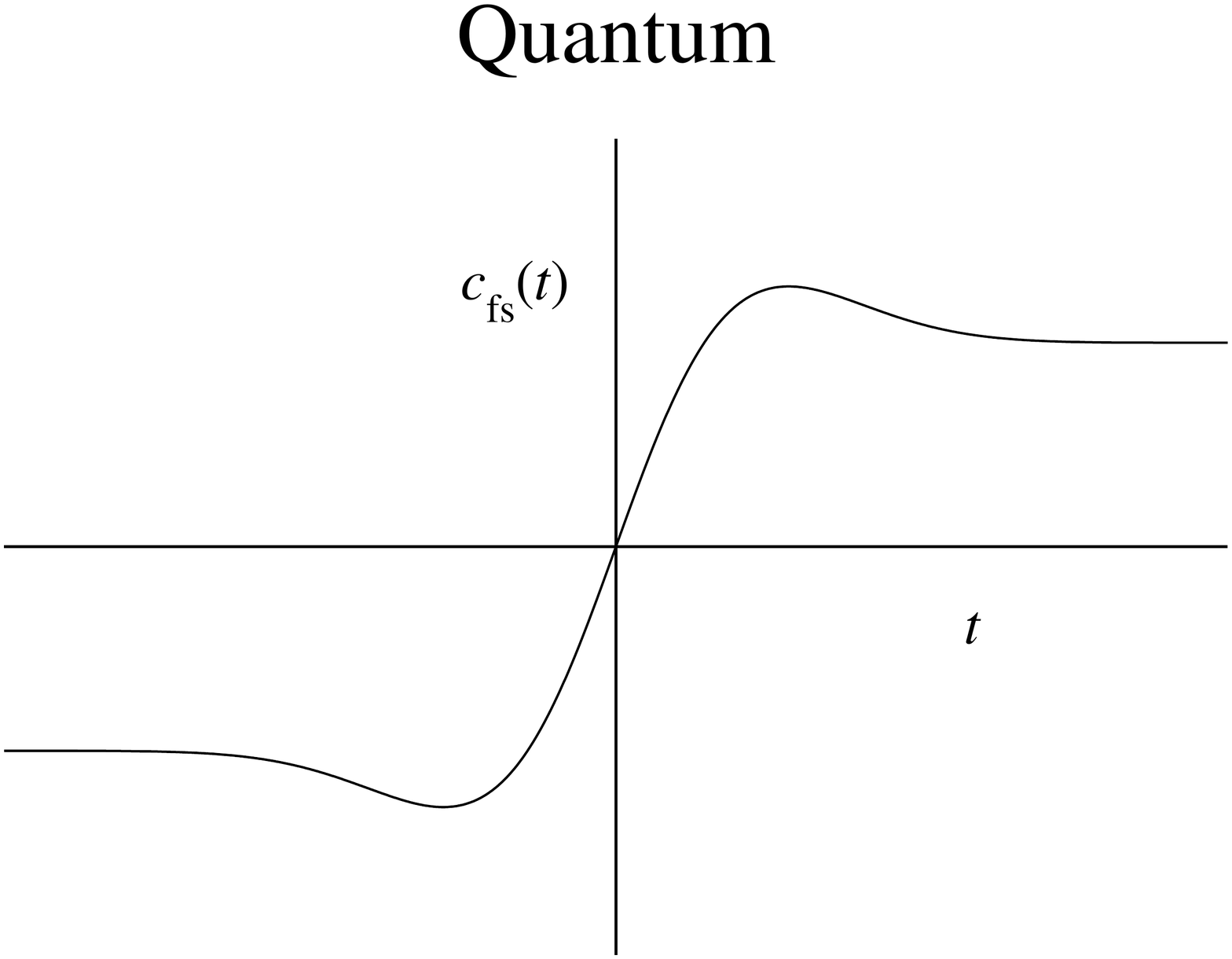}
  \caption{Zero \shortt~quantum TST.}
 \figl{milrate}
% \end{figure}
\end{subfigure}
\caption{Qualitative difference between the classical and quantum (Miller-Schwarz-Tromp) flux-side time correlation functions,
illustrated schematically.}
\figl{clasvqm}
\end{figure}

Nevertheless, many approximate or heuristic QTSTs were proposed\cite{gil87,vot89rig,cal77,ben94,pol98,and09,vot93,tru96,gev01,shi02,mci13}, as well as other methods of obtaining the quantum rate from short-time data \cite{mil03,van05,wan98,wan00,rom11,wol87}. As a consequence, it was often difficult, if not impossible, to discern \emph{a priori} the circumstances in which a given theory would provide a good approximation to the rate, nor how it might be systematically improved.\footnote{Furthermore, the definition of QTST was sometimes relaxed to include virtually any rate theory which accounted for some quantum effects \cite{shi02}. This dissertation concerns itself with the quantum analogue of the original definition of TST by Eyring in 1935 \cite{eyr35rev}, in which the only approximation is the assumption of no recrossing (see chapter~\ref{ch:ltl}).} 

\begin{figure}[tb]
 \centering
 \includegraphics[angle=270,width=0.7\textwidth]{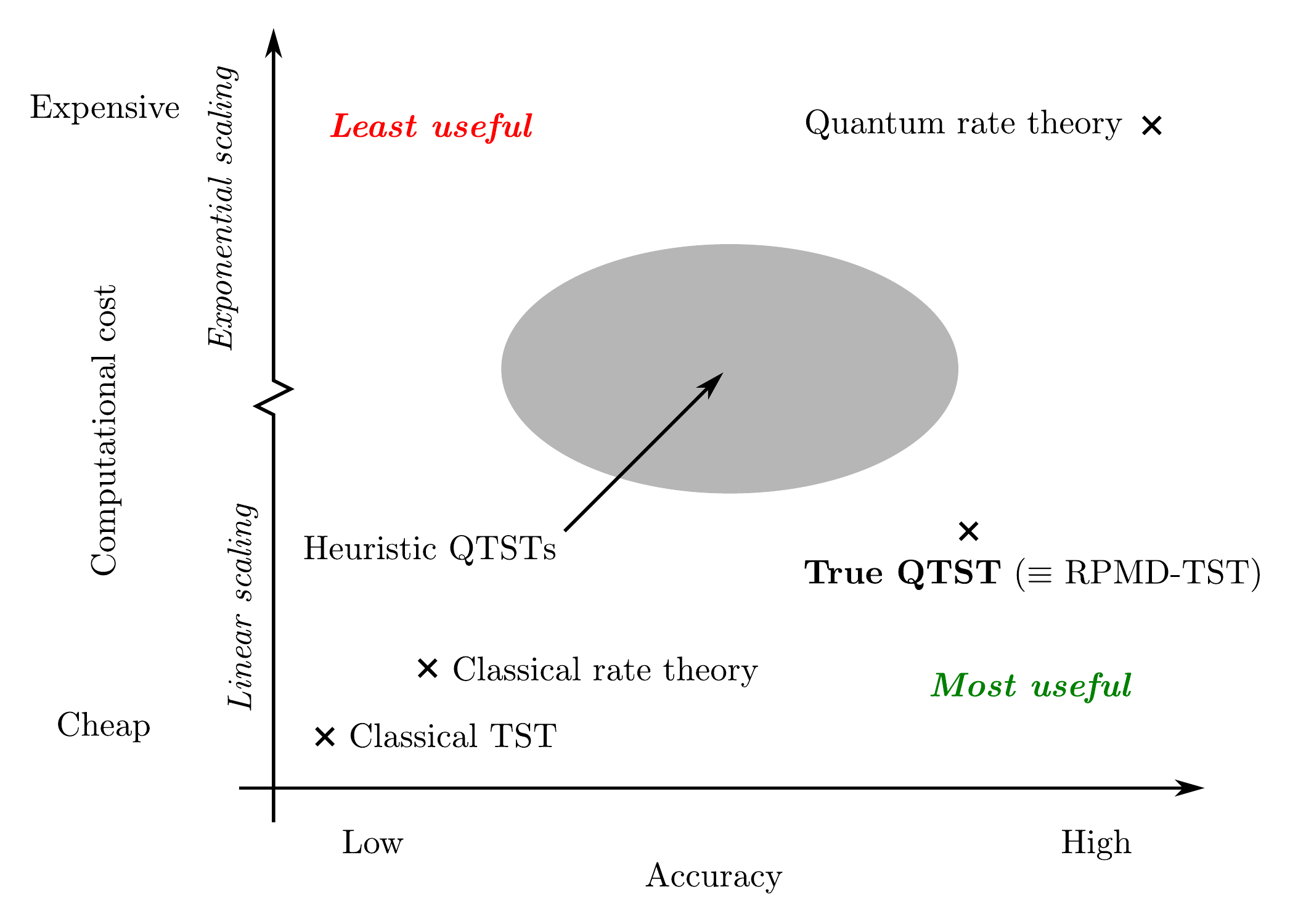}
 \caption{Schematic diagram illustrating competing rate theories. Heuristic rate theories are represented as an indistinct region; their accuracy not known \emph{a priori} without a derivation. RPMD-TST, the true QTST derived in this dissertation, provides high accuracy with computational cost only slightly greater than a classical calculation.}
 \label{fig:qtstd}
\end{figure}

The central object of this dissertation is the derivation of Quantum Transition-State Theory \cite{hel13,alt13}. In so doing we establish a single, pre-eminent method for the practical and accurate calculation of thermal quantum rates (see \figr{qtstd}) \cite{hel13unique}, and validate an existing methodology previously considered heuristic.

We initially review classical rate theory and its associated TST in chapter~\ref{ch:rev}, along with quantum rate theory, the apparent absence of a QTST, and associated heuristic methods. In chapter~\ref{ch:stl} we observe that earlier quantum flux-side time-correlation functions did not have the dividing surfaces in the same location in path-integral space, and therefore vanished in the short-time limit (as for classical TST). Upon alignment of these surfaces a non-zero QTST is obtained which was previously proposed on heuristic grounds by Wigner in 1932\cite{wig32uber}, but which produces poor results at low temperatures as the dividing surface is a function of only one point in imaginary time, leading to non positive-definite statistics. 

By polymerizing the rate expression in path-integral space, we obtain a different QTST which, when the dividing surface is invariant
to permutation of the path-integral beads, possesses positive-definite statistics.\footnote{That is, the rate is guaranteed to be
positive at any finite temperature.} Remarkably, the rate theory thus obtained is identical to an earlier method known as
Ring-Polymer Molecular Dynamics Transition-State Theory (RPMD-TST), which was previously proposed on heuristic grounds
\cite{man04,man05che,man05ref}. Chapter~\ref{ch:ltl} then shows that this ring-polymerized flux-side time-correlation function
produces the exact quantum rate in the absence of recrossing of the dividing surface or those orthogonal to it in path-integral
space, thereby fulfilling the requirements of a QTST.

% A numerical illustration for a model gas-phase scattering system is presented in chapter~\ref{ch:num}, corroborating the algebraic
% derivation by showing that the short-time limit of the polymerized flux-side time-correlation function is identical to RPMD-TST. We
% also observe that there is significant quantum coherent recrossing of the dividing surface at low temperatures, though QTST still
% provides a good estimate of the rate.\footnote{Attributable to recrossing of the planes orthogonal to the dividing surface
% compensating for recrossing of the dividing surface itself.} A full explanation for this, and the quantum dynamics more generally,
% is left as future work.

Given the plethora of competing heuristic QTSTs, the question arises as to whether RPMD-TST is the unique QTST with
positive-definite statistics. In chapter~\ref{ch:unique} we provide very strong evidence that this is the case, and RPMD-TST is
therefore the pre-eminent theory for thermal quantum rate calculation in direct reactions.\footnote{Where `direct reactions'
corresponds to those with a well-defined transition state and no long-lived intermediates.}

% The foregoing derivations are applicable to systems with a single (Born--Oppenheimer) potential energy surface, and the theory would
% be of far greater utility if it could be applied to general non-adiabatic systems. While a general non-adiabatic Quantum
% Transition-State Theory is presently unknown, chapter~\ref{ch:nag} presents a preliminary extension of the QTST derived in
% chapters~\ref{ch:stl} and \ref{ch:ltl} to systems with multiple electronic states where the dividing surface is in position space.
% This is expected to provide a good approximation to the quantum rate when the non-adiabatic coupling is relatively large and there
% is consequently little recrossing of the position-space dividing surface. The feasibility of generalization to weakly-coupled
% systems with an electronic-space dividing surface, arguably an area of greater scientific interest, is also discussed.

Finally, conclusions and avenues for future research are presented in chapter~\ref{ch:con}.

%% file: rev.tex
\chapter{Review}
\label{ch:rev}
Reaction rate theory is a vast discipline and here we confine our attention to rate theories relevant to the derivation of Quantum Transition-State Theory. For a fuller historical overview, the reader is referred to various review articles \cite{han90,tru96,ber88,pol05,nym14}. We begin with classical rate theory and its associated classical transition-state theory, before exploring quantum rate theory and various attempts at heuristic QTSTs.

\section{Classical rate theory}
We consider an $F$-dimensional classical system at inverse temperature $\beta \equiv 1/k_{\rm B}T$ where $k_{\rm B}$ is the Boltzmann constant, with mass $m$ and classical Hamiltonian $H(\bq,\bp)$. Here $\bq$ and $\bp$ are $F$-dimensional vectors of position and momentum respectively, such that\footnote{One can assume without any loss of generality that the masses along each co-ordinate axis are equal, as a mass-scaled co-ordinate system can always be found in which this is the case.}
\begin{align}
 H(\bq,\bp) = V(\bq) + \sum_{i=0}^{F-1} \frac{p_i^2}{2m}, \eql{clasham}
\end{align}
where $V(\bq)$ is the potential energy of the system.
The classical rate is given by the long-time limit of the classical flux-side time-correlation function, \cite{cha78,col08,col09,mil74}
\begin{align}
 k_{\rm clas}(\beta) = \ltti \frac{c_{\rm fs}^{\rm clas}(t)}{Q_{\rm r}^{\rm clas}(\beta)},
\end{align}
where $Q_{\rm r}^{\rm clas}(\beta)$ is the classical partition function in the reactant region and $c_{\rm fs}^{\rm clas}(t)$ is the classical flux-side time-correlation function
\begin{align}
 c_{\rm fs}^{\rm clas}(t) = \tphF \int d\bp \int d\bq \ e^{-\beta H(\bq,\bp)} \delta[s(\bq)] S(\bq,\bp) h[s(\bq_t)],\eql{cfsclas}
\end{align}
where $\int d\bp = \inti dp_0 \ldots \inti dp_{F-1}$, and likewise for $\int d\bq$. The notation $\bq_t$ denotes the position of a trajectory at time $t$, starting from the initial configuration $(\bq,\bp)$ at time $t=0$. The dividing surface is defined to be at $s(\bq) = 0$, such that $s(\bq)>0$ is the product region and $s(\bq) < 0$ the reactant region. Equation~\ref{eq:cfsclas} therefore measures the thermal flux through the classical dividing surface separating products and reactants $s(\bq)$ at $t=0$,
\begin{align}
 S(\bq,\bp) = \frac{1}{m} \sum_{i=0}^{F-1} \ddp{s(\bq)}{q_i}p_i,
\end{align}
and correlates it with the side of the particles $h[s(\bq_t)]$ evolved to some later time $t$ under the classical Hamiltonian [\eqr{clasham}]. Classical rate theory is rigorously independent of the dividing surface location \cite{mil74,col08}, though in practice it is numerically favourable to locate it near to the `Transition State' or bottleneck (the saddle point in the minimum energy path between products and reactants) \cite{van05tst}. 

However, classical rate theory includes no quantum effects, so can be in error by many orders of magnitude at low temperatures\cite{top94}. It also requires computation of the real-time classical dynamics, which for large systems can be computationally expensive. Furthermore, if the dividing surface is at the transition-state of the reaction, the majority (if not all) trajectories initiated on $s(\bq)$ will never recross, such that $c_{\rm fs}^{\rm clas}(t)$ will be constant $\forall t > 0$ and computation of the dynamics will be unnecessary. This is the origin of classical transition-state theory.

\section{Classical Transition-State Theory}
\label{sec:cltst}
If few trajectories initiated at $s(\bq)=0$ recross the flux dividing surface at some later time, and the flux and side dividing surfaces are in the same location, one can take the \shortt~limit of \eqr{cfsclas} \cite{cha78,cha87} and define
\begin{align}
 k_{\rm clas}^{\ddag}(\beta) = \lttz \frac{c_{\rm fs}^{\rm clas}(t)}{Q_{\rm r}^{\rm clas}(\beta)}
\end{align}
as the classical TST rate \cite{cha78,fre02}. In the short-time limit the dividing surface function can be Taylor-expanded, 
\begin{align}
 \lttz \delta[s(\bq_0)]h[s(\bq_t)] = & \lttz \delta[s(\bq_0)]h[s(\bq_0 + \bp t/m)] \\
 = & \lttz \delta[s(\bq_0)] h\!\!\left[s(\bq_0) + \frac{t}{m} \sum_{i=0}^{F-1} \ddp{s(\bq)}{q_i}p_i \right] \eql{momcon} \\% Momentum constribution
 = & \delta[s(\bq_0)] h[S(\bq_0,\bp_0)]
\end{align}
where for clarity I have added a subscript zero for momenta and positions at time $t=0$, and we have noted that the Heaviside function is invariant to the scaling of its argument, leading to
\begin{align}
 \lttz c_{\rm fs}^{\rm clas}(t) = \tphF \int d\bp \int d\bq \ e^{-\beta H(\bq,\bp)} S(\bq,\bp) h[S(\bq,\bp)]\delta[s(\bq)] \eql{clastst}.
\end{align}
Due to the $e^{-\beta H(\bq,\bp)}\delta[s(\bq)]$ term, classical TST is exponentially sensitive to the location of the dividing surface. Since (classical) recrossing can only reduce the rate,  $k_{\rm clas}^{\ddag}(\beta) \geq k_{\rm clas}(\beta)$; i.e.\ classical TST is a rigorous upper bound to the classical rate. Thus in complex multidimensional systems where the location of the dividing surface is not obvious, it can be variationally optimized \cite{tru96,van05tst}. 

Taking the \shortt\ limit is an approximation (otherwise TST would equal the exact reaction rate) and in general physical systems there will be some recrossing of the dividing surface. The TST will break down for systems with significant recrossing, such as diffusive processes (the high-friction Kramers regime being a particular example \cite{kra40}), and those with long-lived intermediates.  Nevertheless, for one-dimensional systems, classical TST is exact (equal to the classical rate) if the dividing surface is at the energy maximum, and for general multidimensional systems where reaction is dominated by a free energy bottleneck, classical TST is a \emph{good approximation} to the exact classical rate \cite{lai83, tru96}.

If the Heaviside dividing surface is in a different location in path-integral space\footnote{Where `path-integral space' is
the configuration space of path integrals.} to the flux dividing
surface, i.e.~
\begin{align}
 c_{\rm fs}^{\rm clas}(t) = \tphF \int d\bp \int d\bq \ e^{-\beta H(\bq,\bp)} \delta[s(\bq)] S(\bq,\bp) h[s'(\bq_t)],
\end{align}
the momentum contribution in \eqr{momcon} would smoothly vanish as \shortt, resulting in 
\begin{align}
 \lttz c_{\rm fs}^{\rm clas}(t) = & \tphF \int d\bp \int d\bq \ e^{-\beta H(\bq,\bp)} \delta[s(\bq)]  S(\bq,\bp) h[s'(\bq)] \nonumber \\
 = &\ 0,
\end{align}
as one has to wait a finite time for the particle, initially constrained at $s(\bq) = 0$ to cross the dividing surface $s'(\bq)$. While well-known in classical TST, we show in chapter~\ref{ch:stl} that the dividing surfaces being in different locations in quantum-mechanical path-integral space caused the apparent absence of QTST.

% As classical TST requires no real-time dynamics and only calculation of the constrained energy, it is computationally feasible, even for very large systems, though exponentially sensitive to the location of the dividing surface. Its omission of any real-time effects (classical or quantum) means it can be in error by many orders of magnitude \cite{ker06} (as shown in \figr{wigrpmdqm}). 

\section{Quantum rate theory}
\label{sec:qmrt}
For algebraic simplicity, we consider a one-dimensional system with coordinate $q$, mass $m$ and Hamiltonian $\hat H$ at an inverse temperature $\beta \equiv 1/k_B T$. 

The quantum rate can, in principle, be computed from the long-time limit of the Miller-Schwartz-Tromp (MST) quantum flux-side time-correlation function \cite{mil74,mil83}\footnote{There exist other correlation functions from which the exact quantum rate rate can be calculated, such as Yamamoto's kubo-transformed flux-flux form\cite{yam59}, and others based on different splitting of the Boltzmann operator around the flux operator\cite{mil74,mil83}, but these all possess the `curse of dimensionality' and a vanishing \shortt~limit.}:
\begin{align}
 k^{\rm QM}(\beta) = \lim_{t\to \infty}c_{\rm fs}^{\rm sym}(t)/\Qrb, \eql{rxnrate}
\end{align}
where $\Qrb$ is the reactant partition function, and
\begin{align}
 c_{\rm fs}^{\rm sym}(t) = \tr \left[ \ebt \hat F \ebt \etb \hat h \etf \right]
\label{eq:milsym}
\end{align}
where $\hat F$ is the quantum-mechanical flux operator
\begin{align}
 \hat F = \frac{1}{2m}\left[\delta(\hat q - q^\ddag) \hat p + \hat p \delta(\hat q - q^\ddag) \right],
\end{align}
and $\hat h$ is the Heaviside operator projecting onto states in the product region, defined relative to the dividing surface
$q^\ddag$, where $h(q-\qdd) = 1$ if $q>\qdd$ and zero otherwise.

As for classical rate theory, the quantum rate is independent of the location of the dividing surface, here due to the quantum mechanical continuity equation \cite{man05ref}. Evaluation of \eqr{milsym}, in particular that of the exact real-time quantum dynamics ($\etf$), scales exponentially with system size. Full-dimensional calculations are limited to a few atoms \cite{hua01} or model systems \cite{top94}.

It would therefore be very useful to have a quantum analogue of classical TST --- a rate theory which did not require real-time dynamics, but included quantum effects such as zero-point energy and tunnelling\footnote{This thesis concerns position-space TST, not the formally-exact phase space TST of, e.g.~Ref.~\cite{wig01}. While formally exact, computation of the phase-space dividing surface is as costly as solving the Schr\"odinger equation for the system and therefore of calculating \eqr{milsym}, so is of little computational utility.}, and would produce the exact quantum rate in the absence of recrossing by the quantum dynamics. However, as depicted in \figr{milrate}, $\cfss$ tends smoothly to zero in the the $t\to 0_+$ limit, discussed more fully in Sec.~\ref{sec:wigmil}. This appears to preclude the existence of a rigorous quantum TST. Other arguments have been advanced against a quantum analogue of TST, particularly the uncertainty principle \cite{tru96}, whereby one is unable to specify simultaneously and precisely the position and momentum of a quantum particle.\footnote{Note that by a careful factorization of \eqr{clastst} momenta can be integrated out, so it is not actually necessary to know position and momentum simultaneously in the classical case, even though one could.}

Nevertheless, many approximate QTSTs have been proposed.

\section{Heuristic quantum TSTs}

\subsection{Wigner rate theory}
This expression was proposed on heuristic grounds by Wigner in 1932 \cite{wig32uber}, on the basis that it corresponds to a classical flux multiplied by a Wigner-transformed \cite{wig32qm} Boltzmann operator and produces the classical rate in the high-temperature ($\beta \to 0$) limit,
\begin{align}
 k_{\rm wig}(\beta) = \frac{1}{\Qrb} \tph \int dq \int dp \ h(p) \delta(q-\qdd) \frac{p}{m} \left[e^{-\beta \hat H}\right]_{\rm W}, \eql{wig}
\end{align}
where
\begin{align}
 \left[e^{-\beta \hat H}\right]_{\rm W} = \int d\Delta \ \bra{q-\Delta/2} \eb \ket{q+\Delta/2} e^{i p\Delta/\hbar}
\end{align}
and integration is performed between $\pm \infty$ unless otherwise stated, a convention used throughout this dissertation.

However, this was known to produce erroneous low-temperature statistics \cite{liu09}, and practical calculation of \eqr{wig} is hindered by the Fourier transform, which would be computationally unfeasible for multidimensional systems with similar dimensionality scaling to solving the exact quantum dynamics in the first place \cite{man05che, wan98}.

\subsection{Voth-Chandler-Miller rate theory}
In 1989 Voth, Chandler and Miller \cite{vot89rig,vot89time,jan99}, augmenting the earlier work of Gillan \cite{gil87,gil87hyd}, proposed a rate theory based on the calculation of a constrained partition function at the dividing surface,
\begin{align}
 k_{\rm VCM}(\beta) = \tfrac{1}{2} \langle |\dot q| \rangle  \frac{Q_{\ddag}}{\Qrb}
\end{align}
where $\tfrac{1}{2} \langle | \dot q | \rangle $ is half the mean magnitude of the thermal velocity\footnote{The factor of $\tfrac{1}{2}$ accounts for only half the trajectories moving in the reactive direction.}, and 
\begin{align}
 Q_{\ddag} = \oint Dq(\tau)\ \delta(\bar q - q^{\ddag}) e^{-S[q(\tau)]/\hbar}
\end{align}
where $\bar q$ is the centroid
\begin{align}
 \bar q = \frac{1}{\beta\hbar} \int_0^{\beta \hbar} d\tau\ q(\tau), % \frac{1}{N} \smiNz q_i
\end{align}
and $S[q(\tau)]$ is the classical action of an imaginary time trajectory of length $\tau = -i\beta \hbar$,
\begin{align}
 S[q(\tau)] = \int_0^{\beta \hbar} d\tau\ \tfrac{1}{2}m \dot q(\tau)^2 + V[q(\tau)].
\end{align}
In practice, the imaginary-time path integral is evaluated using the classical isomorphism, where the partition function of a quantum particle is identical to the classical partition function of $N$ replicas of the system joined by harmonic springs (a `ring polymer'), whose spring constant is $\omega_N = 1/\beta_N \hbar$, and in the \largeN\ limit \cite{fey72,fey65}. Mathematically, the imaginary-time path integral is discretized into $N$ segments of length $\betaN\equiv \beta/N$, 
\begin{align}
 k_{\rm VCM}(\beta) = \frac{\tfrac{1}{2} \langle |\dot q| \rangle}{\Qrb} \int d\bq \ \delta(\bar q - q^{\ddag}) \piNz \bra{q_{i-1}}\ebN \ket{q_i}.
\end{align}
By taking the \largeN~limit, the $\bra{q_{i-1}}\ebN \ket{q_i}$ terms can be evaluated analytically,
\begin{align}
 k_{\rm VCM}(\beta) = \frac{1}{\Qrb} \tphN \int d\bq \int d\bp \ e^{-\beta_N H_N(\bq,\bp)} \frac{\bar p}{m} \delta(\bar q-\qdd) h(\bar p)
\end{align}
where the Hamiltonian for an $N$-bead ring polymer is given by 
\begin{align}
H_N(\bq,\bp) = \smiNz \frac{p_i^2}{2m} + \frac{m(q_{i} - q_{i-1})^2}{2\beta_N^2\hbar^2} + V(q_i), \eql{ringpolham}
\end{align}
and the momentum centroid calculated as
\begin{align}
 \bar p = \frac{1}{N} \smiNz p_i, \eql{centroiddef}
\end{align}
and likewise for $\bar q$. 

Unlike Wigner rate theory, this expression does not produce negative results at low temperatures, produces good results (compared to exact quantum calculations on model systems) for relatively symmetric barriers, and can be applied to real physical systems, such as diffusion of hydrogen on ruthenium \cite{mci13}. No rigorous reason was given for the use of the centroid, which can lead to poor results for asymmetric systems at low temperatures \cite{ric09,ric12}. Nevertheless, the research in this dissertation justifies the Centroid-TST method (as a special case of RPMD-TST) provided that the barrier is symmetric.\footnote{Or asymmetric and above the crossover temperature, see Refs~\cite{hel13} and \cite{ric09}.}

\subsection{Ring-Polymer Molecular Dynamics rate theory}
Combining the classical isomorphism with real-time evolution of the fictitious ring polymer \cite{man04}, in 2005 Craig and Manolopoulos proposed \cite{man05ref,man05che}
\begin{align}
 k_{\rm RPMD}(\beta) = \ltti \frac{c_{\rm RPMD}(t)}{\Qrb} 
\end{align}
where
\begin{align}
 c_{\rm RPMD}(t) = \tphN \int d\bq \int d\bp \ e^{-\beta_N H_N(\bq,\bp)} S_{N}(\bq,\bp) \delta[f(\bq)] h[f(\bq_t)]. \eql{rpmd}
\end{align}
The ring-polymer Hamiltonian is given in \eqr{ringpolham}, $\fq$ is a general dividing surface separating products from reactants, and $S_{N}(\bq,\bp)$ is the ring-polymer flux perpendicular to the dividing surface $\fq$,
\begin{align}
 S_{N}(\bq,\bp) = \frac{1}{m}\smiNz \ddp{\fq}{q_i}p_i. \eql{sndef}
\end{align}

Ring polymers have long been used to calculate statistical properties rigorously, since the fictitious RPMD dynamics offer a method of exploring quantum phase space cheaply while conserving the quantum Boltzmann distribution. This method, known as Path-Integral Molecular Dynamics (PIMD), which predates RPMD \cite{mar96}, has been applied to systems ranging from metallic liquid hydrogen \cite{che13} to the formic acid dimer \cite{miu98}. The heuristic aspect of RPMD rate theory (and the RPMD method in general) is therefore not the quantum statistics, but the use of fictitious, real-time RPMD dynamics as an approximation to the exact real-time quantum dynamics.\footnote{This dissertation does \emph{not} seek to explain RPMD dynamics, instead showing that the instantaneous \shortt\ flux of a ring polymer though a dividing surface is identical to that of a quantum particle, such that RPMD-TST is a true QTST, explained more fully in Secs~\ref{ssec:rpmdimpst} and \ref{ssec:rpmdimp}.} Nevertheless, the RPMD method has also been applied to assess many dynamical properties in addition to thermal rates, such as diffusion \cite{mil05,sul12,mil05water,hab09,hab09com} and X-ray scattering \cite{smi14}.

Braams and Manolopoulos have shown that in the $t\to0$ limit the exact quantum result is obtained when the operators in the RPMD correlation function are linear functions of position \cite{bra06}. However, this does not apply to rates [the ring-polymer flux and side in \eqr{rpmd} being highly non-linear] \cite{man05che} or to other properties of non-linear operators \cite{jan14,hor05,hab13}. 

As RPMD rate theory is in an extended classical phase-space, it shares many properties with classical rate theory, including being rigorously independent of the location of the dividing surface \cite{man05ref}. It also scales linearly with the number of ring-polymer beads $N$ and the dimensionality of the system $F$, allowing simulation of large physical systems such as enzymatic hydride transfer \cite{boe11}. RPMD rate theory therefore generalizes well to multidimensional systems and has been applied to condensed phase \cite{col08,men11,kre13,mar11,vid13}, as well as gas phase \cite{col09,col10,sul11,per12,lix13,lix14,per14}, reactions. It reduces to classical rate theory in the high-temperature, $N=1$ limit, and is exact for a parabolic barrier (at all temperatures for which the parabolic barrier rate is defined)\cite{col09}.

The conservation of the quantum Boltzmann distribution by RPMD is not present in many other competing heuristic QTSTs\cite{liu09}, in which the quantum Boltzmann distribution degrades over time causing spurious effects such as zero-point energy leakage \cite{hab09}.

\subsection{Ring-Polymer Molecular Dynamics Transition-State Theory}
Analogous to classical TST, RPMD-TST is obtained as the short-time limit of the corresponding RPMD flux-side time-correlation function,
\begin{align}
 k_{\rm RPMD}^{\ddag}(\beta) = \lttz \frac{c_{\rm RPMD}(t)}{\Qrb},
\end{align}
and
\begin{align}
  \lttz c_{\rm RPMD}(t) = \tphN \int d\bq \int d\bp \ e^{-\beta_N H_N(\bq,\bp)} S_{N}(\bq,\bp) \delta[f(\bq_0)] h[S_{N}(\bq,\bp)] \eql{rpmd.tst}.
\end{align}
As one might expect, RPMD-TST is a rigorous upper bound to the RPMD rate, such that the optimal ring-polymer dividing surface can be found variationally.

For the case of a centroid dividing surface, $k_{\rm RPMD}^{\ddag}(\beta) = k_{\rm VCM}(\beta)$. Richardson and Althorpe showed that, in the deep tunnelling regime,\footnote{Beneath the crossover temperature [see \eqr{xover}] where the rate is dominated by tunnelling rather than over-the-barrier scattering.} RPMD-TST has a close link with the so-called ``Im $F$'' instanton theory \cite{ric09}, which is widely used as it produces accurate rates for model systems where the exact quantum results are computable for comparison, but has no rigorous derivation \cite{alt11}. 

Consequently, until the work presented in this dissertation, RPMD-TST was regarded as a heuristic QTST with some desirable features,
producing accurate rates for asymmetric systems, and interpolating smoothly between classical TST (the $N=1$, high-temperature
limit) and Im $F$ instanton theory (at low temperatures). 

\section{Summary}
We have explored the properties of classical rate theory, and how classical TST obviates the need for real-time dynamics, but is exponentially sensitive to the location of the dividing surface and fails to account for any quantum effects. Exact quantum rate theory is prohibitively expensive for all but the simplest of systems, and there was considered to be no rigorous version of `quantum transition-state theory', despite numerous efforts to construct one on heuristic grounds. Of the many approximate methods, RPMD-TST appeared to be one of the more promising heuristic QTSTs.

%% file: stl.tex
\chapter[The short-time limit]{The short-time limit: instantaneous thermal flux}
\label{ch:stl}

Having reviewed quantum and classical rate theory, the existence of classical TST and the apparent absence of a QTST, we now show why previous attempts have failed to produce a QTST and how, by alignment of flux and side dividing surfaces in path-integral space, a non-zero QTST can be obtained. 

We initially obtain a QTST corresponding to a rate expression proposed by Wigner on heuristic grounds in 1932, but which produces
poor results at low temperatures. However, by ring-polymerizing the path-integral expression whose short-time limit is the Wigner
rate, and imposing the requirement of positive-definite statistics, we derive RPMD-TST. In doing so we show that RPMD-TST is
equivalent to calculation of the instantaneous thermal quantum flux through a permutationally-invariant dividing
surface.\footnote{By which we mean that the dividing surface is invariant to cyclic permutation of ring-polymer beads.}

While this chapter shows that it is possible to construct a quantum flux-side time-correlation function with a non-zero short-time limit, the demonstration that the resultant expression produces the exact quantum rate in the absence of recrossing by the quantum dynamics, fulfilling the final requirement for a QTST, is presented in chapter~\ref{ch:ltl}.

\section{Apparent absence of QTST}
\label{sec:wigmil}
For algebraic simplicity, we consider a one-dimensional system with coordinate $q$, mass $m$ and Hamiltonian $\hat H$ at an inverse temperature $\beta \equiv 1/k_B T$, as in section~\ref{sec:qmrt}. The results generalize immediately to multi-dimensional systems, as discussed in section~\ref{sec:multid}.

To examine the short-time behaviour of the conventional (Miller-Schwartz-Tromp \cite{mil83}) quantum flux-side time-correlation function, \eqr{milsym}, we expand the trace in the position representation,
\begin{align}
 \cfss =& \int dq \int d\Delta \int dz \ \bra{q-\Delta/2} \ebt \hat F \ebt \ket{q+\Delta/2} \nonumber\\
 &\times \bra{q+\Delta/2} \etb \ket{z}h(z-\qdd)\bra{z} \etf \ket{q-\Delta/2},
 \label{eq:cfss}
\end{align}
and in the short-time limit,
\begin{align}
 \lttz \etf = \etfo e^{-i \hat Vt/\hbar},
\end{align}
where $\hat H_0 = \hat p^2/2m$ is the free particle Hamiltonian and $\hat V$ the potential energy operator. As $\hat V$ is diagonal in the co-ordinate representation,
\begin{align}
 \lim_{t\to 0_+} \bra{y} \etb \kb{z} \etf \ket{x} = \bra{y} \etbo \kb{z} \etfo \ket{x},
 \label{eq:shorttlim}
\end{align}
 and by contour integration,
\begin{align}
 \bra{x} \etfo \ket{y} = & \sqrt{\frac{m}{2\pi i \hbar t}} e^{im(x-y)^2/2\hbar t} \label{eq:etfo}, \\
 \bra{x} \etfo \hat p \ket{y} = & \frac{(x-y)m}{t}\sqrt{\frac{m}{2\pi i \hbar t}} e^{im(x-y)^2/2\hbar t}. \label{eq:etfop}
\end{align}
Inserting \eqr{etfo} (and its complex conjugate) into \eqr{cfss},
\begin{align}
 \lttz \cfss = &  \int dq \int d\Delta \int dz \ \bra{q-\Delta/2} \ebt \hat F \ebt \ket{q+\Delta/2} \nonumber\\
 &\times \frac{m}{2\pi\hbar t} h(z-\qdd) e^{im(z-q)\Delta/\hbar t},
\end{align}
we can define the short-time momentum $p = (z-q)m/t$, such that 
\begin{align}
 \lttz \cfss = & \tph \int dq \int d\Delta \int dp \ \bra{q-\Delta/2} \ebt \hat F \ebt \ket{q+\Delta/2} \nonumber\\
 &\times  h(q+pt/m-\qdd) e^{ip\Delta/\hbar}.
\end{align}
In the short-time limit, the contribution of the momentum to the Heaviside function vanishes, such that $\lttz h(q+pt/m-\qdd) = h(q-\qdd)$. Integrating over $p$ yields a Dirac delta function in $\Delta$, which can itself then be integrated out,
\begin{align}
 \lttz \cfss = & \int dq \ h(q-\qdd) \bra{q} \ebt \hat F \ebt \ket{q},
\end{align}
and a position state (whether acted on by a Boltzmann operator or not) has zero flux, such that
\begin{align}
 \lttz \cfss = 0 \eql{cfssz}
\end{align}
for any system. This result \cite{vot89time,vot93,mil93}, widely recognised since the 1980s, ostensibly precluded a quantum transition-state theory\cite{vot93,mil93,pol05}. 

\section{Derivation of Wigner QTST}
A physical understanding of the vanishing short-time limit in \eqr{cfssz} is possible by inserting further unit operators and dummy time evolution into \eqr{cfss},
\begin{align}
 \cfss =& \int d\bq \int d\bz \int d\bDelta \ h(z_2-\qdd) \hat F(q_1-\qdd) \no \\
 & \times \prod_{i=1}^2 \bra{q_{i-1}-\tfrac{1}{2}\Delta_{i-1}} \ebt \kb{q_i+\tfrac{1}{2}\Delta_i} \etb \ket{z_i} \nonumber \\
 & \qquad \times \bra{z_i} \etf \ket{q_{i}-\tfrac{1}{2}\Delta_{i}},
 \label{eq:twoq}
\end{align}
where $\int d\bq = \inti dq_1 \inti dq_2$, and likewise for $\bz$ and $\bm{\Delta}$.
Taking the short-time limit of \eqr{twoq},
\begin{align}
 \lttz \cfss =& \frac{1}{(2\pi\hbar)^2}\int d\bq \int d\bp \int d\bDelta \ h(q_2 + p_2 t/m - \qdd) \hat F(q_1-\qdd) \no \\
 & \times \prod_{i=1}^2 \bra{q_{i-1}-\tfrac{1}{2}\Delta_{i-1}} \ebt \ket{q_i+\tfrac{1}{2}\Delta_i} e^{i\Delta_i p_i/\hbar}.
 \label{eq:milst} % Miller short-time
\end{align}
Equation~\eqref{eq:milst} demonstrates that the flux and side dividing surfaces are acting at different points in path-integral
space, the Heaviside function at $q_2$ and the flux operator at $q_1$. As with classical TST, a zero result is obtained when the
dividing surfaces are not in the same location, as discussed in section~\ref{sec:cltst}.\footnote{The separation of the dividing
surfaces can also be understood from the delocalization of the Boltzmann operator, which was one of the earliest arguments against
the existence of a QTST \cite{wig38}. Alternatively, by the insertion of position-space identities into \eqr{milst} to form a ring
polymer-like expression [cf.~\eqr{string}], the flux and side dividing surfaces are seen to act in \emph{orthogonal dimensions} in
ring-polymer space.}

\begin{figure}
\includegraphics[angle=270, width=\textwidth]{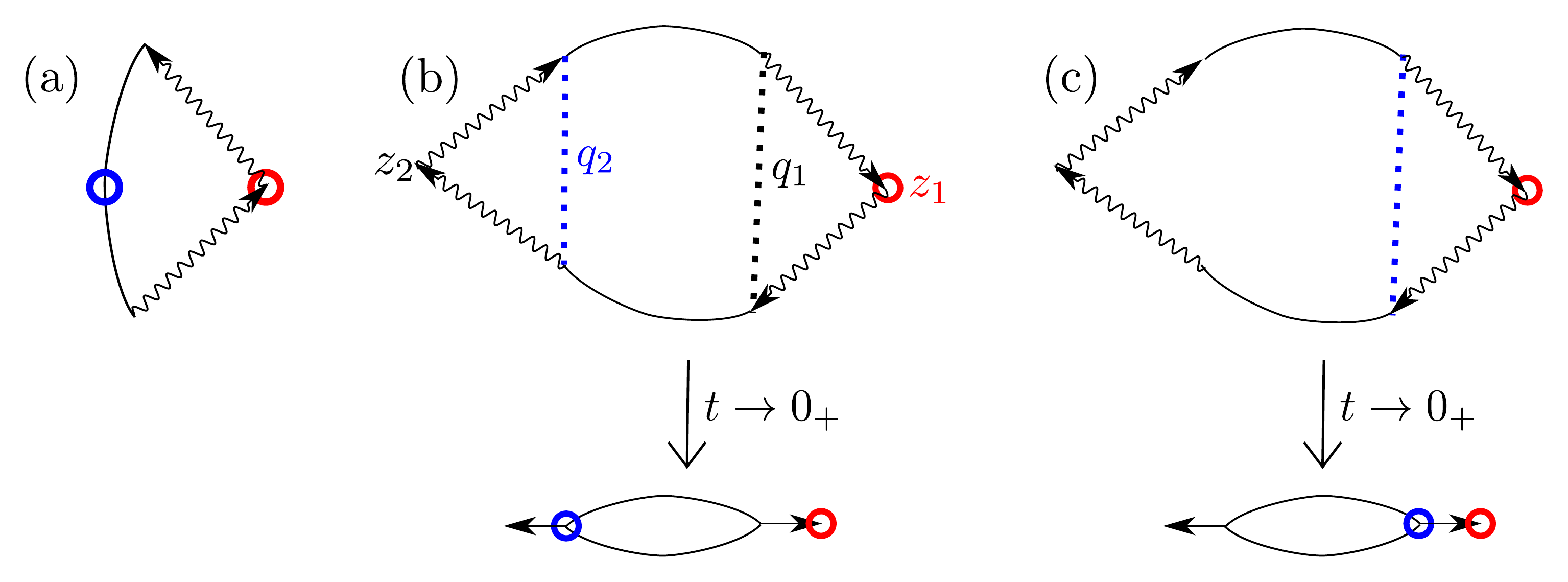}
\caption{Schematic path-integral diagrams demonstrating alignment of the dividing surfaces. (a) corresponds to the MST expression \eqr{cfss}, (b) to \eqr{twoq} and (beneath it) the short time limit, showing how the flux operator (blue circle) is acting at a different point to the side operator (red circle). (c) represents \eqr{cfso} with the dividing surfaces aligned, and the associated short-time form.}
\figl{pidiagram}
\end{figure}

However, if we move the Heaviside dividing surface to be in the same location as the flux dividing surface, such that it becomes $h(z_1 - \qdd)$ (as shown in \figr{pidiagram}), we can integrate out $p_2$, $\Delta_2$ and $q_2$ \cite{hel13}:
\begin{align}
 \Cfsov{t} = & \int dq \int dz \int d \Delta \ h(z-\qdd) \hat F(q-\qdd) \no \\
 & \times \bra{q-\Delta/2} \eb \ket{q+\Delta/2} \bra{q+\Delta/2} \etb \ket{z}\bra{z} \etf \ket{q-\Delta/2},
 \label{eq:cfso} % cfs one
\end{align}
where we have dropped the subscript 1 in the position variables for clarity, and the superscript 1 in $\Cfsov{t}$ corresponds to sampling the flux and side at a single point in imaginary time. The short-time limit of this expression is, using Eqs.~\eqref{eq:etfo} and \eqref{eq:etfop},
\begin{align}
 \lttz \Cfsov{t} = & \tph \int dq \int dp \int d \Delta \ h(p) \delta(q-\qdd) \frac{p}{m} \bra{q-\Delta/2} \eb \ket{q+\Delta/2} e^{i p\Delta/\hbar}.
 \eql{wig1}
\end{align}
This equation is identical to the Wigner rate [\eqr{wig}], introduced in 1932 \cite{wig32uber}, which prior to the research in this dissertation, had no rigorous justification beyond producing the correct rate for a parabolic barrier\footnote{At temperatures above crossover [\eqr{xover}] where the parabolic barrier rate is defined.}. 
The Wigner rate therefore corresponds to the instantaneous thermal quantum flux through a dividing surface, and we demonstrate in chapter~\ref{ch:ltl} that \cite{hel13,alt13}
%Although moving the dividing surface appears to be an \emph{ad hoc} step, it can be justified rigorously by showing that\cite{hel13,alt13}
\begin{align}
 \lim_{t\to \infty}\Cfsov{t} = k^{\rm QM}(\beta)\Qrb \eql{wiglt} % Wigner long time
\end{align}
i.e.~\eqr{cfso} produces the exact quantum rate in the long-time limit (regardless of any recrossing of the dividing surface) and therefore the exact rate in the short time limit in the absence of recrossing (where, by definition, $\Cfsov{t}$ is constant $\forall t>0$), fulfilling the requirement for a QTST.

\begin{figure}[h]
 \centering
 \begin{subfigure}[t]{0.45\linewidth}
 \includegraphics[width=\textwidth]{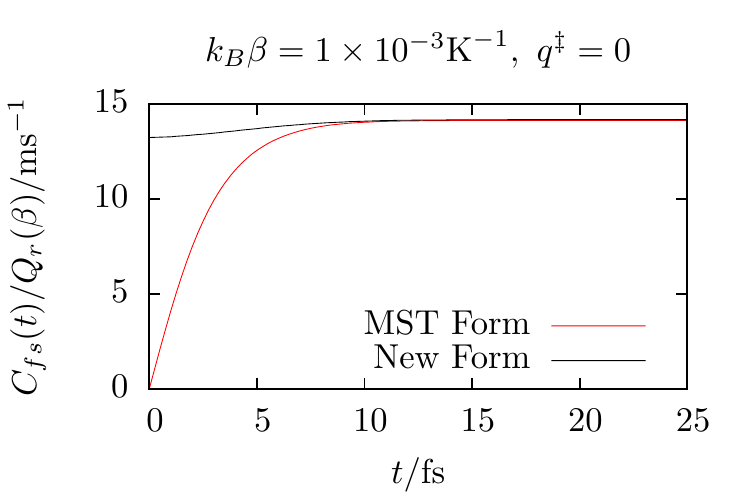}
 \caption{High temperature: Wigner TST is a good approximation to $k_{\rm QM}(\beta)$, though the QTST rate \emph{underestimates} the quantum rate.}
 \figl{b1}
 \end{subfigure}
 \hspace{1mm}
 \begin{subfigure}[t]{0.45\linewidth}
 \includegraphics[width=\textwidth]{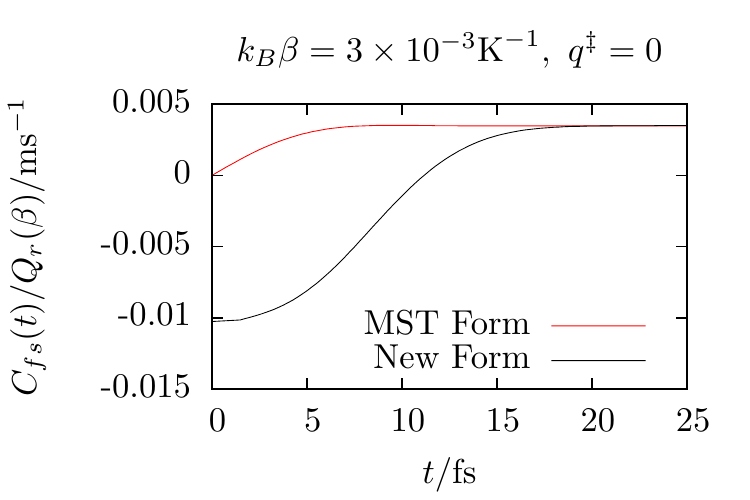}
 \caption{Low temperature (beneath crossover): Spurious statistics lead to erroneous rate. The long-time limit of $\Cfsov{t}$ is still equal to the exact QM rate.}
 \figl{b3}
 \end{subfigure}
 \begin{subfigure}[h]{0.45\linewidth}
 \includegraphics[width=\textwidth]{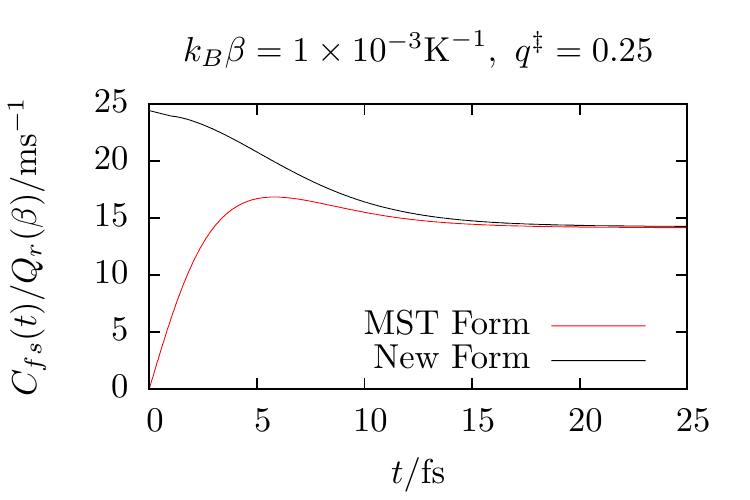}
 \caption{Poor dividing surface: overestimation of the rate and recrossing reduces the QTST result to exact quantum rate.}
 \figl{mqdd}
 \end{subfigure}
 \caption{Illustrative calculations of $\Cfsov{t}$ for the symmetric Eckart barrier.}
 \figl{wigrates}
\end{figure}

\section{Numerical illustration}
To illustrate the QTST we have derived, we evaluate the flux-side function \eqr{cfso} for the symmetric Eckart barrier.\footnote{The
parameters for this system are detailed in Ref.~\cite{hel13}. A numerical calculation will always have a finite
gradient in the $t\to 0_+$ limit due to the impossibility of a `perfect' Dirac delta function, which can only be as narrow as the
spacing of points in the
position-space grid. Consequently, the plots presented here have the numerical simulation of $\Cfsov{t}$ at finite time spliced with
the short-time limit determined from numerically exact evaluation of $\Cfsov{t\to 0_+}$ from \eqr{wig1}.}%, discussed further in
% section~\ref{ssec:stconv}.}. 

We observe in \figr{b1} that at high temperature, the Wigner rate [the QTST of \eqr{wig1}] is a good approximation to the exact quantum rate (given by the long-time limit of $\Cfsov{t}$ and the MST expression). However, unlike classical TST, the exact quantum rate is higher than the QTST rate, such that QTST is not a strict upper bound to the quantum rate; attributable to quantum coherence causing recrossing of the dividing surface. 

Beneath the crossover temperature of $k_{\rm B}\beta = 2.69\times 10^{-3}\textrm{K}^{-1}$ [see \eqr{xover}], \figr{b3} shows the Wigner rate to break down completely, producing a negative result \cite{liu09, and09}. %In fact, the Wigner rate expression is often presented for a parabolic barrier\footnote{The original derivation was for a parabolic barrier as it was obtained by expanding in powers of $\hbar$ and truncating to second order \cite{wig32uber}.}, and as such becomes undefined beneath the crossover temperature, which for this system is $k_B\beta = 2.69\times 10^{-3}\textrm{K}^{-1}$. 
Nevertheless, the exact quantum rate is obtained at long time, as to be expected from \eqr{wiglt}.

Although \figr{b1} shows that the QTST rate can underestimate the exact QM rate, for a general multidimensional system the dividing surface will not be optimal and hence there will be recrossing, such that QTST overestimates the quantum rate. To illustrate this, we calculate \eqr{cfso} for a poor dividing surface,\footnote{The optimal dividing surface is $q^\ddag = 0$.} observing an initial overestimation of the rate, followed by decay to the exact quantum rate as the suboptimal dividing surface is recrossed, as would be expected for the corresponding classical calculation.

% More comprehensive numerical results are presented in chapter~\ref{ch:num}, but at present we explore why a spurious result is obtained at low temperatures, as shown in \figr{b3}.

\section{Non positive-definite statistics}
The quantum transition-state theory we have derived is equivalent to Wigner rate theory and produces the exact result in the absence of recrossing, but is known to fail at low temperatures\cite{liu09,hel13,and09}, as shown in \figr{wigrates}. This is not a fault with the quantum dynamics, as the corresponding flux-side correlation function produces the exact rate at long time [\eqr{wiglt} and \figr{wigrates}]. It is attributable to erroneous quantum statistics.

By a co-ordinate transformation of \eqr{wig1}, where
\begin{align}
 q_0 = &\ q - \Delta/2, \\
 q_N = &\ q + \Delta/2
\end{align}
and inserting unit operators in $q_i,\ i = 1,\ldots,N-1$, we obtain
\begin{align}
 \lttz \Cfsov{t} = & \tph \int d\bq \int dp \ h(p)\delta[\tfrac{1}{2}(q_0+q_{N}) - \qdd] \frac{p}{m} e^{ip(q_{N}-q_0)/\hbar} \nonumber \\
 & \times \prod_{i=0}^{N-1} \bra{q_i} \ebN \ket{q_{i+1}} 
\end{align}
which in the \largeN\ limit becomes
\begin{align}
 \lttz \Cfsov{t} = & \tph \int d\bq \int dp \ h(p)\delta[\tfrac{1}{2}(q_0+q_{N}) - \qdd] \frac{p}{m} e^{ip(q_{N}-q_0)/\hbar} \nonumber \\
 & \times \tphN \int d\bp' e^{-\beta_N \left\{\left[V(q_0) + V(q_N)\right]/2 + \sum_{i=1}^{N-1} V(q_i)\right\}} \no\\
 & \times e^{-\betaN \left[\smiN m(q_i - q_{i-1})^2/2\beta_N^2\hbar^2+ p_{i}^2/2m\right]},\eql{string}
\end{align}
where $\int d\bp' \equiv \int dp_1 \ldots \int dp_{N}$.
Examination of the third line of \eqr{string} shows that we have a \emph{string polymer}, not a ring polymer. There is no section
connecting $q_0$ and $q_N$, which can be as far apart as the springs in $q_i,\ i = 1,\ldots,N-1$ will allow them. 

The value of the integral in \eqr{string} is dominated by the stationary points of the string polymer \cite{ric09}. For a conventional, cyclic ring polymer at temperatures above the crossover temperature $\beta_c$, where
\begin{align}
 \beta < \beta_c \equiv \frac{2\pi}{\hbar\omega_b} \eql{xover}
\end{align}
and $\omega_b$ is the imaginary frequency at the top of the barrier, the stationary point is a collapsed ring polymer (like a single classical bead) at the apex of the barrier. In these high-temperature circumstances, whether one has a polymer string or ring is unlikely to significantly affect the statistics and Wigner rate theory is expected to do well, as seen in \figr{b1}. For conventional ring polymers, when $\beta > \beta_c$, another stationary point emerges, the `instanton', corresponding to a periodic trajectory in imaginary time $\beta \hbar$.\footnote{Equivalent to a periodic classical trajectory of length $\beta \hbar$ on the inverted potential energy surface.} Qualitatively, this corresponds to the springs being sufficiently lax that the polymer `hangs down' off the sides of the barrier.

However, the string polymer in \eqr{string} is not cyclic, but its ends are constrained to be symmetrically distributed around $\qdd$. Furthermore, its spring constant is half that of the conventional ring polymer, so it begins to collapse over the barrier at $\beta = \beta_c/2$. This is the origin of the spurious results for the Wigner rate observed in \figr{b3}; the Boltzmann matrix is dominated by contributions corresponding to a string-polymer hanging over the barrier, as shown in \figr{halfinst}. 

While the Boltzmann matrix itself, $\bra{q}\eb\ket{q'}$, is positive $\forall q,q'$, the momentum-space Fourier transform of the
constrained distribution [$f(\Delta)$ in \figr{halfinst}] contains regions of negative density, which in turn cause the rate to be
negative. Consequently, the nature of the constraint upon the ring polymer, which chooses a single point in imaginary time at which
to sample the flux, leads to statistics which are non positive-definite, such that at sufficiently low temperatures, an erroneous
rate is obtained.

% , due the formation of spurious half-instantons\footnote{Where an instanton is a periodic trajectory in imaginary time of length $\beta \hbar$, and a half-instanton is a trajectory of length $\beta \hbar$ which is half of an instanton at a temperature of $2\beta \hbar$.} beneath the crossover temperature, detailed in \figr{halfinst}. \eqr{wig} represents a ring polymer string (rather than necklace) whose ends are symmetrically constrained around $\qdd$. Above the crossover temperature the dominant contribution to this distribution arises from the (collapsed) instanton centred at the top of the barrier. However, at lower $T$ (higher $\beta$), the instanton stretches across the dividing surface, as is illustrated schematically in \figr{halfinst}. The nature of the constraint therefore makes the distribution non positive-definite.
\begin{figure}[tb]
 \centering
 \includegraphics[width=.6\textwidth]{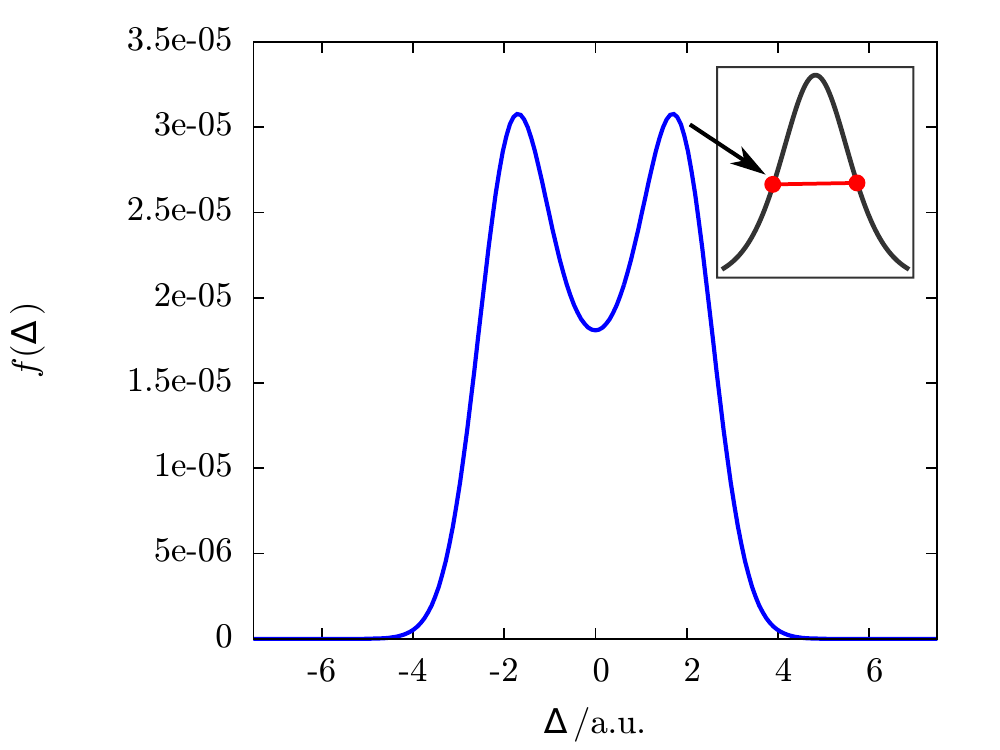}
 \caption{Illustrating the density of the Boltzmann matrix $f(\Delta) = \bra{q-\Delta/2} \eb \ket{q+\Delta/2}$ beneath crossover, where $k_B\beta = 3\times 10^{-3}{\rm K}^{-1}$. Inset is a schematic depiction of the spurious half-instanton producing the bimodal distribution. For the Eckart barrier considered here, $\beta_c = 2.69\times 10^{-3}{\rm K}^{-1}$.}
 \figl{halfinst}
\end{figure}

% The collapse of the half-instanton across the barrier, detailed in the previous section, occurs because the inverse temperature of the bra-ket ($\beta$) is greater than the crossover temperature,
% \begin{align}
%  \beta_c = \frac{2\pi}{\hbar \omega_b}
%  \eql{xover}
% \end{align}
% where $\omega_b$ is the imaginary barrier frequency. For the Eckart barrier considered above, $\beta_c = 2.69\times 10^{-3}{\rm K}^{-1}$. 

\section{Ring-polymerized flux-side form}
\label{sec:rpfsf}
In the previous section we saw how Wigner rate theory was beset by problems at low temperature. However, if one were to ring-polymerize \eqr{cfso} to an expression with $N$ beads, such that the inverse temperature of a Boltzmann bra-ket was $\beta_N \equiv \beta/N$, for any non-zero temperature (finite $\beta$) it would always be possible to increase $N$ to a sufficiently high value that $\beta_N < \beta_c$ and spurious (half) instantons would not occur. 

We therefore construct a ring-polymerized flux-side time-correlation function, and by placing the dividing surfaces in the same place this leads to a non-zero short-time limit and therefore a QTST. Further manipulation shows that, in the limit of infinitely many path-integral beads and when the dividing surface is invariant to their permutation, positive-definite statistics are obtained so the rate is guaranteed to be positive at any finite temperature. Satisfaction of the second requirement of a QTST (producing the exact rate in the absence of recrossing) is reserved for the next chapter.

We begin by taking the side-side form corresponding to \eqr{cfso},
\begin{align}
 C_{\rm ss}^{[1]}(t) = & \int dq \int dz \int d \Delta \ h(z-\qdd) h(q-\qdd) \no \\
 & \bra{q-\Delta/2} \eb \ket{q+\Delta/2} \bra{q+\Delta/2} \etb \ket{z}\bra{z} \etf \ket{q-\Delta/2}
\end{align}
which is ring-polymerized to
\begin{align}
 \Cssnv{t} = & \int d\bq \int d{\bf \Delta} \int d\bz \ h[\fq] h[\fz] \nonumber \\
& \times \prod_{i=0}^{N-1} \bra{q_{i-1}-\tfrac{1}{2}\Delta_{i-1}} \ebN \kb{q_i+\tfrac{1}{2}\Delta_i} \etb \ket{z_i} \nonumber \\
& \qquad \times \bra{z_i} \etf \ket{q_{i}-\tfrac{1}{2}\Delta_{i}} \eql{cssnvt}
\end{align}
where the superscript $N$ in $\Cfsnv{t}$ corresponds to the number of ring-polymer beads. We then differentiate w.r.t.~time, noting that \cite{mil83}
\begin{align}
 \Cfsnv{t} = -\dd{}{t}\Cssnv{t}.
\end{align}
The mathematics is lengthy and presented in full in appendix~\ref{ap:doc}, the eventual result being
\begin{align}
 \Cfsnv{t} = & \int d\bq \int d{\bf \Delta} \int d\bz \ \mathcal{\hat F}[\fq] h[\fz] \nonumber \\
& \times \prod_{i=0}^{N-1} \bra{q_{i-1}-\tfrac{1}{2}\Delta_{i-1}} \ebN \kb{q_i+\tfrac{1}{2}\Delta_i} \etb \ket{z_i} \nonumber \\
& \qquad \times \bra{z_i} \etf \ket{q_{i}-\tfrac{1}{2}\Delta_{i}}, \label{eq:ubfn}
\end{align}
where $\mathcal{\hat F}[\fq]$ is the `ring polymer flux operator',
\begin{align}
 \mathcal{\hat F}[\fq] = \frac{1}{2m} \smiNz \left\{\ddp{\fq}{q_i} \dfs{\fq} \hat p_i + \hat p_i \dfs{\fq} \ddp{\fq}{q_i}\right\}, \label{eq:sfo} % Superflux operator
\end{align}
where the first term in braces is placed between $\ebN \ket{q_i+\tfrac{1}{2}\Delta_i}$ and $\bra{q_i+\tfrac{1}{2}\Delta_i} \etb$,
and the second term between  $\etf \ket{q_{i}-\tfrac{1}{2}\Delta_{i}}$ and $\bra{q_{i}-\tfrac{1}{2}\Delta_{i}} \ebN$.\footnote{There
exist other, equivalent placements of the components of the ring-polymer flux operator, as detailed in appendix~\ref{ap:doc}.} Here
$\fq=0$ defines the dividing surface separating products and reactants, such that
\begin{align}
 \lim_{q\to\infty} f(q,q,\ldots,q) > 0, \eql{divsurfdef} \\
 \lim_{q\to -\infty} f(q,q,\ldots,q) < 0, \eql{divsurfdef2}
\end{align}
and it is also defined to be convergent in the \largeN~limit (in order for the rate to converge).

Equation~\eqref{eq:ubfn}, referred to as the ``Generalized Kubo form'', represents a generalization of a Kubo-transformed\cite{kub57} correlation function, correlating the flux of $N$ imaginary-time paths at time $t=0$ with their side at some later time $t$. To our knowledge, it has not appeared before in the rate theory literature, though the concept of a generalizing the Kubo transform for the computation of correlation functions of non-linear operators has been suggested previously\cite{rei00}.

\subsection{The short-time limit}
Taking the short-time limit of \eqr{ubfn}, we obtain
\begin{align}
 \lttz \Cfsnv{t} = & \tphN \int d\bq \int d{\bf \Delta} \int d\bp \ \dfq S(\bq,\bp) h[f(\bq + \bp t/m)] \nonumber \\
& \times \prod_{i=0}^{N-1} \bra{q_{i-1}-\tfrac{1}{2}\Delta_{i-1}} \ebN \ket{q_i+\tfrac{1}{2}\Delta_i} e^{i\Delta_i p_i/\hbar}
\eql{ubfst}
\end{align}
where we have made the substitution $p_i = (z_i - q_i)m/t$, and 
\begin{align}
 S_N(\bq,\bp) = \frac{1}{m} \smiNz \ddp{\fq}{q_i} p_i \eql{sqpdef}
\end{align}
is the flux perpendicular to $\fq$.

In the short-time limit, $f(\bq + \bp t/m)$ can be Taylor-expanded such that
\begin{align}
 \lttz \dfq h[f(\bq + \bp t/m)] = & \dfq h\left[\fq + \frac{t}{m}\smiNz \ddp{\fq}{q_i} p_i \right] \no \\
 = & \lttz \dfq h\left[\frac{t}{m}\smiNz \ddp{\fq}{q_i} p_i\right] \no\\
 = & \ \dfq h[S_N(\bq,\bp)] \eql{dfhf} %Delta function Heaviside function
\end{align}
where we have noted that the Heaviside function is invariant to the scaling of its argument and that the Dirac delta function holds
$\fq = 0$. Consequently, \eqr{ubfst} produces a finite result in the \shortt~limit, fulfilling one criterion of a QTST (the other
being equivalence to the exact quantum rate in the absence of recrossing, which is explored in the next chapter).\footnote{If the
dividing surfaces were different functions of path-integral space [as was the case for the MST correlation function $\cfss$], the
result in \eqr{dfhf} would not hold and the contribution of the momentum term to the Heaviside function would be switched off
smoothly as \shortt, leading to a zero QTST.} The above will hold for any value of $N$,\footnote{Consequently there appear to be an
infinite number of non-zero QTSTs with different values of $N$; in chapter~\ref{ch:unique} we show that there are an infinity of
QTSTs for every value of $N \ge 1$, though only in the \largeN\ limit is \eqr{ubfn} positive-definite and therefore of practical
use.} so we can therefore define a quantum transition-state theory as
\begin{align}
 k_{\rm QM}^{\ddag}(\beta) = \lNti \lttz \Cfsnv{t}/\Qrb \eql{qtstdef}
\end{align}
where the purpose of the \largeN\ limit will become apparent later.

\subsection{Normal mode transformation}
\label{ssec:nmt}
Equation~\eqref{eq:ubfst} possesses an $N$-dimensional Fourier transform, so, \emph{prima facie}, is even more expensive to compute than the Wigner expression [\eqr{wig}] that we started from. However, $N-1$ of the Fourier transforms can be eliminated by using a normal mode transformation 
\begin{align}
 (\bp, \bDelta) & \to ({\bf \tilde p}, \bm{\tilde \Delta}), 
\end{align}
where
\begin{align}
 \tilde p_j = & \smiNz p_i T_{ij} \\
 \tilde \Delta_j = & \smiNz \Delta_i T_{ij}
 \end{align}
and
\begin{align}
 T_{i0} = & \frac{1}{\sqrt{B_N}} \ddp{\fq}{q_i} \eql{tdef} \\
 B_N = & \smiNz \left[ \ddp{\fq}{q_i} \right]^2 \eql{bndef}.
\end{align}
The other normal modes are defined to be orthogonal to $T_{i0}$ and their exact form need not concern us further. Applying the transformation and noting that the Jacobian is unity,
\begin{align}
\lttz \Cfsnv{t} = & \tphN \int d\bq \int d\btDelta \int d\btp \ \dfq \frac{\tilde p_0}{m} h(\tilde p_0) \sqrt{B_N}\nonumber \\
& \times \prod_{i=0}^{N-1} \bra{q_{i-1}-\tfrac{1}{2}\smjNz T_{i-1\ j}\tilde \Delta_{j}} \ebN \ket{q_i+\tfrac{1}{2}\smjNz T_{ij}\tilde \Delta_{j}} e^{i \tilde \Delta_i \tilde p_i/\hbar}.
\end{align}
The momenta $\tilde p_i,\ i=1,\ldots,N-1$ can be integrated out, leading to $N-1$ Dirac delta functions in $\tilde \Delta_i,\ i=1,\ldots,N-1$, which themselves are integrated over,
\begin{align}
\lttz \Cfsnv{t} = & \tph \int d\bq \int d \tilde \Delta_0 \int d\tilde p_0 \ \dfq \frac{\tilde p_0}{m} h(\tilde p_0) e^{i \tilde \Delta_0 \tilde p_0/\hbar} \sqrt{B_N}\nonumber \\
& \times \prod_{i=0}^{N-1} \bra{q_{i-1}-\tfrac{1}{2} T_{i-1\ 0}\tilde \Delta_{0}} \ebN \ket{q_i+\tfrac{1}{2} T_{i0}\tilde \Delta_{0}} .
\eql{onep}
\end{align}
One Fourier transform remains in the `ring-opening' mode $\tilde \Delta_0$, and in Appendix~\ref{ap:rom} we show that, in the
\largeN\ limit and when $\fq$ is invariant with respect to (w.r.t.)~permutation of the ring-polymer beads, this can be integrated
out, yielding
\begin{align}
  \lttz \Cfsnv{t} = & \tph \int d\bq \int d\tilde p_0 \ \delta[\fq] \frac{\tilde p_0}{m} h(\tilde p_0) e^{-\beta_N \tilde p_0^2/2m} \sqrt{B_N} \nonumber \\
& \times \sqrt{\frac{2\pi\beta_N\hbar^2}{m}}\prod_{i=0}^{N-1} \bra{q_{i-1}} \ebN \ket{q_i}. \eql{qponly}
\end{align}
The only linear permutationally-invariant dividing surface is the centroid \cite{hel13}, defined in \eqr{centroiddef}. However, for systems beneath the crossover temperature the optimal dividing surface may involve other normal modes of the ring polymer and take a conical form \cite{hel13, ric09}.

\subsection{Emergence of RPMD-TST}
We now reinstate $N-1$ momentum integrals to \eqr{qponly}, transform back from normal modes, and expand the Boltzmann bra-kets as
\begin{align}
 \bra{q_{i-1}} \ebN \ket{q_i} = \sqrt{\frac{m}{2\pi\beta_N\hbar^2}} e^{-\beta_N\{[V(q_{i-1}) + V(q_i)]/2 + m(q_{i} - q_{i-1})^2/2\beta_N^2\hbar^2\}},
\end{align}
leading to
\begin{align}
 \lttz \Cfsnv{t} = \tphN \int d\bq \int d\bp \ e^{-\beta_N H_N(\bq,\bp)} \dfq S_N(\bq,\bp) h[S_N(\bq,\bp)] \eql{rpmdtst}
\end{align}
where
\begin{align}
 H_N(\bq,\bp) = \smiNz \frac{p_i^2}{2m} + \frac{m(q_{i} - q_{i-1})^2}{2\beta_N^2\hbar^2} + V(q_i)
\end{align}
is the classical ring-polymer Hamiltonian and $S_N(\bq,\bp)$ is the ring-polymer velocity perpendicular to the dividing surface
$\fq$ given in \eqr{sqpdef}. Remarkably, \eqr{rpmdtst} is \emph{identical} to RPMD-TST \cite{hel13}
\begin{align}
 k_{\rm QM}^\ddag (\beta) \myeq & \lttz \lNti \Cfsnv{t} / \Qrb \no\\
 \equiv & \ k_{\rm RPMD}^\ddag(\beta) \eql{identical}
\end{align}
where $k_{\rm RPMD}^\ddag(\beta)$ is defined in \eqr{rpmd.tst} and $k_{\rm QM}^\ddag (\beta)$ in \eqr{qtstdef}.

It is also possible to integrate out momenta completely from \eqr{qponly}, to obtain an expression similar to centroid-TST\cite{vot89rig}, but with a generalized dividing surface, 
\begin{align}
\lttz \Cfsnv{t} = & \frac{1}{\sqrt{2\pi\betaN m}}\int d\bq \ \sqrt{B_N} \dfq \prod_{i=0}^{N-1} \bra{q_{i-1}} \ebN
\ket{q_i}.\eql{simpqtst}
\end{align}

\section{Multidimensional generalization}
\label{sec:multid}
Here we sketch how the results from earlier in the chapter can be generalized to multidimensional systems, and thereby the condensed phase, provided that there is sufficient separation of timescales between reaction and equilibration \cite{cha78}.
For a system with $F$ dimensions, there are $N$ copies of the system with co-ordinates $\bq = \{ \bq_1,\ldots,\bq_N\}$, where $\bq_i = \{ q_{i,1},\ldots,q_{i,F} \}$. Here $q_{i,j}$ is the scalar co-ordinate of the $j$th dimension of the $i$th bead, with $\bDelta,\ \bz$ and so on similarly defined.

The bra-ket states then become $F$-co-ordinate \cite{hel13};
\begin{align}
 \ket{q_i - \Delta_i/2} \to \ket{q_{i,1} - \Delta_{i,1}/2,\ldots, q_{i,F} - \Delta_{i,F}/2}
\end{align}
as does the ring-polymer flux operator,
\begin{align}
 \mathcal{\hat F}[\fq] =  \sum_{j = 0}^{F-1}\frac{1}{2m_j} \smiNz \left\{\ddp{\fq}{q_{i,j}} \dfs{\fq} \hat p_{i,j} + \hat p_{i,j} \dfs{\fq} \ddp{\fq}{q_{i,j}}\right\} \eql{multidflux}
\end{align}
where $m_j$ is the mass in the $j$th dimension.

One takes the short-time limit as before and finds that in the \largeN~limit, and with a dividing-surface which is invariant to imaginary-time translation, RPMD-TST in $F$ dimensions is obtained \cite{hel13,man05ref}. 

\section{Interpretation}

The central result of this chapter is that it is possible to construct a quantum flux-side time-correlation function with a non-zero limit, which was previously considered not to exist and cited as one of the main reasons for the absence of a QTST \cite{mil93,vot89time,vot93}. 

The key step in obtaining a non-zero QTST was the alignment of the dividing surfaces in path-integral space. Previously (in the MST and other flux-side time-correlation functions) the flux and side dividing surfaces were in different places, leading to a vanishing rate in the short-time limit, as would also be expected for the classical case. Performing this to the standard MST flux-side correlation function led to a QTST expression previously introduced by Wigner in 1932\cite{wig32uber}, but fails in the low-temperature regime, where a QTST would be of most interest\cite{and09,liu09,hel13}. 

By ring-polymerizing the resulting expression and choosing the $N\to\infty$ limit, we prevent the formation of spurious half-instantons which lead to negative rates, and also allow the Boltzmann bra-kets to be expanded analytically, from which we observe that the dividing surface function $\fq$ must be permutationally invariant. If not, then one is effectively privileging a point in imaginary time arbitrarily, leading to non positive-definite statistics.\footnote{The earliest attempt at reaction rate calculation from RPMD \cite{man05che} also produced poor statistics, which were removed by the use of a permutationally invariant dividing surface \cite{man05ref}.} Further algebraic manipulation then leads to RPMD-TST.

\subsection{The uncertainty principle}
Other arguments for the absence of QTST have centred on the uncertainty principle \cite{wig38, tru96}, namely the difficulty of knowing the location and momentum of a quantum particle simultaneously and exactly. Classical TST was conceived as measuring the momentum, and thereby flux, of a particle constrained to the top of the potential barrier. This appeared to require simultaneous specification of position and momentum, shown in \figr{cltst}.\footnote{However, by projecting out motion perpendicular to the dividing surface, it is possible to integrate momenta out of \eqr{clastst} leading to a term corresponding to the classical flux of a free particle, such that, even in the classical case, one need not specify the position and momentum of a single particle simultaneously.}

QTST, given by \eqr{simpqtst}, corresponds to the thermal reactive flux at inverse temperature $\beta$ multiplied by the free energy
of the quantum particle constrained to the dividing surface by $\dfq$. The flux of a free particle with momentum $p$ is $p/m$
% \begin{align}
%  \bra{p} \hat F \ket{p} = \frac{p}{m}
% \end{align}
and its momentum is known precisely; the normalized thermal reactive flux therefore being $1/\sqrt{2\pi\beta m}$.
% \begin{align}
%  F(\beta) = & \frac{\int_0^\infty dp \ e^{-\beta p^2/2m} \bra{p} \hat F \ket{p}}{\inti dp \ e^{-\beta p^2/2m} \bra{p} p\rangle} \\
%  = & \frac{1}{\sqrt{2\pi\beta m}}
% \end{align}
However, the position of the free particle is completely undefined, thereby satisfying the uncertainty principle for the flux term.
Concerning the free energy term, the quantum particle is not constrained to a single point in phase space, but its representation as
a ring-polymer is confined to an $N-1$-dimensional surface, being constrained there by $\dfq$. Consequently, there is uncertainty
caused by fluctuations of the ring polymer, both in the beads' positions and momenta, shown in \figr{qtst}.\footnote{As the
temperature is lowered and $\beta$ rises, the spring constant of the ring polymer $\omega_N = 1/\beta_N\hbar$ decreases and the
ring-polymer stretches, increasing the delocalization at lower temperatures, as to be expected from increased delocalization of the
quantum Boltzmann operator.}

Alternatively, one could consider $\tilde q_0/\sqrt{N}$ and $\tilde p_0/\sqrt{N}$ to represent the position and momentum respectively of a classical-like particle, for which one is calculating the TST rate. However, the underlying $q_i$ and $p_i$ which constitute the ring polymer are still subject to quantum mechanical uncertainty.

\newlength{\figwidth}
\setlength{\figwidth}{0.45\columnwidth}
\begin{figure}[tb]
 \centering
 \begin{subfigure}[t]{\figwidth}
  \includegraphics[angle=270,width=\figwidth]{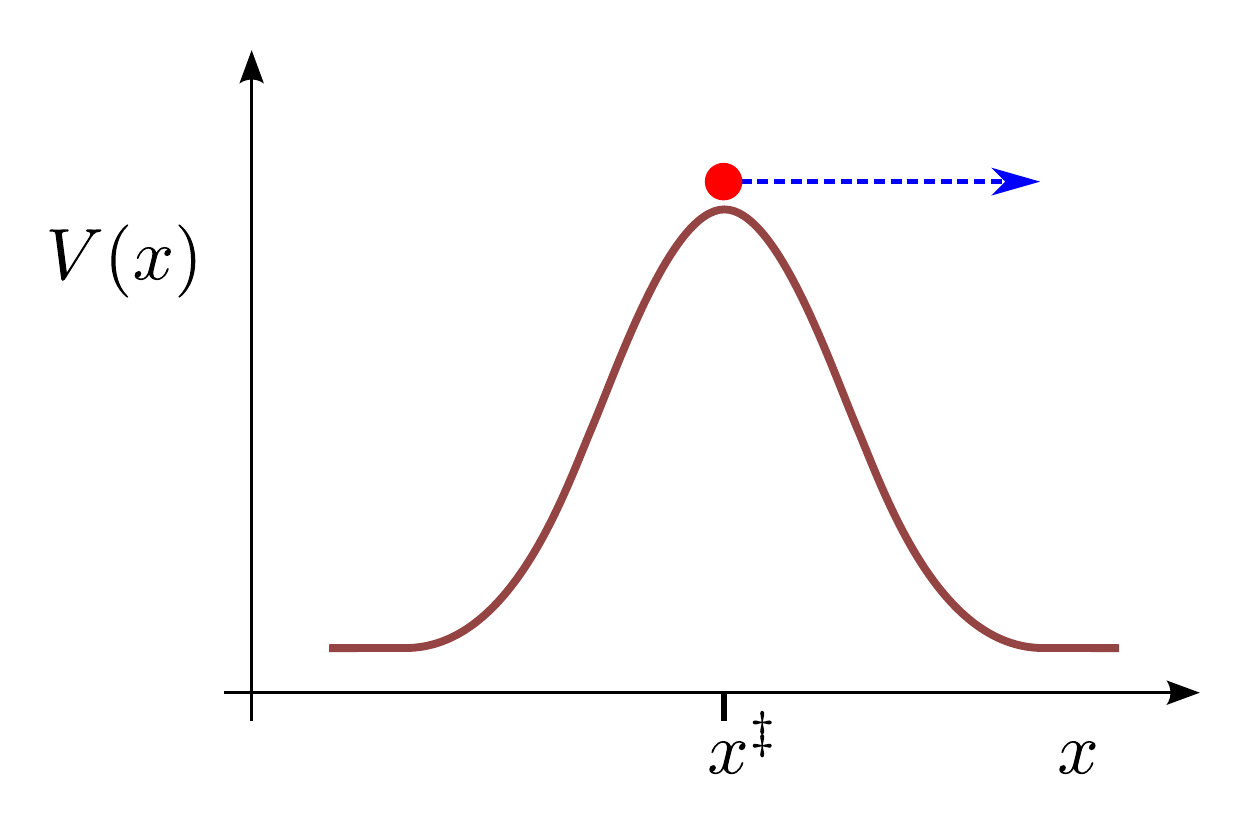}
  \caption{Classical TST: well-defined position of classical particle (filled red circle) at the dividing surface and well-defined momentum along reaction co-ordinate. The classical TST rate is instantaneous classical flux past~$x^\ddag$.}
  \label{fig:cltst}
 \end{subfigure}
 \hspace{0.5cm}
%
%  \subfloat[Classical TST: well-defined position of classical particle (filled red circle) at the dividing surface and well-defined momentum along reaction co-ordinate. The classical TST rate is instantaneous classical flux past~$x^\ddag$.]{
%  \resizebox{\figwidth}{!}{\includegraphics[angle=270]{cltstcol.pdf}}
%  \label{fig:cltst}
%  }
\begin{subfigure}[t]{\figwidth}
 \includegraphics[angle=270,width=\figwidth]{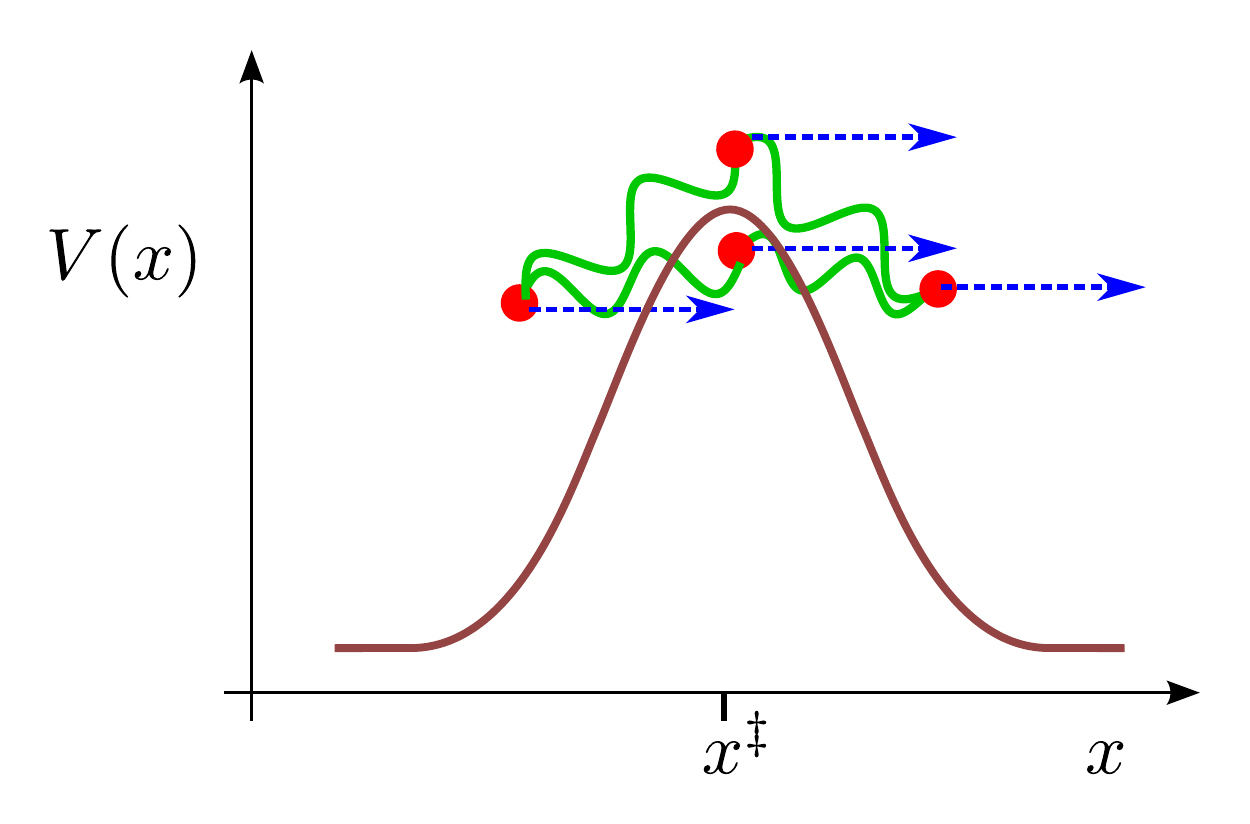}
 \caption{Quantum TST: Quantum particle represented as a ring polymer, the classical `beads' (filled red circles) connected by harmonic springs (wavy green lines). The QTST rate is the instantaneous, collective flux of the beads from configurations constrained at $x^\ddag$.}
 \label{fig:qtst}
\end{subfigure}

%  \hspace{0.5cm}
%  \subfloat[Quantum TST: Quantum particle represented as a ring polymer, the classical `beads' (filled red circles) connected by harmonic springs (wavy green lines). The QTST rate is the instantaneous, collective flux of the beads from configurations constrained at $x^\ddag$.]{
%  \resizebox{\figwidth}{!}{\includegraphics[angle=270]{qtstcol.pdf}}
%  %\label{fig:qtst}
%  }
 \caption{Schematic illustrations of classical and quantum TST for a one-dimensional potential $V(x)$ (thick brown line), with co-ordinate $x$ and a dividing surface located at $x = x^{\ddag}$. Momentum is represented by a dashed blue arrow in both cases.}
 \label{fig:qmcltst}
\end{figure}

\subsection{Implications for RPMD}
\label{ssec:rpmdimpst}
In numerical simulations, the RPMD rate is calculated whereby the ring-polymer is evolved under its fictitious Hamiltonian in order to calculate the `transmission coefficient', the ratio between the RPMD-TST and RPMD rate:
\begin{align}
 k_{\rm RPMD}(\beta) = \ltti k_{\rm RPMD}^{\ddag}(\beta) \kappa(t).
\end{align}
As recrossing by the ring-polymer dynamics can only reduce the rate, $\kappa(t) \leq 1$ for any system.
Defining the optimal dividing surface \cite{ric09} as the one which minimizes recrossing and therefore maximises $\kappa(t)$, and denoting this with an asterisk,
\begin{align}
 k_{\rm RPMD}(\beta) = \ltti k_{\rm RPMD}^{\ddag *}(\beta) \kappa^*(t) \eql{optds}
\end{align}
and for systems where the optimal dividing surface has minimal recrossing $\kappa^*(t) \simeq 1$.

This chapter has not sought to justify the fictitious RPMD dynamics, which are generally regarded as \emph{ad hoc} \cite{man04,hab09,jan14}, it being sufficient to know that they preserve the quantum Boltzmann distribution. Instead, we have shown that the instantaneous thermal flux of a ring polymer is identical to the instantaneous thermal flux of a quantum particle. Consequently, by combining \eqr{identical} and \eqr{optds}, 
\begin{align}
 k_{\rm QM}^{\ddag}(\beta) = \frac{k_{\rm RPMD}(\beta)}{\ltti \kappa^*(t)}
\end{align}
i.e.~provided that there is minimal recrossing of the optimal dividing surface by the (fictitious) RPMD dynamics [$\kappa^*(t) \simeq 1$], the RPMD simulation will be a good approximation (and a strict lower bound) to the instantaneous thermal quantum flux past the statistical bottleneck.\footnote{The region on the potential surface which has the greatest potential of mean force along the minimum energy path from reactants to products. Classically, this would be the saddle point.}

Relating $k_{\rm RPMD}(\beta)$ and $k_{\rm QM}(\beta)$ is discussed in section~\ref{ssec:rpmdimp}, after demonstrating that the QTST derived above produces the exact quantum rate in the absence of recrossing (by the exact quantum dynamics).

\subsection{Connection with alternative rate theories}
\label{ssec:art}
The derivation in this chapter has explained the origin of Wigner rate theory and RPMD-TST (along with its precursors Voth-Chandler-Miller rate theory and RPMD rate theory). It can also suggest the utility of other rate theories, such as rate theories obtained from the linearized semiclassical initial value representation (LSC-IVR) \cite{liu09,shi03} which, while useful (and arguably superior to RPMD for the calculation of spectra \cite{hab09}), employ dynamics which do not conserve the quantum Boltzmann distribution and whose accuracy is likely to degrade at lower temperatures as longer periods of time evolution are required for the flux-side function to reach the plateau region \cite{hel13}.

Prior to the publication of the work presented in this chapter, the best explanation for the success of RPMD rate theory and RPMD-TST at low temperatures arose from its connection to semiclassical instanton theory, which itself has no rigorous derivation \cite{alt11}. Richardson and Althorpe showed that \cite{ric09}
\begin{equation}
 k_{\rm inst}(\beta) = \alpha(\beta) k_{\rm RPMD}^{\ddag}(\beta) \eql{inst}
\end{equation}
where 
\begin{align}
 \alpha(\beta) = \frac{2 \pi}{\beta \hbar} \sqrt{\frac{m}{F''(0)}} \eql{alpha}
\end{align}
and $F''(0)$ is the double derivative of the free energy along the unstable degree of freedom (the saddle point). Amongst other insights, their work suggested that for model 1-dimensional systems, the instanton rate was superior to RPMD-TST; i.e.~it was a closer approximation to the exact quantum mechanical rate than RPMD-TST. 

They also showed that RPMD-TST underestimated the instanton rate for symmetric systems and overestimated it for asymmetric systems,
explaining to some extent the numerically observed tendency for RPMD to underestimate exact quantum rates for symmetric systems, and
the converse for asymmetric systems. The numerical illustration in this chapter 
%(and the more detailed results in chapter~\ref{ch:num}) 
corroborate this for the symmetric Eckart barrier, since quantum recrossing can cause the QTST rate to
underestimate the exact quantum rate.\footnote{Strictly speaking, the results presented in \figr{wigrates} are for the $N=1$ limit
of $\Cfsnv{t}$ whereas RPMD-TST only emerges in the \largeN\ limit, but at high temperatures such as in \figr{b1} the Wigner rate is
very close to that of RPMD-TST.} %(see \figr{wigrpmdqm}).}

Having derived RPMD-TST, derivation of the proportionality factor in \eqr{inst}, or some other explanation for the success of instanton theory is a matter for future research.

\section{Conclusions}
The key result of this chapter is the demonstration that a quantum flux-side time-correlation function exists with a non-zero short-time limit, which represents the instantaneous thermal quantum flux through a dividing surface and therefore is a true QTST, despite previous assertions that one did not exist\cite{mil93,vot89time,vot93,wig39,sma05}. The initial result led to Wigner rate theory, whose spurious low-temperature results can be avoided by constructing a Generalized Kubo form, and taking the limit of an infinite number of path-integral beads. In doing so we obtain a positive-definite QTST that, remarkably, is identical to RPMD-TST, which was previously regarded as an interpolative theory which produced the correct rate in the classical and parabolic barrier limits\cite{man05ref} and had a link to semiclassical instanton theory \cite{ric09, alt11}. 

In the following chapter we show how $\Cfsnv{t}$ produces the exact quantum rate in the absence of recrossing of the dividing
surface (nor of surfaces orthogonal to it in path-integral space), thereby fulfilling the final requirement for a QTST. 

%% file: ltl2.tex
\chapter[The long-time limit]{The long-time limit: effects of no recrossing}
\label{ch:ltl}
Having constructed a positive-definite quantum flux-side time-correlation function which possesses a non-zero short-time limit [\eqr{ubfn}], in this chapter we demonstrate that in the absence of recrossing of the dividing surface $\fq$ by the exact quantum dynamics, and of any dividing surfaces orthogonal to it in path-integral space, this is equal to the exact quantum rate. 

In doing so we also show that the expression leading to the Wigner rate, $\Cfsov{t}$ in \eqr{cfso} also produces the exact rate in
the absence of recrossing, a result stated without proof in chapter~\ref{ch:stl}.

Our task is therefore to prove
\begin{align}
 k_{\rm QM}^{\ddag}(\beta) = k_{\rm QM}(\beta)_{\rm NR}
\end{align}
where the NR subscript denotes No Recrossing, and $k_{\rm QM}^{\ddag}(\beta)$ is defined from \eqr{qtstdef} as $\lNti \lttz \Cfsnv{t}/\Qrb $. As in classical rate theory, no recrossing is (by definition) no net flux across the dividing surface \cite{sun98,hel13unique},
\begin{align}
 C_{\rm ff}^{[N]}(t)_{\rm NR} = 0 \qquad \forall t>0_+ \eql{ffz} % flux-flux zero
\end{align}
where $C_{\rm ff}^{[N]}(t)_{\rm NR}$ is the flux-flux correlation function, and since \cite{mil74,mil83}
\begin{align}
 \Cfsnv{t} = \int_0^{t} dt' C_{\rm ff}^{[N]}(t'), \eql{ffi} % flux-flux integral
\end{align}
this is equivalent to 
\begin{align}
  \lttz \Cfsnv{t}_{\rm NR} = \ltti \Cfsnv{t}_{\rm NR}, \eql{nrzi} % No Recrossing Zero Infinity
\end{align}
so our task can be equivalently stated as proving
\begin{align}
 \lNti \ltti \Cfsnv{t}_{\rm NR}/\Qrb = k_{\rm QM}(\beta)_{\rm NR}. \eql{tobeproven}
\end{align}

We begin by detailing the scattering theory used in this chapter and obtaining the exact quantum rate $k_{\rm QM}(\beta)$ from the long-time limit of the Miller-Schwartz-Tromp flux-side expression, \eqr{cfss} \cite{mil83}. 
We then take the long-time limit of $\Cfsnv{t}$, which is represented as an integral over an $N$-dimensional hypercube of scattering momenta. We demonstrate that the hypercube is composed of a series of Dirac delta function spikes running along paths corresponding to equal energies of the scattering eigenstates, and residues whose contribution vanishes in the $N\to\infty$ limit. 

In general systems with recrossing, these spikes mean that
\begin{align}
  \lNti \ltti \Cfsnv{t}/\Qrb \neq k_{\rm QM}(\beta),
\end{align}
but when there is no recrossing of the dividing surface $\fq$ by the quantum dynamics nor of any dividing surfaces orthogonal to it in path-integral space, all spikes except those corresponding to all scattering eigenstates moving with equal sign and magnitude of momentum vanish, such that the entire density in the hypercube is localized along the `centroid' axis.\footnote{Defined as the axis running through the hypercube where all long-time scattering momenta $p_i$ are equal.} This allows us to rotate the dividing surface anywhere, so long as it cuts out the half of the centroid spike corresponding to positive (product) momenta, and we choose a dividing surface which leads to a hybrid between the MST expression and the generalized flux-side form $\Cfsnv{t}$. This `hybrid' equation can be shown to produce the exact quantum rate in the long-time limit, regardless of recrossing or not, thereby completing the proof.

The chapter then explains how the theory may be generalized to multidimensional systems and discusses implications for RPMD rate theory and QTST before conclusions are presented.

\section{Preliminary quantum scattering theory}
\label{sec:scat}
By taking the long-time limit of the Miller-Schwartz-Tromp form we obtain the exact quantum rate expression. Initially expanding \eqr{cfss} in the position representation,
\begin{align}
 \cfss = \int dx \ \bra{x} \etf \ebt \hat F \ebt \etb \ket{x} h(x) \eql{posrep}
\end{align}
we note that, using \eqr{etfo},
\begin{align}
 \int dx \ \ket{x} h(x) \bra{x}  = \ltti \int dp \ \etfo \ket{p} h(p) \bra{p} \etbo \eql{xtop}
\end{align}
and that \cite{tay06,alt13}
\begin{align}
\int dp \ltti \etb \etfo \ket{p}h(p) \bra{p} \etbo \etf = & \int dp \ \hat \Omega_- \ket{p} h(p) \bra{p} \hat \Omega_-^\dag \no \\
 = & \int dp \ h(p) \kb{\psi_p} \eql{molint}
\end{align}
where $\hat \Omega_-$ is the M\o ller operator\cite{mol45},
\begin{align}
 \hat \Omega_- = \ltti \etb \etfo
\end{align}
corresponding to the scattering eigenstate with outgoing conditions and asymptotic momentum $p$ \cite{tay06},
\begin{align}
 \lim_{x\to\infty} \langle x \ket{\psi_p} = \bk{x}{p} + R(p) \bk{x}{-p}
\end{align}
where $R(p)$ is the anticausal reflection coefficient and 
\begin{align}
 \bk{x}{p} = \frac{1}{\sqrt{2\pi\hbar}} e^{ipx/\hbar}.
\end{align}
Applying \eqsr{xtop}{molint} to \eqr{posrep},
\begin{align}
 \ltti \cfss = \int dp \ h(p) \bra{\psi_p} \ebt \hat F \ebt \ket{\psi_p},
\end{align}
and as $\ket{\psi_p}$ are eigenstates of the Boltzmann operator,
\begin{align}
 \ebt \ket{\psi_p} = \ket{\psi_p} e^{-\beta p^2/4m} \eql{boltzeigen}
\end{align}
such that
\begin{align}
 \ltti \cfss = \int_0^\infty dp \ e^{-\beta p^2/2m} \bra{\psi_p} \hat F \ket{\psi_p} %\eql{exactlt}
\end{align}
and the exact quantum mechanical rate is given by \cite{mil74}
\begin{align}
 k_{\rm QM}(\beta) = \frac{1}{\Qrb}\int_0^\infty dp \ e^{-\beta p^2/2m} \bra{\psi_p} \hat F \ket{\psi_p}. \eql{exactlt}
\end{align}

\section{Long-time limit of the Generalized Kubo Form}
For generality, we consider here the case of a different dividing surface in the flux and side [$\fq$ and $\gz$ respectively]. From the arguments in the previous chapter, unless $\fq \equiv \gq$ the corresponding flux-side function will possess a zero short-time limit, but recrossing of $\gz$ at finite time can cause the long-time limit to be non-zero.

Taking the long-time limit of \eqr{ubfn} with this modification, we obtain 
\begin{align}
\ltti \Cfsnv{t} = & \int d\bq \int d{\bf \Delta} \int d\bp \ \mathcal{\hat F}[\fq] h[\bar g(\bp)] \nonumber \\
& \times \prod_{i=0}^{N-1} \bra{q_{i-1}-\tfrac{1}{2}\Delta_{i-1}} \ebN \kb{q_i+\tfrac{1}{2}\Delta_i} \ppi \rangle \nonumber \\
& \qquad \times \langle \ppi \ket{q_{i}-\tfrac{1}{2}\Delta_{i}}, \eql{cfslt}
\end{align}
where we define\footnote{For the special case of a centroid, $g(\bp) = \bar g(\bp)$ but for a more general curvilinear dividing surface this will not be the case. However, the function $\gz$ [or $\fq$] must converge in this limit to adequately separate reactants and products [see \eqsr{divsurfdef}{divsurfdef2}].}
\begin{align}
 \bar g(\bp) = \ltti g(\bp t/m)
\end{align}
and $\int d\bp = \inti dp_0 \ldots \inti dp_{N-1}$ and likewise foe $\bq$ and $\bDelta$.
\begin{figure}[tb]
 \centering
 \includegraphics[angle=270,width=0.5\textwidth]{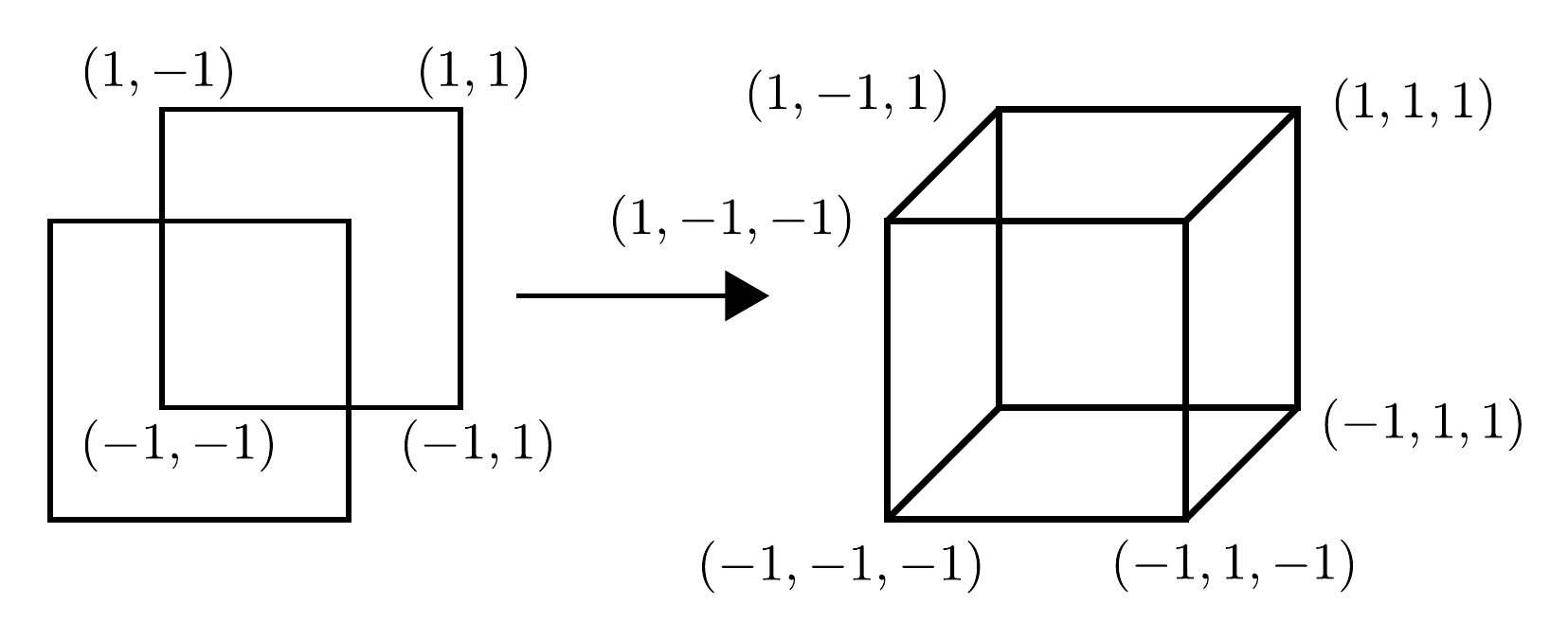}
 \caption{Forming an $N$-dimensional hypercube from two ($N-1$)-dimensional hypercubes, illustrated for $N=3$ (forming a cube by
connecting two squares at their vertices). The notation is discussed in section~\ref{ssec:sysimp} using the axes in
\figr{hexsquare}.}
 \figl{spinenote}
\end{figure}

\subsection{The $\ap$ function}
From \eqr{cfslt} we can define 
\begin{align}
\ap= & \int d\bq \int d{\bf \Delta} \ \mathcal{\hat F}[\fq] \prod_{i=0}^{N-1} \bra{q_{i-1}-\tfrac{1}{2}\Delta_{i-1}} \ebN \kb{q_i+\tfrac{1}{2}\Delta_i} \ppi \rangle \nonumber \\
& \qquad \times \langle \ppi \ket{q_{i}-\tfrac{1}{2}\Delta_{i}}, \eql{apdef}
\end{align}
such that 
\begin{align}
 \ltti \Cfsnv{t} = & \int d\bp \ h[\bar g(\bp)] \ap \eql{cfsap}.
\end{align}

$\ap$ can be considered as an $N$-dimensional hypercube\cite{coe73}, where the $i$th dimension corresponds to the long-time momentum
of the $i$th bead, $p_i$, and the size of the hypercube is in the limit $p_i\to\pm \infty$. We also choose to define $2^N$
`subcubes', such that within a subcube a particular $p_i$ value is exclusively positive or negative. A hypercube in $N$ dimensions
can be formed by connecting the vertices of a hybercube in $N-1$ dimensions, such as a cube ($N=3$ hypercube) being formed by
connecting two squares, illustrated in \figr{spinenote}.

For a symmetric system, a scattering eigenstate with asymptotic momentum $p$ has equal energy to one with $-p$. For an asymmetric system, one must account for the asymmetry of the barrier,
\begin{align}
 \acute{p} = 
 \left\{
 \begin{array}{ll}
 -\sqrt{p^2 + 2m(V_p - V_r)} & p > 0 \\
 +\sqrt{p^2 + 2m(V_r - V_p)} & p < 0
 \end{array}
 \right.\eql{pacute}
\end{align}
where 
\begin{align}
 V_p = & \lim_{x\to\infty} V(x) \\
 V_r = & \lim_{x\to-\infty} V(x),
\end{align}
such that if $p>0$ (forward reaction), $\acute{p}$ corresponds to a backward reacting momentum of the same energy, as sketched in \figr{asymeck}. If $p=0$, there is no corresponding scattering eigenstate, and states which have insufficient energy to react do not contribute to the rate calculation (neither do bound states), such that the square roots in \eqr{pacute} are always real \cite{alt13}. 
\begin{figure}[tb]
 \centering
 \includegraphics[angle=270,width=.6\textwidth]{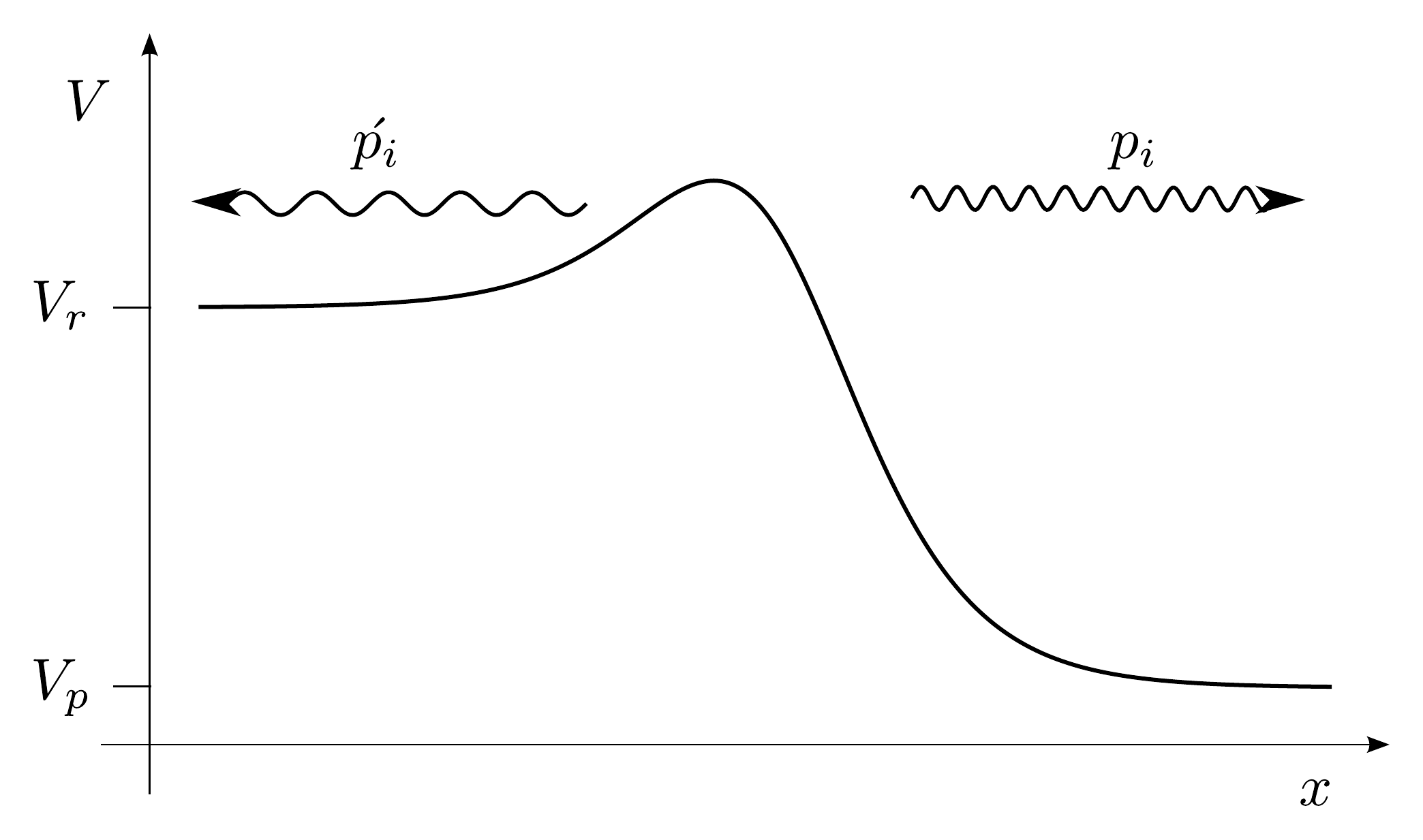}
 \caption{The relationship between $p$ and $\acute{p}$ schematically illustrated for the asymmetric Eckart barrier \cite{ric09,man05ref}.}
 \figl{asymeck}
\end{figure}

In appendix~\ref{ap:aps} we show that the density inside the $\ap$ function consists of delta-function spikes\footnote{For a
symmetric
system, the spikes will be straight, corresponding to lines in the hypercube along which the magnitude of each $p_i$ is equal,
but for a general asymmetric system they represent hyperbolae\cite{alt13}.} running along momentum states with equal energies, and a
residue term,
\begin{align}
 \bm{A}(\bp) = \bm{a}(\bp) \left\{ \prod_{i=1}^{N-1} \delta[E(p_i) - E(p_{i-1})] \right\} + \mathcal{R}(\bp) \eql{astruc}
\end{align}
where the residue $\mathcal{R}(\bp)$ is of alternating sign in adjacent subcubes as detailed in \eqr{ressym}.
\subsection{Integral over residues}
The integral in \eqr{cfsap} can be computed by summing over the contributions from adjacent subcubes in an iterative fashion, as illustrated in \figr{cubesum} for the case of $N=3$.
If there were no heaviside function present in \eqr{cfsap}, the integrals over adjacent subcubes in \eqr{astruc} would cause the
residues to cancel completely [and the resulting flux function (without the Heaviside function) would also be zero]. 
\begin{figure}[tb]
 \centering
 \includegraphics[width=0.5\textwidth]{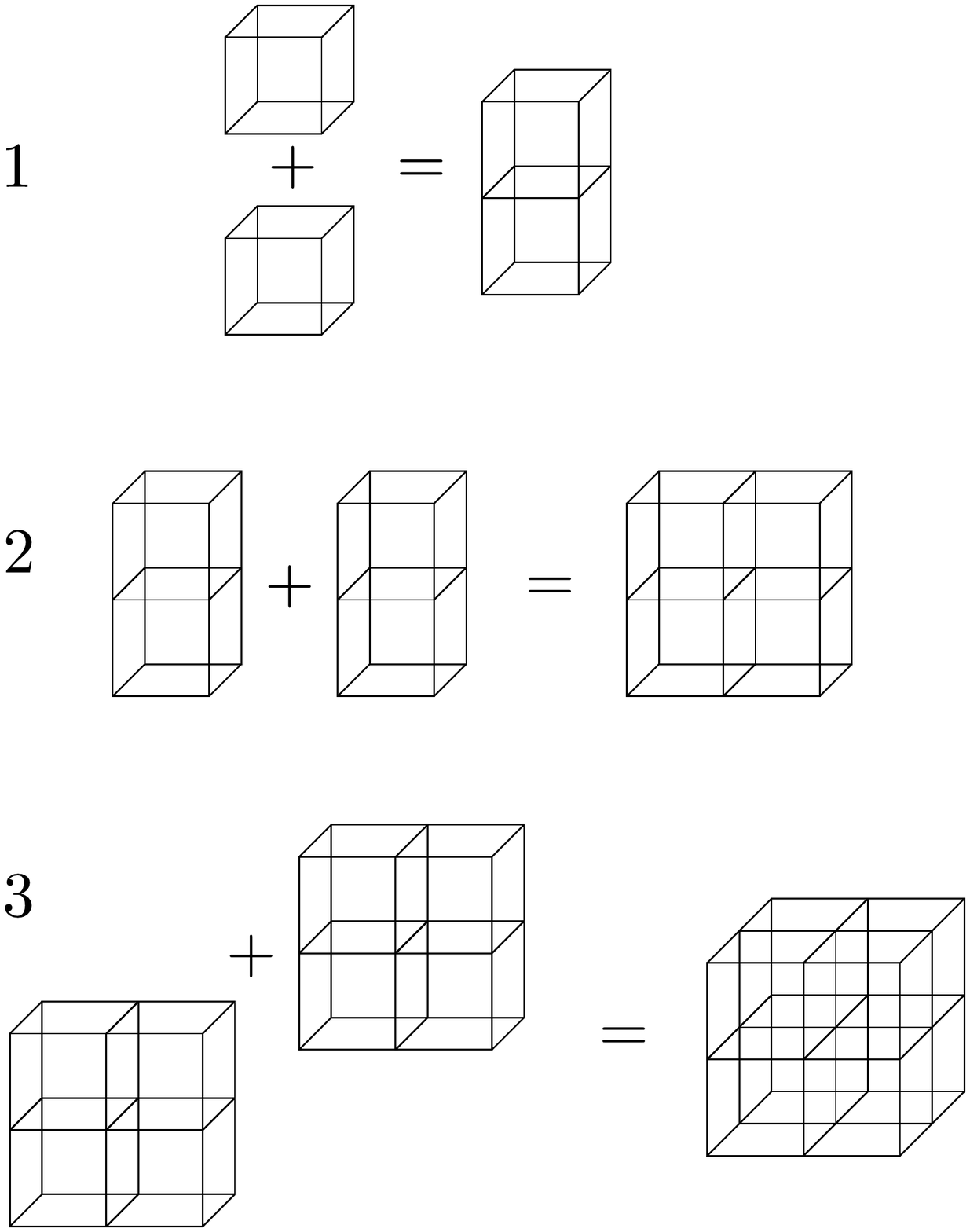}
 \caption{Constructing a cube ($N=3$ hypercube) from adding together a pair of subcubes, then a pair with another pair, and then four with another four, thereby illustrating evaluation of \eqr{cfsap} through successive summation over subcubes.}
 \figl{cubesum}
\end{figure}

However, for a general dividing surface one will not be summing over pairs of adjacent subcubes completely, because some of the subcubes will be cut through by the dividing surface function $\bar g(\bp)$. Nevertheless, in the \largeN~limit, we show in Appendix \ref{ap:ior} that the portion of the residue remaining as one sums over successive subcubes (equivalent to integrating of successive dimensions in $\{p_i\}$) becomes a sliver whose volume vanishes as $N^{-N}$. Consequently, in the $N\to\infty$ limit, the residues need not be considered further.\footnote{For the case of a symmetric system and even $N$, one can also demonstrate from a geometric argument that the residues will vanish for a centroid dividing surface.}

\subsection{Integral over spikes}
From \eqsr{astruc}{cfsap} and the results of the previous section,
\begin{align}
 \ltti \lNti \Cfsnv{t} = \int d\bp\ h[\bar g(\bp)] \bm{a}(\bp) \prod_{i=1}^{N-1} \delta[E(p_i) - E(p_{i-1})], \eql{ape}
\end{align}
such that the long-time limit of the flux-side function, with an infinite number of beads, is dictated entirely by the integral over
delta-function spikes, of which there are $2^N$. 

We now explore the constraints on the spikes when there is no recrossing of the dividing surfaces orthogonal to $\fq$. One can, of course, construct $N-1$ independent orthogonal surfaces $g_\perp(\bz)$ satisfying\footnote{The orthogonal planes must converge in the \largeN~limit in order for $\ltti\lNti D_{\rm fs \perp}^{[N]}(t)$ to converge and the flux through them to be well-defined (see section~\ref{ap:ior}), but by construction they will not satisfy \eqsr{divsurfdef}{divsurfdef2} as the centroid axis will lie along the dividing surface of any orthogonal plane.}
\begin{align}
 \smiNz \ddp{\fq}{q_i}\ddp{g_\perp(\bq)}{q_i} = 0 \ \forall \bq.
\end{align}
Using Eqs.~\eqref{eq:ffz} and \eqref{eq:ffi}, no recrossing is defined as the flux-side time-correlation function 
\begin{align}
 D_{\rm fs \perp}^{[N]}(t) = & \int d\bq \int d{\bf \Delta} \int d\bz \ \mathcal{\hat F}[\fq] h[g_\perp(\bz)] \nonumber \\
& \times \prod_{i=0}^{N-1} \bra{q_{i-1}-\tfrac{1}{2}\Delta_{i-1}} \ebN \kb{q_i+\tfrac{1}{2}\Delta_i} \etb \ket{z_i} \nonumber \\
& \qquad \times \bra{z_i} \etf \ket{q_{i}-\tfrac{1}{2}\Delta_{i}}.
\end{align}
being constant $\forall t >0$ \cite{sun98}.

From chapter~\ref{ch:stl} we know that $\lttz \Dfsnv{t} = 0$ as the dividing surfaces are not in the same location, such that the no recrossing criterion enforces
\begin{align}
 \ltti D_{\rm fs \perp}^{[N]}(t)_{\rm NR} = & \int d\bp \ h[\bar g_\perp(\bp)] \bm{a}(\bp) \prod_{i=1}^{N-1} \delta[E(p_i) - E(p_{i-1})] \no \\
  = & \ 0 \eql{enforce}.
\end{align}
For $N\geq3$ there will be an infinite number of ways of constructing orthogonal surfaces, which we can choose as required. We initially consider the case of a centroid dividing surface, which is later generalized to any permutationally invariant one.\footnote{The QTST derivation of chapter~\ref{ch:stl} holding only when the dividing surface $\fq$ is permutationally invariant.} Transforming to normal modes $\tilde \bq$, as detailed in \ref{ssec:nmt}, we obtain $N-1$ normal modes $\tilde q_i,\ i=1,\ldots N-1$. We can define the first such surface as emanating radially out from the centroid spike (defined when all $p_i$ are equal),
\begin{align}
 g_r(\bq) = \sqrt{\smiNz \tilde q_i^2} - r^\ddag \eql{radial}
\end{align}
and can then define other dividing surfaces orthogonal to this, which are of the form to sweep out angles in the hypercube,
\begin{align}
 g_G(\bq) = G[\phi(\tilde q_j,\tilde q_k)], \eql{gphi}
\end{align}
where
\begin{align}
 \phi(\tilde q_j,\tilde q_k) = \arctan(\tilde q_j/\tilde q_k).
\end{align}
The normal modes $\tilde q_j$ and $\tilde q_k$ can be chosen as desired, and (in higher dimensions) $G$ could be a function of more than one angle. For the case of $N=3$, a depiction of $\phi$ is schematically illustrated in \figr{anglesweep}.
\begin{figure}[tb]
\centering 
 \includegraphics[width=0.4\textwidth]{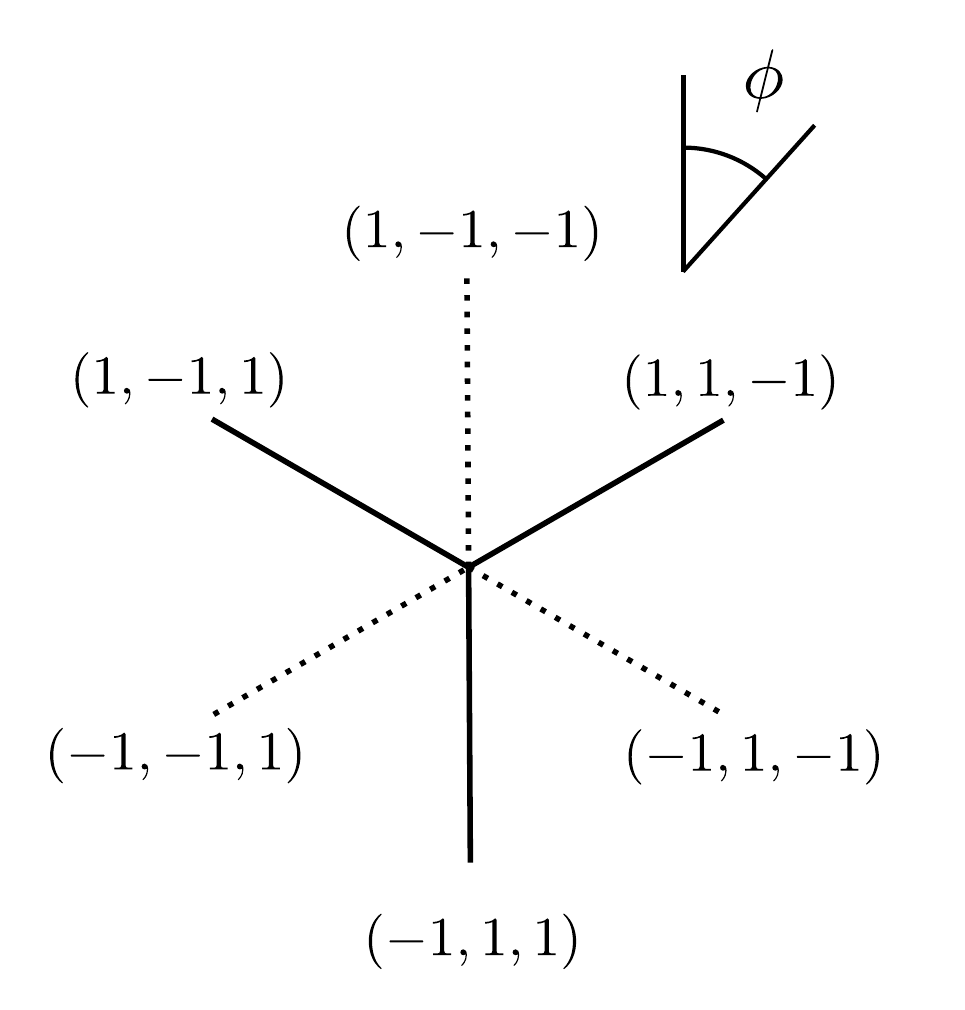} 
 \caption{Looking down the centroid spike [red arrow in \figr{hexsquare}b] and constructing angles for the dividing surfaces to sweep through. The solid lines represent spikes pointing out of the plane of the page, the dashed lines spikes pointing backwards. The labelling of the spikes is described in section~\ref{ssec:sysimp}.}
 \figl{anglesweep}
\end{figure}
Taking the long-time limit we find
\begin{align}
 \bar g_r(\bp) = \lim_{\epsilon\to 0} \sqrt{\smiNz \tilde p_i^2} - \epsilon\\
 \bar g_G(\bq) = G[\phi(\tilde p_j,\tilde p_k)], 
\end{align}
such that $\bar g_r(\bp)$ encloses an infinitesimally thin cylinder surrounding the centroid axis.

One can then construct $g_G$ to pick out each spike in turn, as no two spikes differ only in their position along the centroid
axis\footnote{Except for the spike along the centroid axis, which is the sole contributor to the rate in the absence of
recrossing.}. Enforcing the no-recrossing conditions on each of the spikes via \eqr{enforce} means that all spikes, except the
centroid spike, must vanish. Note that these spikes need not necessarily be mutually orthogonal, although they are all orthogonal to
$\fq$ and $g_r$.

This reasoning can be applied to a non-centroid dividing surface which is permutationally invariant. Near the centroid axis the function will reduce to the centroid, and one can therefore create a radial surface similar to \eqr{radial}. By defining dividing surfaces orthogonal to this [akin to \eqr{gphi}], one can enclose each spike in turn, the end result being that all such spikes, except that of the centroid, must be zero.

It therefore follows that, when there is no recrossing of dividing surfaces orthogonal to $\fq$, the only density in the hypercube
will be along the centroid axis. Consequently, any dividing surface which separates products and reactants in the long-time limit
[satisfies \eqsr{divsurfdef}{divsurfdef2}] must pick out the centroid spike.\footnote{By construction, this excludes all orthogonal
surfaces $\bar g_\perp(\bp)$.} We therefore choose $\gz = z_1$ (or any other individual $z_i$), and defining the corresponding
flux-side function as $\Cfsbnv{t}$:
\begin{align}
  \Cfsbnv{t} = & \int d\bq \int d{\bf \Delta} \int d\bz \ \mathcal{\hat F}[\fq] h(z_1) \nonumber \\
& \times \prod_{i=0}^{N-1} \bra{q_{i-1}-\tfrac{1}{2}\Delta_{i-1}} \ebN \kb{q_i+\tfrac{1}{2}\Delta_i} \etb \ket{z_i} \nonumber \\
& \qquad \times \bra{z_i} \etf \ket{q_{i}-\tfrac{1}{2}\Delta_{i}} \eql{cfsb}.
\end{align}
From the foregoing argument, the flux-side function in \eqr{cfsb} will, in the absence of recrossing, be equal to the general
flux-side function,\footnote{This is derived in the context of the long-time limit, but in the absence of recrossing the
corresponding flux-side functions must be constant for all time $t>0$.}
\begin{align}
 \Cfsbnv{t}_{\rm NR} = \Cfsnv{t}_{\rm NR} \eql{cfsbnr}
\end{align}
where the subscript NR denotes No Recrossing. Equation~\eqref{eq:cfsb} corresponds to a hybrid of the generalized-Kubo flux-side time correlation function \eqr{ubfn} where the flux dividing surface is a function of many points in imaginary time, and the Miller-Schwartz-Tromp form \eqr{cfss} where the side dividing surface is only a function of a single point in path-integral space. 

For $N>1$ the dividing surfaces will cut through different regions in path-integral space and the hybrid form will not have a TST limit. However, from its corresponding (and equivalent) side-flux form $\Csfbnv{t}$ we show in Appendix~\ref{ap:eoc},
\begin{align}
  \ltti \Csfbnv{t}  = & k_{\rm QM}(\beta) \Qrb \eql{anaint} % Analogue integral
\end{align}
under all circumstances (whether there exists any recrossing or not), and combining Eqs.~\eqref{eq:nrzi}, \eqref{eq:cfsbnr} and \eqref{eq:anaint},
\begin{align}
 \ltti \Cfsnv{t}_{\rm NR} = k_{\rm QM}(\beta)_{\rm NR} \Qrb
\end{align}
as was to be proven from \eqr{tobeproven}.
For the case of $N=1$, we observe
\begin{align}
 C_{\rm fs}^{[1]}(t) \equiv \bar C_{\rm fs}^{[1]}(t)
\end{align}
such that 
\begin{align}
 \ltti C_{\rm fs}^{[1]}(t) = k_{\rm QM}(\beta) \Qrb.
\end{align}
In the absence of recrossing of the dividing surface,\footnote{As this function only samples a single point in path-integral space, there are no orthogonal surfaces whose recrossing requires consideration.} the short-time limit of $\Cfsov{t}$ will equal its long-time limit, and we have therefore also shown that
\begin{align}
 \lttz C_{\rm fs}^{[1]}(t)_{\rm NR} = k_{\rm QM}(\beta)_{\rm NR} \Qrb
\end{align}
a result stated without proof in chapter~\ref{ch:stl}.\footnote{The Wigner rate is therefore equal to the exact quantum rate when their is no recrossing of its dividing surface, but we have seen in chapter~\ref{ch:stl} that its dividing surface is poor at low temperatures, exhibiting significant recrossing (\figr{b3}).}

\section{Orthogonal planes}
\label{ssec:sysimp}
Here we consider in more detail the nature of the $\ap$ matrix and how, by accounting for the non-centroid spikes it is possible to construct a function which smoothly interpolates between $k_{\rm QM}^{\ddag}(\beta)$ in the \shortt~limit and $k_{\rm QM}(\beta)$ in the long-time limit, allowing the construction of correction terms to RPMD-TST.
\begin{figure}[tb]
\centering 
 \includegraphics[angle=270,width=0.7\textwidth]{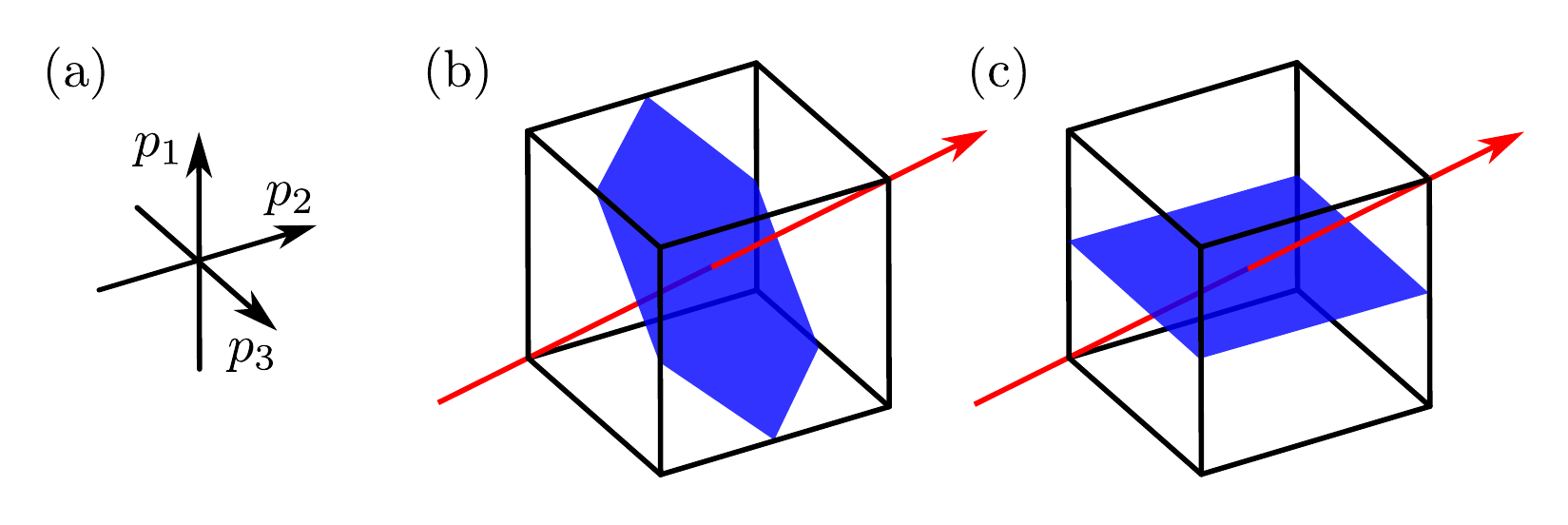}
 \caption{Illustrating the different sections of the hypercube, with axes labelled as in (a), cut out by a permutationally-invariant dividing surface (the centroid) (b) and the dividing surface used in the hybrid form \eqr{cfsb} (c), which produces the exact rate.}
 \label{fig:hexsquare}
\end{figure}

Deviations of the long-time limit of $\Cfsnv{t}$ from the exact quantum rate are due to the presence of non-centroid spikes in the $\ap$ matrix possessing finite density, which corresponds to overlap between scattering eigenstates of equal energy but momenta of different sign.\footnote{For a symmetric system, this corresponds to momenta of equal magnitude but differing sign.}

Conversely, the flux-side function $\Cfsbnv{t}$ produces the exact quantum rate in the long-time limit, regardless of whether there is any recrossing or not, as shown in appendix \ref{ap:eoc}. The side-dividing surface in this expression, $h(z_1)$, evidently cuts out a different part of the hypercube to the generalized, permutationally-invariant dividing surface $h[\gz]$ and therefore encloses a different set of non-centroid spikes. 

This can be observed graphically in \figr{hexsquare} by the centroid dividing surface (b) enclosing a different set of vertices to the $h(p_1)$ dividing surface (c). For the $N=3$ case and a centroid dividing surface used here, $h(z_1)$ encloses $(1,1,1)$, $(1,1,-1)$, $(1,-1,1)$ and $(1,-1,-1)$ whereas the centroid cuts out $(1,1,1), \ (-1,1,1),\ (1,-1,1)$ and $(1,1,-1)$, where we label the spikes by the vertex of the cube which they point to from \figr{spinenote}. Given that this selection of a different set of spikes causes the deviation from the exact quantum rate, it is therefore possible to write a modified flux-side function which is a linear combination of $\Cfsnv{t}$, and flux-side functions involving planes orthogonal to the dividing surface,
\begin{align}
 G_{\rm fs}^{[N]}(t) = \Cfsnv{t} + \sum_{l = 0}^{L-1} c_l D_{\textrm{fs} \perp }^{[N]}(t)_l, \eql{gdef}
\end{align}
where the coefficients $c_l$ and orthogonal surfaces (``orthoplanes'') in $D_{\textrm{fs} \perp }^{[N]}(t)_l$ can be determined by geometric considerations, such that in the long-time limit the same spikes are enclosed as in $\Cfsbnv{t}$. Consequently,
\begin{align}
 \lttz G_{\rm fs}^{[N]}(t) = & \lttz \Cfsnv{t} + \sum_{l = 0}^L c_l \lttz D_{\textrm{fs} \perp }^{[N]}(t)_l \nonumber \\
 = & k_{\rm RPMD}^{\ddag} \Qrb
\end{align}
as the orthogonal planes result in zero instantaneous flux, and by construction
\begin{align}
 \ltti G_{\rm fs}^{[N]}(t) = & \ltti \left\{ \Cfsnv{t} + \sum_{l = 0}^L c_l D_{\textrm{fs} \perp }^{[N]}(t)_l \right\} \nonumber \\
 = & \ltti \Cfsbnv{t} \nonumber \\
 = & k_{\rm QM}(\beta) \Qrb. \eql{orthoplanes}
\end{align}
As an illustrative example, let us return to the case of an $N=3$ system with a centroid dividing surface such that the hypercube is easily visualised as a cube.\footnote{For a general asymmetric system the residues would only cancel in the \largeN\ limit, but the case presented here can be extended to any $N$ and is illustrated using $N=3$ for graphical simplicity.} We can define normal modes
\begin{align}
 \tilde q_0  & = \frac{1}{\sqrt{3}}(q_0 + q_1 + q_2) \\
 \tilde q_1  & = \frac{1}{\sqrt{2}}(q_1 - q_2) \\
 \tilde q_2  & = \frac{1}{\sqrt{6}}(2q_0 - q_1 - q_2)
\end{align}
and likewise in $\tilde \bz$. The flux dividing surface is therefore given by $\fq = \tilde q_0/\sqrt{N}$.
By a geometric argument\footnote{Computing the vertices cut through by each dividing surface, and solving the simultaneous equations in $c_l$ so that the expansion in \eqr{orthoplanes} produces the same number of each type of vertex as that cut out by $h(\tilde z_0)$.} and noting that the hypercube has $N$-fold rotational symmetry along the centroid axis due to the permutational invariance of the flux dividing surface, the parameters for \eqr{orthoplanes} can be determined as:
\begin{table}[h]
\centering
 \begin{tabular}{c|c|c}
  Flux-side function & Coefficient $c_l$ & $g(\bz)$ \\ \hline
  $\Cfsnv{t}$ & 1 & $\tilde z_0$ \\
  $D_{\textrm{fs} \perp }^{[N]}(t)_1$   & 2     & $\tilde z_1$ \\
  $D_{\textrm{fs} \perp }^{[N]}(t)_2$   & $-1$    & $\tilde z_2$
 \end{tabular}
\end{table}
\noindent and further geometry can generalize this to more complex dividing surfaces and higher $N$.

One can therefore construct a function using \eqr{gdef} which smoothly interpolates between the RPMD-TST rate and the exact quantum rate. Realistically, computation of the long-time limit of $\Cfsnv{t}$ and all the $D_{\textrm{fs} \perp }^{[N]}(t)_l$ would be considerably more expensive than direct evaluation of the Miller-Schwartz-Tromp equation \eqr{cfss}, so it would not be advocated as a computational tool. Significantly, at least in a theoretical framework, it is possible to systematically improve RPMD-TST towards the exact quantum rate. 

\section{Multidimensional generalization}
\label{sec:mdscat}
Here we sketch how the above results can be generalized to multidimensional systems, and thereby the liquid phase, provided that there is sufficient separation of timescales between reaction and equilibration \cite{cha78}.
For a system with $F$ dimensions, there are $N$ copies of the system with co-ordinates $\bq = \{ \bq_1,\ldots,\bq_N\}$, where $\bq_j = \{ q_{j,1},\ldots,q_{j,F} \}$. Here $q_{j,k}$ is the scalar co-ordinate of the $k$th dimension of the $j$th bead, with $\bz,\ \bDelta$ and so on similarly defined.

The bra-ket states then become $F$-co-ordinate;
\begin{align}
 \ket{q_i - \Delta_i/2} \to \ket{q_{i,1} - \Delta_{i,1}/2,\ldots, q_{i,F} - \Delta_{i,F}/2}
\end{align}
as does the ring polymer flux operator, whose multidimensional form is given in \eqr{multidflux}.
The cyclic permutation properties discussed earlier of $\fq$ apply to collective permutation of the $N$ path-integral replicas of the system, not of the $F$ classical dimensions.
\subsection{Multidimensional quantum scattering theory}
Fortunately, it suffices to know that a scattering state is separable into its outgoing (or incoming) momentum contribution $\pi_i$ and its internal state $v_i$, \cite{tay06}
\begin{align}
 \ltti \etf \etbo \ket{\pi_i; v_i} & = \bm{\hat \Omega_-} \ket{\pi_i; v_i} \no \\
 & = \ket{\psi_{\pi_i,v_i}}
\end{align}
where $\bm{\hat \Omega_-}$ is the multidimensional M\o ller operator, such that
\begin{align}
\int d(\bq_j + \bDelta_j/2) \ltti \etf \ket{\bq_j + \bDelta_j/2} = \int d\bm{\pi_j} \sum_{v_j} \ket{\psi_{\pi_j,v_j}}
\end{align}
where the internal states of a bound molecule $v_j$ are all discrete. Furthermore, the multidimensional dividing surface function possesses the long-time limit
\begin{align}
 \ltti \fq & = \ltti f(\bm{\pi}t/m; \textbf{v}) \no \\
 & = \bar f(\bm{\pi}) \eql{multidfq}
\end{align}
otherwise it would not successfully separate different product channels in the \longt~limit.

\subsection{Exact rate in the absence of recrossing}
From the scattering theory outlined above, the hybrid form \eqr{cfsb} still produces the exact rate in the \longt~limit, where its side dividing surface becomes $h(\pi_1)$. By taking the long-time limit of the multidimensional form of $\Cfsnv{t}$, one generates 
\begin{align}
 \ltti \Cfsnv{t} = \int d\bpi \ h[\bar f(\bpi)] \api.
\end{align}
As before, one can demonstrate that $\api$ contains spikes and residues, and that the residues vanish in the \largeN~limit. There will be many more spikes than before, each corresponding to a different $\delta[E(\bpi_j,\mathbf{v}_j) - E(\bpi_{j-1},\mathbf{v}_{j-1})]$, where the $j$ indices correspond to different bead numbers, not classical dimensions. Nevertheless, one can still construct sufficient orthogonal dividing surfaces to show that all spikes must vanish (as for each extra degree of freedom producing a spike, one has an extra dimension in which to form an orthogonal dividing surface). Consequently in the absence of recrossing the only density in the $\api$ matrix is found along the centroid axis, i.e.~when all path-integral beads proceed down the product channel with identical momenta. Therefore any dividing surface separating products from reactants, such as \eqr{multidfq} or that of the hybrid, will produce the exact rate. 

As in one dimension, we finally find that the exact quantum rate is produced in the absence of recrossing of the dividing surface nor of any of the $N-1$ surfaces orthogonal to it in path-integral space.

\section{Implications}
The primary aim of this chapter has been the proof that \eqr{ubfn}, which reduces to RPMD-TST in the \shortt~limit, will produce the exact quantum rate in the absence of recrossing, by the exact quantum dynamics, of the dividing surface $\fq$ and any surfaces orthogonal to it in $(N-1)$-dimensional path-integral space. This is satisfied automatically for a parabolic barrier with the dividing surface at the apex of the barrier \cite{alt13},\footnote{At all temperatures above crossover ($\beta<\beta_c$), where a rate for the parabolic barrier is defined.} and therefore also for a free particle, which is a limiting case of the parabolic barrier where its imaginary frequency $\omega_b = 0$.

The no recrossing criteria impose the requirement that the only density in the momentum-space hypercube $\ap$ is along the so-called `centroid axis', where all momenta are of the same magnitude and sign, and the corresponding scattering eigenstates of the same energy. Physically, this means that no recrossing is equivalent to the path-integral beads, constrained at time $t=0$ by the Boltzmann operator, moving in concert with equal momentum. 

\subsection{Classical limit}
In the high temperature, classical limit, the ring-polymer at $t=0$ shrinks to a point and there will be no quantum coherence
effects in the recrossing. RPMD-TST reduces to classical-TST in this limit (as RPMD rate theory reduces to classical rate theory
\cite{man05che}), such that the only recrossing consideration is of the dividing surface in the flux function and not that of the
orthogonal planes.%\footnote{A rigorous proof that the density of the non-centroid spikes vanishes in the high temperature limit
% (and the dynamics more generally) is an area for future research, but preliminary numerical results such as those in
% chapter~\ref{ch:num} strongly suggest this is the case.}

\subsection{RPMD}
\label{ssec:rpmdimp}
The present chapter has demonstrated that RPMD-TST will produce the exact quantum rate when there is no recrossing of the
permutationally invarariant dividing surface (and those orthogonal to it in path-integral space) by the exact quantum dynamics. In
the previous chapter we showed that an RPMD simulation will calculate a good approximation to the instantaneous thermal quantum flux
through the statistically optimal dividing surface, provided that the TST assumption holds in the space of the fictitious ring
polymer dynamics. Combined, these mean that an RPMD simulation will compute the exact quantum rate past the statistically optimal
dividing surface, provided that there is no recrossing of the dividing surface by the exact quantum dynamics,\footnote{By which we
mean action of $\etf$ and \emph{not} the fictitious dynamics of the ring-polymer Hamiltonian.} nor of the (fictitious) ring-polymer
dynamics. 

For general physical systems it is extremely difficult to locate the optimal dividing surface \emph{a priori}; even for a classical calculation it is an $(F-1)$-dimensional manifold in $F$-dimensional Cartesian space. As transition-state theory is exponentially sensitive to the location of the dividing surface (see chapter~\ref{ch:rev}), this can diminish the utility of such methods in multidimensional systems. However, RPMD surmounts both these hurdles; by dynamics which conserve the quantum Boltzmann distribution it will locate the optimal dividing surface (the `bottleneck'), and return the instantaneous thermal quantum flux past this surface (scaled by any ring-polymer recrossing). In the event that there is little recrossing of this surface by the exact quantum dynamics and by the ring-polymer dynamics (and numerical simulations suggest this is the case \cite{sul11,man05ref}) RPMD will provide a good approximation to the rate, \emph{without} requiring prior knowledge of the optimum dividing surface location.

\section{Conclusions}
In this chapter we have shown that the QTST obtained from $\Cfsnv{t}$ has satisfied the second requirement of a QTST, namely that it produces the exact quantum rate in the absence of recrossing.

For a real physical system it is difficult to find the optimal dividing surface, and even if it is found there may still be some recrossing. However, under these circumstances QTST represents a \emph{good approximation} to the exact quantum rate, just as classical TST represents a good approximation to the exact classical rate.

The results in this chapter are derived using quantum scattering theory, which is exact in the gas phase. They can then
be extended to the condensed phase provided that there is a sufficient separation of timescales between reaction and equilibration
\cite{cha78, alt13}. Future work might include a derivation based on linear response theory\cite{yam59}, which would not rely on the
plateau in $\Cfsnv{t}$ extending to infinity.

Of course, there are some systems where there exists significant recrossing of the dividing surface, pronounced quantum coherence
effects, or no meaningful position-space dividing surface. These include the inverted regime in Marcus theory\cite{mar85,men11},
some diffusive processes (where classical TST also breaks down), and low temperature gas-phase scattering systems.
%(as is observed numerically in chapter~\ref{ch:num})
In these circumstances, a QTST of the form described above would not be
expected to provide a good approximation to the rate and other methodologies are required. 

We now investigate numerical results for the Generalized Kubo expression in order to validate the algebra in the past two chapters.

%% file: uni.tex
\chapter{Uniqueness}
\label{ch:unique}
Having seen that a true \shortt~QTST exists (chapter~\ref{ch:stl}) and that this gives the exact rate in the absence of recrossing in (chapter~\ref{ch:ltl}), we now present strong evidence that RPMD-TST is the only positive-definite QTST; that it is \emph{unique}. 

There exist a large number of heuristic QTSTs \cite{wig32uber,gil87,vot89rig,cal77,ben94,pol98,and09,vot93,tru96,gev01,shi02,mci13}, and, given that RPMD-TST was considered a heuristic guess before the derivation in chapters~\ref{ch:stl}--\ref{ch:ltl} was produced, the question arises as to whether there exist any other quantum transition-state theories which could also be of practical benefit. 

In chapter~\ref{ch:stl} we showed that Wigner TST \cite{wig32uber,mil75} satisfies the requirements for a QTST, but does not give positive-definite statistics, an essential requirement for a practical rate theory. We therefore consider whether there exist any other QTSTs which produce positive-definite statistics, and are not equivalent to RPMD-TST.

Naturally, any claim of uniqueness is subject to the definition of QTST, and here we use the original premise of Eyring
\cite{eyr35}, namely that all trajectories which cross the barrier react (rather than recross). For classical TST, this was
subsequently recognized as being equivalent to taking the short-time limit of a classical flux-side time-correlation function
\cite{cha78,fre02}. We confine ourselves to the quantum mechanical analogue of this, namely whether there exists another quantum
flux-side time-correlation function which possesses a non-zero (and positive-definite) short-time limit, produces the exact rate in
the absence of recrossing\footnote{Strictly speaking, there is the extra requirement in QTST for there to be no recrossing by the
quantum dynamics of the planes orthogonal to the dividing surface in path-integral space, which does not exist in classical TST
(where a path-integral dividing surface is unnecessary due to the locality of the Boltzmann operator), discussed further in
chapter~\ref{ch:ltl}.}, and is not equivalent to RPMD-TST. With this definition, the QTST represents the instantaneous thermal
quantum flux through the dividing surface, and the rate is guaranteed to be positive at any temperature.\footnote{The Wigner rate
also represents the instantaneous thermal quantum flux through a dividing surface, but the nature of the dividing surface,
privileging a single point in imaginary time, causes the non positive-definite statistics.}

Here we give very strong evidence (though not a conclusive proof) that RPMD-TST is indeed the unique positive-definite QTST. In section~\ref{sec:gqfs} we construct an extremely general quantum flux-side time correlation function \eqr{gencfs}; we cannot prove that a more general one does not exist, but \eqr{gencfs} is sufficiently general that it includes all known flux-side functions as special cases. By taking the \shortt~limit in section~\ref{sec:stl}, and imposing the conditions that the expression thus obtained is non-zero and positive-definite, RPMD-TST emerges.

\section{General quantum flux-side time-correlation function}
\label{sec:gqfs}

One can observe that $\Cfsnv{t}$ [\eqr{ubfn}] is not the most general flux-side time-correlation function because one can modify \eqr{cfso} to give a `split Wigner flux-side time-correlation function':
\begin{align}
 {C_{\rm fs}^{[1]}}'(t) = & \int dq \int dz \int d \Delta \int d \eta \ h(z) \mathcal{\hat F}(q) \nonumber \\
 & \times \bra{q - \Delta/2} \ebt \kb{q+\Delta/2} \etb \ket{z-\eta/2} \nonumber \\
 & \times \bra{z-\eta/2} \ebt \kb{z+\eta/2} \etf \ket{q-\Delta/2},
\label{eq:opff} 
\end{align}
which can be shown to give the exact quantum rate in the \longt~limit and to have a non-zero \shortt~limit. This limit is not positive-definite, but one could generalize \eqr{opff} in the analogous way to which \eqr{ubfn} is obtained by ring-polymerizing \eqr{cfso}. 

A form of flux-side time-correlation function which does include \eqr{opff}, as well as a ring-polymerized generalization of it, is 
\begin{align}
 \Cfsun{t}= \int & d\bq \int d\bz \int d\bDelta \int d\bmeta\ \mathcal{\hat F}[\fq]h[\gz] \nonumber \\
 \times & \piNz \bra{q_{i-1}-\Delta_{i-1}/2} e^{-\beta \xi_i^- \hat H} \kb{q_i+\Delta_i/2} e^{i\hat H t /\hbar} \ket{z_i - \eta_i/2} \nonumber \\
 & \times  \bra{z_i - \eta_i/2} e^{-\beta \xi_i^+ \hat H} \kb{z_i+\eta_i/2} e^{-i\hat H t /\hbar} \ket{q_i - \Delta_i/2}.
 \label{eq:gencfs} % General cfs
\end{align}
Here the imaginary time-evolution has been divided into pieces of varying lengths $\xi_i^\pm\beta\hbar$, which are interspersed with forward-backward real-time propagators. To set the inverse temperature $\beta$, we impose the requirement
\begin{align}
 \smiNz \xi_i^- + \xi_i^+ = 1, \label{eq:sumtoone}
\end{align}
where $\xi_i^{\pm} \ge 0\ \forall i$. 
The only restrictions, at present, on the dividing surface $\fq$ are
\begin{align}
 \lim_{q\to\infty} f(q,q,\ldots,q) > 0, \\
 \lim_{q\to -\infty} f(q,q,\ldots,q) < 0,
\end{align}
and similarly for $\gq$, as discussed in section~\ref{sec:rpfsf}. The subscript $\neq$ symbolises that the dividing surfaces are not necessarily equivalent functions of path-integral space.
Equation~\eqref{eq:gencfs} is represented diagrammatically in \figr{xi}a.

The function  $\Cfsun{t}$ correlates the flux averaged over a set of imaginary-time paths with the side averaged over another set of imaginary-time paths at some later time $t$.
Every form of quantum flux-side time-correlation function (known to the author) can be obtained either directly from $\Cfsun{t}$, using particular choices of $\fq$, $\gq$ and $\bm{\xi}$, or as linear combinations of such functions, as shown in Table~\ref{tab:cfs} on page~\pageref{tab:cfs}. We believe that $\Cfsun{t}$ is the most general expression yet obtained for a quantum flux-side time-correlation function (before taking linear combinations), although we cannot prove that a more general expression does not exist.\footnote{It would be possible to generalize $\Cfsxv{t}$ yet further by specifying the time-evolution of each bead separately, but as a QTST is defined as an instantaneous, \shortt~thermal quantum flux it would be of no use in the following argument.} 

\begin{sidewaystable}[h]
\centering
\begin{tabular}{l|c|c|c|c|c|c}
 Flux-side t.c.f.			& $N$ 	& $\xi_i^-$ 	& $\xi_i^+$ 	& $\mathcal{\hat F}[\fq]$	& $h[\gz]$ & \shortt~limit \\ \hline
 Miller-Schwarz-Tromp \cite{mil83} 			& 2 	& 1/2		& 0 		& $\mathcal{\hat F}(q_1)$ 	& $h(z_0)$ & 0 \\
 Asymmetric MST \cite{mil83}			& 2	& $\xi_0^- = 1, \ \xi_1^- = 0$& 0 & $\mathcal{\hat F}(q_1)$	& $h(z_0)$ & 0  \\
 Kubo-transformed \cite{man05che}				&$\infty$& $1/N$	& 0 		& $\mathcal{\hat F}(q_0)$	& $\sum_{i=1}^{N-1} h(z_i)$ & 0  \\
 Wigner [$C_{\rm fs}^{[1]}(t) $ of Eq.~\eqref{eq:cfso}]			& 1	& 1		& 0		& $\mathcal{\hat F}(q_0)$	& $h(z_0)$ & Wigner TST \cite{wig32uber}  \\
 $C_{\rm fs}^{[1]}(t)'$ of Eq.~\eqref{eq:opff}		& 1	& 1/2		& 1/2		& $\mathcal{\hat F}(q_0)$	& $h(z_0)$ & Double-Wigner TST  \\
 Hybrid [Eq.~7 of Ref.~\cite{alt13}]					& $>\!1$	& $1/N$		& 0		& $\mathcal{\hat F}[\fq]$	& $h(z_0)$ & 0  \\
Ring-polymer [$\Cfsnv{t}$ of Eq.~\eqref{eq:ubfn}]			& $\infty$	& $1/N$		& 0		& $\mathcal{\hat F}[\fq]$	& $h[\fz]$ & RPMD-TST  \\
 \end{tabular}
 
\caption{Parameters for every (known) form of flux-side time-correlation function as a special case of Eq.~\eqref{eq:gencfs}.
 The terms $\xi_i^-$,  $\xi_i^+$,  $\mathcal{\hat F}[\fq]$ and $h[\gz]$ are defined in Eq.~\eqref{eq:gencfs}. Double-Wigner TST is the
  generalization of Wigner-TST that results from the \shortt~limit of Eq.~\eqref{eq:cfso}. In the hybrid and ring-polymer expressions, $\fq$ is chosen to be invariant under cyclic permutation of the coordinates $q_i$; Centroid-TST is a special case of RPMD-TST obtained when $\fq=\smiNz q_i/N$.} %Note that only RPMD-TST gives the correct positive-definite quantum statistics.}
\label{tab:cfs}
\end{sidewaystable}

\section{The short-time limit}
\label{sec:stl}

We now take the \shortt~limit of \eqr{gencfs}, and determine the conditions under which this limit is non-zero and possesses positive-definite quantum statistics.

\subsection{Non-zero QTST}

\begin{figure}[t]
\resizebox{\textwidth}{!}{\includegraphics[angle=270]{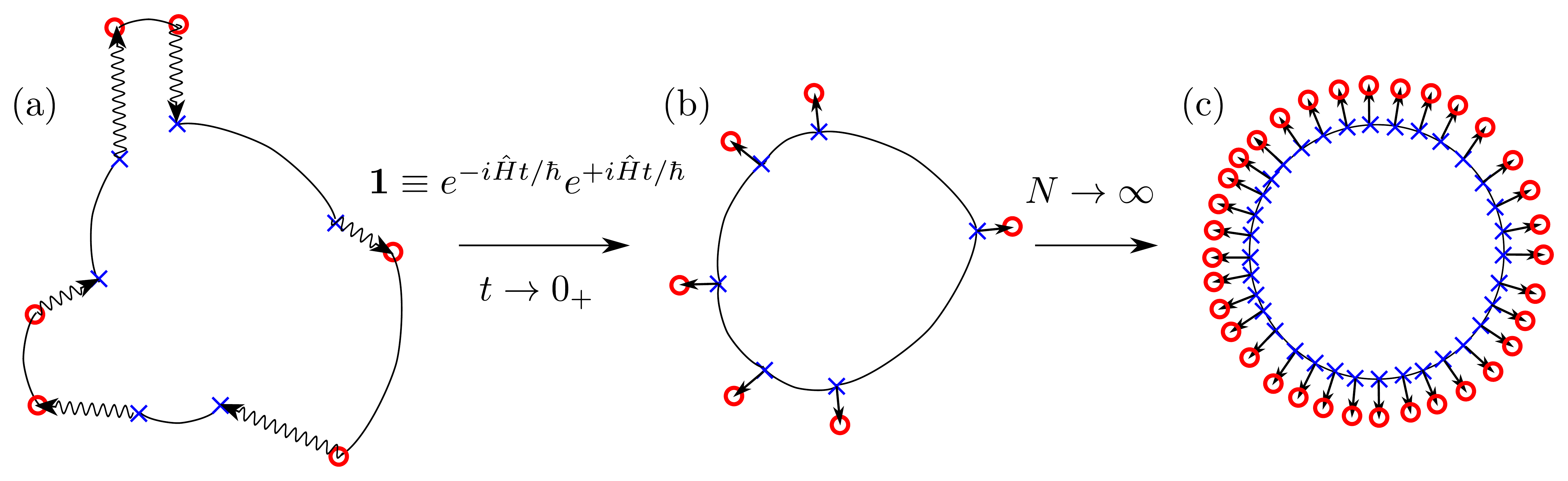}}
\caption{Diagrams showing
 (a) the generalized flux-side time-correlation function  $\Cfsun{t}$ of Eq.~\eqref{eq:gencfs};
  (b) the \shortt~limit of $\Cfsxv{t}$, Eq.~\eqref{eq:likenf2}; 
  (c)  the latter for a large value
  of $N$. Sinusoidal lines represent real-time evolution, curved lines imaginary-time evolution, and the symbols indicate the places  acted on by the
   flux operator $\mathcal{\hat F}[\fq]$ (blue crosses) and the side operator $h[\gz]$ (red circles).}% When $f\equiv g$, and when $f$ is invariant under imaginary-time translation, 
%   the \largeN~limit of (c) gives RPMD-TST; in any other case, it gives either non-positive-definite quantum statistics, or zero.}
\figl{xi}
\end{figure}

In order to calculate the short-time limit of \eqr{gencfs} we substitute the identity
\begin{align}
 e^{-\beta \xi_i^+ \hat H} \equiv & \int dy_i \int d \zeta_i \ \etf \kb{y_i - \zeta_i/2} e^{-\beta \xi_i^+ \hat H} \ket{y_i + \zeta_i/2} \bra{y_i + \zeta_i/2} \etb
\end{align}
 into \eqr{gencfs}, to obtain
\begin{align}
  \Cfsun{t\to 0_+} 
  = \lttz & \int d\bq \int d\bz \int d\bDelta \int d\bmeta \int d\by \int d\bm{\zeta}\ \mathcal{\hat F}[\fq]h[\gz] \nonumber \\
 \times & \piNz \bra{q_{i-1}-\Delta_{i-1}/2} e^{-\beta \xi_i^- \hat H} \kb{q_i+\Delta_i/2} \etb \ket{z_i - \eta_i/2} \nonumber \\
 & \times \bra{z_i - \eta_i/2} \etf \kb{y_i - \zeta_i/2} e^{-\beta \xi_i^+ \hat H} \ket{y_i + \zeta_i/2} \nonumber \\
 & \times \bra{y_i + \zeta_i/2} \etb \kb{z_i+\eta_i/2} \etf \ket{q_i - \Delta_i/2} \label{eq:nfs}. % New Form with Six variables
\end{align}

The imaginary-time propagators in \eqr{nfs} alternate with pairs of forward-backward real-time propagators, which allows us to use Eqs.~\eqref{eq:shorttlim}--\eqref{eq:etfop} to take the \shortt~limit.\footnote{One can evaluate the short-time limit of {\protect \eqr{gencfs}} without the insertion of the unit operators, but their use greatly simplifies the subsequent algebra.} The mathematics is lengthy and presented in full in appendix~\ref{ap:st}; here we sketch the main steps.

% This procedure is straightforward, but algebraically lengthy, so we give only the main steps here, relegating the details to Appendix~\ref{ap:st}.
% 
The first modification (Sec.~\ref{ssec:cot}) is to transform \eqr{nfs} to 
\begin{align}
  \Cfsun{t} = \int  & d{\bf Q} \int d {\bf Z} \int d {\bf D} \ \mathcal{\hat F}[f({\bf Q,D})]h[g({\bf Z})] \nonumber \\
 \times & \prod_{j=0}^{2N-1} \bra{Q_{j-1}-D_{j-1}/2} e^{-\beta \xi_j \hat H} \kb{Q_j+D_j/2} \etb \ket{Z_j} \nonumber \\
 & \times \bra{Z_j} \etf \ket{Q_j - D_j/2}
 \label{eq:simpt} % simp in text
\end{align}
where $\bQ\equiv\{Q_j\}$, $j=0\dots2N-1$, and similarly for $\bZ$, $\textbf{D}$, 
and
\begin{align}
\xi_{2i}&=\xi^-_i\\
\xi_{2i+1}&=\xi^+_i,\ \ i=0,\dots,N-1.
\end{align}
We have halved the number of bra-kets in each imaginary time-slice, by doubling the number of polymer beads.
Equation~\eqref{eq:simpt} is superficially similar to \eqr{ubfn}, but the dividing surface $\fq$ now depends on the coordinate $\bD$
(in the way described in Sec.~\ref{ssec:cot}). As a result the flux and side dividing surfaces are in general {\em different}
functions of path integral space, even if $\fq \equiv \gq$. From the results of chapter~\ref{ch:stl}, one might therefore expect the
\shortt~limit of \eqr{simpt} to be zero, except for the special cases corresponding to Wigner TST and RPMD-TST  (given in Table I).
However, we show in Sec.~A2 that the \shortt~limit of \eqr{simpt} is {\em always} non-zero when $\fq \equiv \gq$, because the
$\bD$-dependence of $f({\bf Q,D})$ can be integrated out in this limit, so we can define a QTST to be 
\begin{align}
\lttz\frac{\Cfsxv{t}}{\Qrb} = & \frac{1}{(2\pi\hbar)^N \Qrb} \int d{\bf Q} \int d {\bf P^+} \int d {\bf D^+} \ \dfQ S_f({\bf
Q,P^+})h[S_f({\bf Q,P^+})]  \nonumber \\
  & \times \prod_{i=0}^{N-1} \bra{Q_{2i-1}-\otrt D_{i-1}^+} e^{-\beta \xi_j \hat H} \ket{Q_{2i}+\otrt D_i^+} \nonumber \\
  & \qquad \times \bra{Q_{2i}-\otrt D_i^+} e^{-\beta \xi_j \hat H} \ket{Q_{2i+1}+\otrt D_i^+} e^{i D^+_j P^+_j/\hbar}, \label{eq:nzq} % Non-Zero QTST
\end{align}
where $\bP^+$ and $\bD^+$ are the $N$-dimensional vectors defined in section~\ref{ssec:stl}, $S_f({\bf Q,P^+})$ is the flux perpendicular to $\fQ$, and the absence of a subscript $\neq$ in $\Cfsxv{t}$ indicates $\fq \equiv \gq$. 
Consequently, in general $f({\bf Q,D})$ is a function of both $\bQ$ and $\bD$, but in the \shortt~limit, the $\bD$ dependence can be integrated out if $\fq \equiv \gq$. In the special case of $\xi_i^-=1/N,\xi_i^+=0$ in which $\Cfsxv{t}\equiv\Cfsnv{t}$, $\fq$ is time-independent (not a function of $\bD$).
%Thus, in general, $f({\bf Q,D})$ acts as a time-dependent flux-dividing surface, which becomes the same as the side-dividing surface in the limit \shortt~if $\fq \equiv \gq$. Clearly $\fq$ is time-independent in the special case that $\xi_i^-=1/N,\xi_i^+=0$, in which \eqr{gencfs} reduces to \eqr{ubfn} (see Table I).

Equation~\eqref{eq:nzq} can be simplified by integrating out $(N-1)$ of the integrals in $\bP^+$ and $\bD^+$ (see Sec.~\ref{ssec:intout}), to obtain
\begin{align}
 \lttz \Cfsxv{t} = & \frac{1}{2\pi\hbar} \int d{\bf Q} \int d \tilde P_0 \int d \tilde D_0 \ h(\tilde P_0) \frac{\tilde P_0}{m} 
\sqrt{B_N} \dfQ e^{i \tilde D_0 \tilde P_0/\hbar} \nonumber \\
 & \times \prod_{j=0}^{2N-1} \bra{Q_{j-1}- T_{j-1\ 0} \tilde D_0/2} e^{-\beta \xi_j \hat H} \ket{Q_j+ T_{j0} \tilde D_0/2}.
\label{eq:likenf2} % Like the New Form
\end{align}
where $\tpo$ is the momentum perpendicular to the dividing surface $\fQ$, $\tdo$ describes a collective ring-opening mode, $T_{j0}$ is the weighting of the $j$th path-integral bead in the dividing surface $\fQ$ [see Eq.~\eqref{eq:tdef2}], and $\sqrt{B_N}$ is a normalization constant associated with $\tpo$.

\subsection{Positive-definite Boltzmann statistics}

Having shown that the \shortt~limit of \eqr{gencfs} is non-zero if $\fq \equiv \gq$, we now determine the conditions on $\fq$ that give rise to
 positive-definite quantum statistics. The special case $\xi_i^-=1/N,\xi_i^+=0$ was discussed in chapter~\ref{ch:stl} and we use the same approach here for the more general case, namely finding the condition on $\fq$ which guarantees that the integral over $\tdo$ in \eqr{likenf2} is positive in the \largeN~limit. We first express the Boltzmann operator in ring polymer form,
\begin{align}
\lim_{N\to\infty} \prod_{j=0}^{2N-1} & \bra{Q_{j-1}- T_{j-1\ 0} \tilde D_0/2} e^{-\beta \xi_j \hat H} \ket{Q_j+ T_{j0} \tilde D_0/2} \nonumber \\
  = & \prod_{j=0}^{2N-1} \sqrt{\frac{m}{2\pi\beta\xi_j\hbar^2}} e^{-\beta \xi_j[ V(Q_{j-1}- T_{j-1\ 0} \tilde D_0/2) + V(Q_j+ T_{j0} \tilde D_0/2)]/2} \nonumber \\
 & \times e^{-m [ Q_{j} - Q_{j-1} + \tilde D_0(T_{j-1\ 0} + T_{j0})/2]^2/2\beta \xi_j \hbar^2} \label{eq:bbk} % Boltzmann bra-ket
\end{align}
and note that $T_{j0} \sim N^{-1/2}$, which ensures that the potential energy terms are independent of $\tilde D_0$ in the limit \largeN.\footnote{For equally spaced imaginary-time intervals, the leading non-zero term in the potential goes as $T_{j0}^2 \sim N^{-1}$, giving more rapid convergence with respect to $N$ than the general case of unequally-spaced intervals discussed in the text.}
Expanding the spring term, 
\begin{align}
\lim_{N\to\infty}  \sum_{j=0}^{2N-1} m & [Q_{j} - Q_{j-1} + \td D_0(T_{j-1\ 0} + T_{j0})/2]^2/2\beta \xi_j \hbar^2 \nonumber \\
 & = \lim_{N\to\infty} \sum_{j=0}^{2N-1} m [Q_{j} - Q_{j-1}]^2/2\beta \xi_j \hbar^2 \nonumber \\
 & \qquad + m [Q_{j} - Q_{j-1}]\td D_0(T_{j-1\ 0} + T_{j0})/2\beta \xi_j \hbar^2 \nonumber \\
 & \qquad + m \td D_0^2 (T_{j-1\ 0} + T_{j0})^2/8\beta \xi_j \hbar^2 \label{eq:ttterm},
\end{align}
we see that the integral over the Boltzmann operator is guaranteed to be positive if and only if the cross-terms vanish. In other words the condition
\begin{align}
\lim_{N\to\infty} \sum_{j=0}^{N-1} m [Q_{j} - Q_{j-1}]\tdo(T_{j-1\ 0} + T_{j0})/2\beta \xi_j \hbar^2 = 0 \label{eq:zerocon} % Zero Condition
\end{align}
must be satisfied for the Boltzmann statistics to be positive-definite.
In Appendix~\ref{app:piiti}, we show that this condition is equivalent to requiring the dividing surface $\fQ$ to be invariant under imaginary-time translation. For the special case of equal imaginary time discretization, this reduces to requiring cyclic permutational invariance of the path-integral beads, as found in chapter~\ref{ch:stl}.

%This is the generalized case of that obtained in chapter~\ref{ch:stl}, where we showed that $\fq$ must be invariant to cyclic permutation of the ring polymer beads.
%This was the same conclusion reached in Paper I, starting from the special case of $\xi_i^-=1/N,\xi_i^+=0$.

\subsection{Emergence of RPMD-TST}

When $\fq$ is invariant under imaginary-time translation we can integrate out $\tdo$ and $\tpo$ (see Appendix~\ref{ap:uneq}), to obtain
\begin{align}
  \lttz\lim_{N\to\infty}\Cfsxv{t} = & \int d{\bf Q} \ \dfQ \sqrt{\frac{\ntn}{2\pi m \beta}} \prod_{j=0}^{2N-1} \bra{Q_{j-1}} e^{-\beta\xi_j \hat H} \ket{Q_j} \label{eq:nop} % No P (momentum)
\end{align}
with
\begin{align}
 \ntn 
  & =  \lim_{N\to\infty} \sum_{j=0}^{2N-1}\frac{1}{4\xi_j}\left[\ddp{\fQ}{Q_{j-1}} + \ddp{\fQ}{Q_{j}}\right]^2.
  %\sum_{j=0}^{2N-1} \left[ \ddp{\fQ}{Q_j} \right]^2 \left(\frac{1}{2\xi_j} + \frac{1}{2\xi_{j+1}}\right)
\end{align}
The integral in \eqr{nop} is the generalisation of the RPMD-TST integral of \eqr{simpqtst} to unequally spaced imaginary time-slices $\xi_j$, where we note that for $\xi_i^- = 1/N,\ \xi_i^+ = 0$,
\begin{align}
 {\cal N}_N  
  & = N \smiNz \left[ \ddp{\fQ}{Q_i} \right]^2  \no \\
  & = NB_N.
\end{align}
% \begin{align}
%  \nn = NB_N.
% \end{align}
% 
% \begin{align}
%   \lttz\lim_{N\to\infty}\Cfsxv{t} = & \int d{\bf Q} \ \dfQ \sqrt{\frac{{\cal N}_N}{2\pi m \beta}} \piNz \bra{Q_{j-1}} e^{-\beta_N \hat H} \ket{Q_j} \nonumber\\
%   =& k_{Q}^\ddag(\beta)\Qrb,
%   \label {eq:nopeq} % No P (momentum) equally spaced
% \end{align}
% where
% \begin{align}
%  {\cal N}_N  
%   & = N \smiNz \left[ \ddp{\fQ}{Q_i} \right]^2 
% \end{align}
% which can be obtained from integrating out momenta from \eqr{rpmdtst}.
Both \eqr{nop} and \eqr{simpqtst} converge to the same result in the \largeN\ limit,\footnote{As they correspond to different quadratures of the same imaginary-time integral.} i.e.
 \begin{align}
  \lttz\lim_{N\to\infty}\Cfsxv{t} &= k_{\rm QM}^\ddag(\beta)\Qrb\nonumber\\
  &\equiv k_{\rm RPMD}^\ddag(\beta)\Qrb
\end{align}  
provided that $\fq \equiv \gq$ and that $\fq$ is invariant under imaginary-time translation. In other words, a positive-definite \shortt~limit can arise from the general time-correlation function \eqr{gencfs} only if $\fq$ is invariant under imaginary-time-translation (in the limit \largeN), in which case this limit is identical to that obtained from the simpler time-correlation function \eqr{ubfn} in chapter~\ref{ch:stl}, namely RPMD-TST.
 
The above derivation can easily be generalized to multiple dimensions by following the same procedure as that applied to \eqr{ubfn}
in section~\ref{sec:multid}. Similarly, as \eqr{nop} and \eqr{simpqtst} converge in the \largeN~limit where the
results of chapter~\ref{ch:ltl} hold, the long-time limit of $\Cfsxv{t}$ will produce the exact rate in the absence of recrossing of
the path-integral dividing surface or those orthogonal to it in path-integral space.

\section{Conclusions}
\label{sec:cc}
This chapter has provided strong evidence that RPMD-TST is the only QTST with positive-definite statistics. This is achieved by constructing an extremely general flux-side time-correlation function, of which all known exact flux-side functions are special cases (detailed in table~\ref{tab:cfs}), and demonstrating that its short-time limit is non-zero if and only if the flux and side dividing surfaces are the same function of path integral space in the \shortt~limit. We then demonstrate that, with the requirement of positive-definite statistics, RPMD-TST naturally emerges.

These results build upon the earlier work presented in chapter~\ref{ch:stl} where we demonstrated that RPMD-TST satisfied the requirements for a positive-definite QTST using a flux-side correlation function \eqr{ubfn} which is a special case of that introduced here, \eqr{gencfs}. 

The evidence is not a conclusive proof, but in order for there to be some other positive-definite QTST which was not equivalent to RPMD-TST, it would be necessary to construct a quantum flux-side time-correlation function which could not be written as a special case of \eqr{gencfs}, and demonstrate that it possessed a non-zero positive-definite short-time limit, that this produced the exact rate in the absence of recrossing, and that it was not equivalent to RPMD-TST. Given that \eqr{gencfs} includes all known quantum mechanical flux-side time-correlation functions, I believe this to be unlikely.

Consequently, this chapter presents strong evidence that the QTST derived in chapter~\ref{ch:stl}, namely RPMD-TST, is unique: that it is the pre-eminent theory for the calculation of thermal quantum rates in direct reactions.

% We have introduced an extremely general quantum flux-side time-correlation function, and found that its \shortt\ limit is non-zero {\em only} when the flux and side dividing surfaces are the same function of path-integral space, and that it gives 
%  positive-definite quantum statistics {\em only} when the common dividing surface is invariant
% to imaginary-time translation. This \shortt\ limit is identical to the one that was derived in Paper I starting from a simpler form of flux-side time-correlation function (a special case of the function introduced here),
% where it was shown to give a true \shortt\ quantum TST which  is identical to RPMD-TST.
%  
%  We cannot prove that a yet more general flux-side time-correlation function does not exist  (than the one introduced here) which might
%   support a {\em different} non-zero \shortt\ limit, which nevertheless gives positive-definite quantum statistics. However, given that the function introduced here includes all known flux-side time-correlation functions as special cases, we think that this is unlikely.
%  
% This article therefore provides strong evidence (although not conclusive proof) that the quantum TST of Paper I is unique, in the sense that there is no other \shortt~limit which gives a non-zero quantum TST containing positive-definite quantum statistics. In other words, if one wishes to obtain an estimate of the thermal quantum rate by taking the instantaneous flux through a dividing surface, then RPMD-TST cannot be bettered.

% \appendix

%% file: con.tex
\chapter{Conclusions}
\label{ch:con}

\section{Scope of the dissertation}
The central task of this dissertation has been to derive a true \shortt\ Quantum Transition-State Theory, despite a previous consensus that one did not exist \cite{wig38,wig39,mil93,vot93,pol05}. In chapter~\ref{ch:stl} we demonstrated that the previous absence of a QTST was due to the dividing surfaces being different functions of path-integral space. Upon alignment, a non-zero QTST was derived which was previously proposed by Wigner on heuristic grounds in 1932\cite{wig32uber}, but which produced poor results at low temperatures as it possessed non positive-definite statistics, evaluating the flux and Heaviside functions at only a single point in imaginary time. 

By polymerizing the rate expression in path-integral space, we obtained a different QTST which, when the dividing surface was invariant to permutation of the path-integral beads, produced positive-definite statistics. Remarkably, this is identical to RPMD-TST, which was previously proposed on heuristic grounds \cite{man04,man05che,man05ref}. Chapter~\ref{ch:ltl} then showed that this QTST produced the exact quantum rate when there is no recrossing of the dividing surface by the quantum dynamics, nor of any surfaces perpendicular to it in ring-polymer space.

The primary significance of this work to the wider scientific community has been the validation of the RPMD method for the computation of reaction rates, and therefore of the large and growing field of numerical rate calculations using RPMD rate theory \cite{man05ref,man05che,men11,kre13,col08,col09,col10,sul11,sul12,hab13,ric12,boe11}. Prior to this, RPMD was regarded as interpolative between various regimes, producing the correct rate for a free particle, parabolic barrier and in the high-temperature limit \cite{man05ref}, as well as a connection to widely-used semiclassical instanton theory \cite{ric09,alt11}, which itself has no rigorous derivation. Many other competing theories (such as instanton theory itself \cite{hel13}) could also be considered interpolative between numerous regimes and there was no rigorous reason to choose RPMD rate theory or RPMD-TST. 

% Numerical results in chapter~\ref{ch:num} corroborate the algebraic derivation in showing that the short-time limit of the Generalized Kubo form [\eqr{ubfn}] is identical to the RPMD-TST rate. They also demonstrate, for the model gas phase scattering system used, that the Generalized Kubo Form is equal to the RPMD-TST rate for finite real time ($\sim 10$fs) and that, at low temperatures, there can be significant quantum coherence effects.

Chapter~\ref{ch:unique} presents strong evidence that RPMD-TST is the only expression for the instantaneous thermal quantum flux through a position-space dividing surface which (i) possesses positive-definite statistics and (ii) produces the exact quantum rate in the absence of recrossing by the exact quantum dynamics. This strongly suggests that RPMD-TST is therefore the \emph{unique} QTST with positive-definite statistics, and therefore the superlative method for thermal quantum rate calculation in direct reactions.

% We also present a derivation of non-adiabatic QTST with a position-space dividing surface, which is expected to perform well in
% systems with large diabatic coupling [\eqr{sc}], and leave a generalization to reactions with an electronic-space dividing
% surface as future research.

The QTST thus derived is similar to its classical counterpart in a number of respects. It is exponentially sensitive the location of the dividing surface (although the RPMD method can elegantly circumvent this problem\cite{man05ref}), and can be written as a thermal flux multiplied by a free energy through a dividing surface [see \eqr{simpqtst}]. 

In other respects QTST differs from its classical counterpart. It is not a rigorous upper bound to the rate, since coherent quantum recrossing can cause the exact quantum rate to be greater than the QTST rate. Nevertheless, QTST therefore provides a good approximation to the upper bound  of the exact quantum rate when the amount of coherence is small, which is expected to be the case in systems not too far beneath the crossover temperature, and in condensed-phase systems which decohere rapidly, where RPMD has been particularly ground-breaking\cite{col08,men11,kre13,boe11}. 
In the classical limit, the transition-state theory rate is equal to the exact classical rate in the absence of recrossing of the dividing surface, whereas in quantum systems, there must also be no recrossing of surfaces orthogonal to the path-integral dividing surface by the quantum dynamics for the QTST rate to equal the exact quantum rate.

Like classical TST, quantum TST has limitations. It will fail for reactions lacking a barrier which is significant compared to the
thermal energy (in which case the notion of a reaction rate is often ill-defined),\footnote{Although recent research by Li
\emph{et.~al.}~\cite{lix14} has shown RPMD rate theory to be accurate in some low-barrier gas-phase insertion reactions.}
and it will also fail for diffusive or high-friction systems where there is significant recrossing of the barrier, and where
classical TST also fails \cite{alt13}. It is also expected to be a poor estimate of the rate in quantum systems exhibiting
significant coherent recrossing, such as low-temperature gas-phase reactions. %The numerical results in sections~\ref{ssec:ltr}
% show,
% surprisingly, that even in systems with significant quantum coherence, QTST can be a reasonable estimate of the rate, discussed
% further in section~\ref{ssec:rtqd}.

There already exist many numerical applications of RPMD rate theory which have produced excellent results
\cite{hab13,men11,kre13,col08,col09,col10,sul11,sul12,ric12,boe11,sul13} that, until the publication of the work in this
dissertation, had no rigorous justification. The evidence presented herein allows \emph{a priori} knowledge that RPMD rate theory
will provide a good approximation to the exact quantum rate provided that there is little recrossing (by exact quantum dynamics) of
the statistically optimal dividing surface.\footnote{Discussed more fully in section~\ref{ssec:rpmdimp}.} Besides justifying a large
corpus of previous work, this increases the utility of the RPMD method as it is generally straightforward to tell in advance if a
reaction has a well-defined barrier and little quantum coherence effects, and therefore whether RPMD or a similar method would be
suitable.
% \footnote{Even if RPMD rate theory were to produce erroneous results compared to an exact calculation (as Menzeleev,
% Ananth and Miller showed for the inverted Marcus region \cite{men11}) this illustrates that there is significant recrossing of the
% dividing surface or no meaningful dividing surface at all, as discussed in section~\ref{sec:nagimp}.}

The success of RPMD-TST, which has now been shown to be true QTST, also shows that for direct reactions, the value of the rate is entirely dominated by quantum statistics (which RPMD-TST captures exactly) and has little dependence on the quantum dynamics (which RPMD-TST does not account for). In doing so we have shown that, although quantum reaction rates appear \emph{prima facie} to be a consequence of the dynamics of the system, in many cases computation of the dynamics is simply unnecessary as the rate can be accurately described by the statistics alone.

\section{Future research}
While the derivation of a rigorous QTST has, after 70 years of debate, solved a major problem in reaction rate theory and quantum
mechanics, it also opens many avenues for future research, both for further theoretical development and numerical applications of
the theory.

\subsection{Derivation of alternative rate theories}
The derivations presented in this dissertation explain the origin of a number of rate theories, in particular Wigner rate theory and RPMD-TST. Nevertheless, as mentioned in section~\ref{ssec:art}, there are other widely-used theories with useful properties whose derivation or explanation would allow \emph{a priori} knowledge of the systems where they are likely to be valid and possible methods for systematic improvement. 

In particular, so called ``Im $F$'' instanton theory has a close connection to RPMD-TST [see \eqr{inst}] and in some circumstances appears to give superior results \cite{ric09}. It is therefore a matter of future research to derive the mysterious `alpha factor' [\eqr{alpha}] which seems to improve upon the RPMD-TST rate. Conversely, were further numerical study to show that, in most cases, RPMD-TST outperformed Im $F$ in proximity to the exact quantum rate, one could regard Im $F$ as an approximation to rigorous QTST with $\alpha(\beta) \simeq 1$.

% \subsection{Real-time quantum dynamics}
% \label{ssec:rtqd}
% % This dissertation has not sought to explain or analyse RPMD dynamics, which are generally regarded as \emph{ad hoc} \cite{man04,hab09,jan14}, it being sufficient that they preserve the quantum Boltzmann distribution\footnote{Discussed further in sections~\ref{ssec:rpmdimpst} and \ref{ssec:rpmdimp}.}.
% 
% The numerical results in chapter~\ref{ch:num} show significant quantum coherence at low temperatures, causing the long-time limit of
% the Generalized Kubo form to greatly exceed the exact quantum rate, though RPMD-TST remains reasonably accurate, attributable to
% recrossing of the planes orthogonal to the dividing surface in \eqr{gdef} cancelling the recrossing of the dividing surface. The
% cause of this, and the explanation of why QTST continues to be accurate even when the central premise of TST, that of no recrossing,
% breaks down, is an active area of research.
% 
% Furthermore, an explanation of RPMD dynamics more generally would be of great practical use, due to its wide applicability and
% accurate results in many
% systems\cite{mil05,mil05water,hab09,hab09com,smi14,hab13,men11,kre13,col08,col09,col10,sul11,sul12,ric12,boe11,sul13}, despite the
% dynamics generally being regarded as \emph{ad hoc} \cite{man04,hab09,jan14}. In particular, a guide to accuracy in a given system,
% or a quantifiable systematic improvement would allow \emph{a priori} knowledge of RPMD's applicability.

\subsection{Efficient implementation}
Computation of accurate rates for real physical systems is limited by the accuracy of the potential energy surface (PES), in addition to a computationally feasible and accurate rate theory (which we have presented with the derivation of QTST). Even for adiabatic systems where one can separate electronic and nuclear motion, accurately calculating the potential energy of the nuclei and the force acting upon them from a given set of nuclear co-ordinates remains extremely challenging. Standard implementations of RPMD-TST (where the real-time, fictitious classical dynamics of the ring polymer are used) require $10^6$--$10^7$ evaluations of the potential, increasing computational cost yet further.\footnote{By comparison, a classical rate theory calculation would require $10^4$--$10^5$ potential evaluations, assuming RPMD required $\sim 100$ beads.} 

Consequently, constructing an implementation of QTST which economizes on the number of potential evaluations would increase the speed, and thereby possible accuracy of rate calculations.

% A collaboration with Professor Ali Alavi at the University of Cambridge, who has developed very fast methods for electronic structure computation\cite{boo09,cle09}, has already been discussed. 

\section{Summary}
The research presented in this dissertation has resolved a 70-year debate on the existence of quantum transition-state theory and established a single, pre-eminent method for the computation of thermal quantum rates for direct reactions. The avenues of future research thereby opened should lead to a quantitative understanding of a plethora of chemical, biological and physical systems that can be examined with previously unobtainable precision.

%% file: css.tex
\chapter{Differentiation of $C_{\rm ss}^{[N]}(t)$}
\label{ap:doc}

\newcommand{\eqn}[1]{Eq.~(\ref{#1})}

\newcommand{\expect}[1]{\langle  #1 \rangle}

\newcommand{\cm}{\mathrm{cm}^{-1}}

% Tim's commands
% \newcommand{\shortt}{$t \to 0_+$\ }
% \newcommand{\css}{C_{\rm ss}(t)}
% \newcommand{\cfs}{C_{\rm fs}(t)}
\newcommand{\kQ}{k_{\rm Q}(\beta)}
\newcommand{\largen}{$N \to \infty$\ }

Differentiating $C_{\rm ss}^{[N]}(t)$ [\eqr{cssnvt}] with respect to time generates a sum of $2N$ terms,

\begin{align}
\dd{}{t} \Cssnv{t} = &\int\! d{\bf q}\, \int\! d{\bf z}\,\int\! d{\bf \Delta}\,h[f({\bf q})]h[f({\bf z})]\nonumber\\ 
\times&\sum_{i=1}^{N} \Bigg[ \expect{q_{i-1}-\Delta_{i-1}/2|e^{-\beta_N{\hat H}}|q_i+\Delta_i/2} \nonumber\\ 
\quad &\times  \Big\{ \expect{q_i+\Delta_i/2|\frac{i}{\hbar} \hat H e^{i{\hat H}t/\hbar}|z_i} \expect{z_i|e^{-i{\hat H}t/\hbar}|q_i-\Delta_i/2} \nonumber \\
& \quad + \expect{q_i+\Delta_i/2|e^{i{\hat H}t/\hbar}|z_i} \expect{z_i|e^{-i{\hat H}t/\hbar}\frac{-i}{\hbar} \hat H|q_i-\Delta_i/2} \Big\} \nonumber \\
& \times \prod_{j=1,j\neq i}^{N}\expect{q_{j-1}-\Delta_{j-1}/2|e^{-\beta_N{\hat H}}|q_j+\Delta_j/2}
\nonumber\\ 
& \quad \times \expect{q_j+\Delta_j/2|e^{i{\hat H}t/\hbar}|z_j}
\expect{z_j|e^{-i{\hat H}t/\hbar}|q_j-\Delta_j/2} \Bigg].
\label{eq:cssd} % css differentiated
\end{align}
We then note
\begin{align}
h[f({\bf q})] & \expect{q_i+\Delta_i/2|\frac{i}{\hbar} \hat H e^{i{\hat H}t/\hbar}|z_i} \nonumber \\
 & = \expect{q_i+\Delta_i/2|\frac{i}{\hbar} \hat h_i[f({\bf q})]\hat H e^{i{\hat H}t/\hbar}|z_i} \nonumber \\
 & = \expect{q_i+\Delta_i/2|\frac{i}{\hbar} \left(\hat H \hat h_i[f({\bf q})] - [\hat H,\hat h_i[f({\bf q})]\right)e^{i{\hat H}t/\hbar}|z_i},
 \label{eq:Hh}
\end{align}
and that
\begin{align}
 \expect{z_i| & e^{-i{\hat H}t/\hbar}\frac{-i}{\hbar} \hat H|q_i-\Delta_i/2}h[f({\bf q})] \nonumber \\
%  & = \expect{z_i| e^{-i{\hat H}t/\hbar}\frac{-i}{\hbar} \hat H \hat h_i[f({\bf q})]|q_i-\Delta_i/2} \nonumber \\
 & = \expect{z_i| e^{-i{\hat H}t/\hbar}\frac{-i}{\hbar} \left(\hat h_i[f({\bf q})] \hat H + [\hat H,\hat h_i[f({\bf q})] \right)|q_i-\Delta_i/2},
 \label{eq:hH}
\end{align}
where $\hat h_i[f({\bf q})]$ is the Heaviside function in a bra-ket corresponding to bead $i$. Strictly speaking, $\hat h_i[f({\bf q})]$ refers to $h[f({\bf q})]$ where every instance of $q_i$ has been replaced with the position operator $\hat q$, and likewise for $\hat \delta_i[f({\bf q})]$. 
Now the Heaviside function has been drawn inside the bra-ket, one can then collapse and reform the bra-kets in $q_i\pm\Delta_i/2$,
\begin{align}
\frac{i}{\hbar} \expect{q_{i-1}-\Delta_{i-1}/2|e^{-\beta_N{\hat H}}|q_i+\Delta_i/2} \expect{q_i+\Delta_i/2| \hat H \hat h_i[f({\bf q})] e^{i{\hat H}t/\hbar}|z_i} \nonumber \\
 = \frac{i}{\hbar}\expect{q_{i-1}-\Delta_{i-1}/2|e^{-\beta_N{\hat H}}\hat H \hat h_i[f({\bf q})]|q_i+\Delta_i/2} \expect{q_i+\Delta_i/2| e^{i{\hat H}t/\hbar}|z_i} \nonumber \\
 =  \frac{i}{\hbar}\expect{q_{i-1}-\Delta_{i-1}/2|\hat H e^{-\beta_N{\hat H}} |q_i+\Delta_i/2}h[f({\bf q})] \expect{q_i+\Delta_i/2| e^{i{\hat H}t/\hbar}|z_i} 
\end{align}
and by also doing so for \eqn{eq:hH} show that the $\hat H \hat h_i[f({\bf q})]$ term in \eqn{eq:Hh} cancels with the $\hat h_i[f({\bf q})] \hat H $ term in \eqn{eq:hH}.

%Denoting the Hamiltonian in a bra-ket corresponding to bead $i$ as $\hat H_i$ and likewise for $\hat h_i[f({\bf q})]$,  
The Heisenberg time derivative of the multidimensional side operator can be evaluated in many different but equivalent ways by manipulation of the momentum operators,
\begin{align}
 \frac{i}{\hbar}\left[\hat H_i,\hat h_i[f({\bf q})]\right] = & \frac{1}{4m} \left\{ \hat p_i \hat \delta_i[f({\bf q})] \frac{\partial f({\bf q})}{\partial q_i} + 
 \frac{\partial f({\bf q})}{\partial q_i}\hat \delta_i[f({\bf q})]\hat p_i \right\} \eql{pddp}\\
 = & \frac{1}{2m} \left\{ \frac{-i\hbar}{2} \frac{\partial}{\partial q_i}\left(\hat \delta_i[f({\bf q})] \frac{\partial f({\bf q})}{\partial q_i}\right) +  \frac{\partial f({\bf q})}{\partial q_i}\hat \delta_i[f({\bf q})]\hat p_i  \right\} \label{eq:dp} \\
 = & \frac{1}{2m} \left\{\hat p_i \hat \delta_i[f({\bf q})] \frac{\partial f({\bf q})}{\partial q_i} + \frac{i\hbar}{2} \frac{\partial}{\partial q_i}\left(\hat \delta_i[f({\bf q})] \frac{\partial f({\bf q})}{\partial q_i}\right)
 \right\}.
 \label{eq:pd}
\end{align}
% where \eqr{dp} and \eqr{pd} are constructed by changing the direction in which a momentum operator acts in \eqr{pddp}.
%where $\hat \delta_i[f({\bf q})]$, strictly speaking, refers to $\delta[f({\bf q})]$ where every instance of $q_i$ has been replaced with the position operator, and likewise for $\hat H_i$ and $\hat h_i[f({\bf q})]$
Choosing to insert \eqn{eq:dp} into \eqn{eq:Hh} and \eqn{eq:pd} into \eqn{eq:hH}, and these equations into \eqn{eq:cssd}, we find that the first term in \eqn{eq:dp} cancels with the second term in \eqn{eq:pd}, forming
\begin{align}
\Cfsnv{t} = \frac{1}{2m}&\int\! d{\bf q}\, \int\! d{\bf z}\,\int\! d{\bf \Delta}\,\delta[f({\bf q})]h[f({\bf z})]\nonumber\\ 
\times&\sum_{i=1}^{N} \Bigg[ \expect{q_{i-1}-\Delta_{i-1}/2|e^{-\beta_N{\hat H}}|q_i+\Delta_i/2} \nonumber\\ 
\quad &\times  \Big\{ \expect{q_i+\Delta_i/2|\frac{\partial f({\bf q})}{\partial q_i}\hat p e^{i{\hat H}t/\hbar}|z_i} \expect{z_i|e^{-i{\hat H}t/\hbar}|q_i-\Delta_i/2} \nonumber \\
& \quad + \expect{q_i+\Delta_i/2|e^{i{\hat H}t/\hbar}|z_i} \expect{z_i|e^{-i{\hat H}t/\hbar}\hat p\frac{\partial f({\bf q})}{\partial q_i}|q_i-\Delta_i/2} \Big\} \nonumber \\
& \times \prod_{j=1,j\neq i}^{N}\expect{q_{j-1}-\Delta_{j-1}/2|e^{-\beta_N{\hat H}}|q_j+\Delta_j/2}
\nonumber\\ 
& \quad \times \expect{q_j+\Delta_j/2|e^{i{\hat H}t/\hbar}|z_j}
\expect{z_j|e^{-i{\hat H}t/\hbar}|q_j-\Delta_j/2} \Bigg],
\end{align}
from which one can construct the `ring-polymer flux operator' in accordance with \eqr{sfo}.

If we had instead inserted \eqn{eq:pd} into \eqn{eq:Hh} and \eqn{eq:dp} into \eqn{eq:hH}, then by a judicious placement of bra-kets one would form the equivalent expression
\begin{align}
\Cfsnv{t} = \frac{1}{2m}&\int\! d{\bf q}\, \int\! d{\bf z}\,\int\! d{\bf \Delta}\,\delta[f({\bf q})]h[f({\bf z})]\nonumber\\ 
\times&\sum_{i=1}^{N} \Bigg[ \expect{q_{i-1}-\Delta_{i-1}/2|\frac{\partial f({\bf q})}{\partial q_{i-1}}\hat p e^{-\beta_N{\hat H}}
 + e^{-\beta_N{\hat H}}\hat p\frac{\partial f({\bf q})}{\partial q_i}
|q_i+\Delta_i/2} \nonumber\\ 
\quad &\times \expect{q_i+\Delta_i/2|e^{i{\hat H}t/\hbar}|z_i} \expect{z_i|e^{-i{\hat H}t/\hbar}|q_i-\Delta_i/2} \nonumber \\
& \times \prod_{j=1,j\neq i}^{N}\expect{q_{j-1}-\Delta_{j-1}/2|e^{-\beta_N{\hat H}}|q_j+\Delta_j/2}
\nonumber\\ 
& \quad \times \expect{q_j+\Delta_j/2|e^{i{\hat H}t/\hbar}|z_j}
\expect{z_j|e^{-i{\hat H}t/\hbar}|q_j-\Delta_j/2} \Bigg].
\end{align}
which corresponds to the ring polymer flux operator acting on the imaginary time rather than real time evolution.

%% file: rom.tex
\chapter{Integration of the ring-opening mode}
\label{ap:rom}
Integration of the ring-opening mode $\tilde \Delta_0$ is achieved when $\fq$ is invariant to cyclic permutation of the ring-polymer beads, and in the \largeN~limit, which allows the Boltzmann bra-kets to be expanded analytically,
\begin{align}
 \lNti & \prod_{i=0}^{N-1} \bra{q_{i-1}-\tfrac{1}{2} T_{i-1\ 0}\tilde \Delta_{0}} \ebN \ket{q_i+\tfrac{1}{2} T_{i0}\tilde \Delta_{0}} \no\\
 = &
 \left(\frac{m}{2\pi\beta_N \hbar^2}\right)^{N/2} \piNz e^{-\beta_N [V(q_{i}- T_{i0}\tilde \Delta_{0}/2)+V(q_i+ T_{i0}\tilde \Delta_{0}/2)]/2} \no \\
  & \qquad \times e^{-m[q_i - q_{i-1} + \tilde \Delta_0(T_{i0} + T_{i-1\ 0})/2]^2/2\beta_N \hbar^2}.
  \eql{bexpand} %Boltzmann expand
\end{align}
From \eqsr{tdef}{bndef}, $T_{i0} \sim N^{-1/2}$ and therefore
\begin{align}
 \lNti V(q_{i}- T_{i0}\tilde \Delta_{0}/2)+V(q_i+ T_{i0}\tilde \Delta_{0}/2)]/2 = V(q_i) + O(\tilde \Delta_0^2 N^{-1}),
\end{align}
so the contribution from $\tilde \Delta_0$ to the potential term vanishes in the \largeN~limit. Expanding the spring term on the third line of \eqr{bexpand} leads to three terms,
\begin{align}
 \smiNz & m[q_i - q_{i-1} + \tilde \Delta_0(T_{i0} + T_{i-1\ 0})/2]^2/2\beta_N \hbar^2 \no \\
 = & \smiNz \frac{m}{2\beta_N \hbar^2} (q_i - q_{i-1})^2 \eql{qq} \\
 & + \frac{m\tilde \Delta_0}{2\beta_N\hbar^2} (q_i - q_{i-1}) (T_{i0} + T_{i-1\ 0}) \eql{qt} \\
 & + \frac{m\tilde \Delta_0^2}{8\beta_N\hbar^2}(T_{i0} + T_{i-1\ 0})^2 \eql{tt}.
\end{align}
The first term \eqref{eq:qq} is the standard ring-polymer spring term\cite{man05ref}, but the cross term \eqref{eq:qt} must vanish to obtain positive-definite statistics\footnote{If this term is non-zero, the resultant integral over $\tilde \Delta_0$ results in an expression which is not guaranteed to be positive-definite.}. In appendix~\ref{ssec:eitd} we show that this is equivalent to requiring permutational invariance of the ring polymer beads in the dividing surface function. 
As $\fq$ must be smooth, converging in the \largeN~limit (or the rate would not be defined),
\begin{align}
 \lNti T_{i\pm1\ 0} = T_{i0} \pm \beta_N\hbar \dot T_{i0},
\end{align}
and the third term simplifies,
\begin{align}
 \frac{m\tilde \Delta_0^2}{8\beta_N\hbar^2}(T_{i0} + T_{i-1\ 0})^2 = \frac{m\tilde \Delta_0^2}{2\beta_N\hbar^2}.
\end{align}
Combining these results, \eqr{onep} becomes 
\begin{align}
 \lttz \Cfsnv{t} = & \tph \int d\bq \int d \tilde \Delta_0 \int d\tilde p_0 \ \dfq \frac{\tilde p_0}{m} h(\tilde p_0) e^{i \tilde \Delta_0 \tilde p_0/\hbar - m\tilde \Delta_0^2/2\beta_N\hbar^2} \nonumber \\
& \times \sqrt{B_N}\prod_{i=0}^{N-1} \bra{q_{i-1}} \ebN \ket{q_i}.
\end{align}
One can now integrate out the Gaussian in $\tilde \Delta_0$, creating one in $\tilde p_0$ instead, leading to \eqr{qponly}.

%% file: aps.tex
\chapter{Properties of the long-time Generalized Kubo Form}

\section{Structure of the $\ap$ function}
\label{ap:aps}
To explicitly derive the $\ap$ function we consider the long-time limit of a different flux-side function
\begin{align}
 \ltti L_{\rm fs}^{[N]}(t) = & \ltti \int d\bq \int d{\bf \Delta} \int d\bz \ \mathcal{\hat F}[\fq] \left[ \piNz h(z_i) \right] \nonumber \\
& \times \prod_{i=0}^{N-1} \bra{q_{i-1}-\tfrac{1}{2}\Delta_{i-1}} \ebN \kb{q_i+\tfrac{1}{2}\Delta_i} \etb \delta(E_i - \hat H) \ket{z_i} \nonumber \\
& \qquad \times \bra{z_i} \etf \ket{q_{i}-\tfrac{1}{2}\Delta_{i}} \\
= & \int d\bq \int d{\bf \Delta} \ \mathcal{\hat F}[\fq] \nonumber \\
& \times \prod_{i=0}^{N-1} \bra{q_{i-1}-\tfrac{1}{2}\Delta_{i-1}} \ebN \ket{q_i+\tfrac{1}{2}\Delta_i} \bk{q_i+\tfrac{1}{2}\Delta_i}{\ppi} \nonumber \\
& \qquad \times \bk{\ppi}{q_{i}-\tfrac{1}{2}\Delta_{i}} \frac{m}{|p_i|} \\
= & \ap \piNz \frac{m}{|p_i|}.
\end{align}
Here we have chosen the subcube with $p_i>0\ \forall i$ but the product over Heaviside functions could be altered to pick out other subcubes. The factor $\piNz m/|p_i|$ arises from the scaling of the Dirac delta function, but the integral over $\ap$ still converges due to the presence of the Boltzmann terms.
However, we could equivalently evaluate the side-flux form 
\begin{align}
  \ltti L_{\rm sf}^{[N]}(t) = & \ltti \int d\bq \int d{\bf \Delta} \int d\bz \ h[\fq] \left[ \smiNz \hat F(z_i) \prod_{j=0,\ j\neq i}^{N-1} h(z_j) \right] \nonumber \\
& \times \prod_{i=0}^{N-1} \bra{q_{i-1}-\tfrac{1}{2}\Delta_{i-1}} \ebN \kb{q_i+\tfrac{1}{2}\Delta_i} \etb \delta(E_i - \hat H) \ket{z_i} \nonumber \\
& \qquad \times \bra{z_i} \etf \ket{q_{i}-\tfrac{1}{2}\Delta_{i}} \\
= & \int d\bs \int d\bs' \int d\bz \ h[f(\tfrac{1}{2}(\bs + \bs'))] \left[ \smiNz \hat F(z_i) \prod_{j=0,\ j\neq i}^{N-1} h(z_j) \right] \nonumber \\
& \times \prod_{i=0}^{N-1} \bra{\psi_{{s_{i-1}}'}} \ebN \kb{\psi_{s_i}} \delta(E_i - \hat H) \ket{z_i} \bk{z_i}{\psi_{{s_i}'}} \\
= & \int d\bs \int d\bz \ h[f(\bs)] \left[ \smiNz \hat F(z_i) \prod_{j=0,\ j\neq i}^{N-1} h(z_j) \right] \nonumber \\
& \times \prod_{i=0}^{N-1} e^{-\beta_N s_i^2/2m} \bra{\psi_{s_{i-1}}} \delta(E_i - \hat H) \ket{z_i} \bk{z_i}{\psi_{s_i}} \\
= & \int d\bs \int d\bz \ h[f(\bs)] \left[ \smiNz \hat F(z_i) \prod_{j=0,\ j\neq i}^{N-1} h(z_j) \right] \nonumber \\
& \times \prod_{i=0}^{N-1} e^{-\beta_N s_i^2/2m}\delta[E_i - E(s_{i-1})] \bk{\psi_{s_{i-1}}}{z_i} \bk{z_i}{\psi_{s_i}}\eql{sidefluxd}
\end{align}
Where $\bs$ and $\acute \bs$ correspond to the momenta in the side-flux representation, so as to distinguish them from $\bp$ and $\acute \bp$ in the flux-side form. As with $\bp$ and $\acute \bp$, $\acute s$ corresponds to the momentum of a scattering eigenstate in the product/reactant direction whose corresponding eigenstate with equal energy, but in the reactant/product direction would have momentum $s$ [see \eqr{pacute}]. If we wish to sum two $\ap$ values corresponding to two subcubes which differ only in the sign of $p_l$, 
\begin{align}
 \bm{A}(p_0,\ldots&,p_l,\ldots,p_{N-1}) + \left|\frac{p_{l-1}}{\acute p_{l-1}}\right|\bm{A}(p_0,\ldots,\acute p_l,\ldots,p_{N-1}) \nonumber \\
  = &\ltti \left[ \piNz \frac{|p_i|}{m}\right] \int d\bq \int d{\bf \Delta} \int d\bz \ h[\fq] \left[ \sum_{i = 0,\ i\neq l}^{N-1} \hat F(z_i) \prod_{j=0,\ j\neq i,l}^{N-1} h(z_j) \right] \nonumber \\
& \times \prod_{i=0}^{N-1} \bra{q_{i-1}-\tfrac{1}{2}\Delta_{i-1}} \ebN \kb{q_i+\tfrac{1}{2}\Delta_i} \etb \delta(E_i - \hat H) \ket{z_i} \nonumber \\
& \qquad \times \bra{z_i} \etf \ket{q_{i}-\tfrac{1}{2}\Delta_{i}}. \eql{aplusa2}
\end{align}
The absence of a Heaviside (or flux) operator in the $l$th dimension causes
\begin{align}
 \int ds_l \int dz_l \bra{\psi_{s_{l-1}}} \delta(E_l - \hat H) \ket{z_l} \bk{z_l}{\psi_{s_l}} = \int dE(s_l) \frac{m}{|s_l|} \delta[E(s_l) - E(s_{l-1})],
\end{align}
and noting $E(s_i) \equiv E(p_i)$, we find 
\begin{align}
 |p_l|^{-1}\bm{A}&(p_0,\ldots,p_l,\ldots,p_{N-1}) + |\acute p_l|^{-1} \bm{A}(p_0,\ldots,\acute p_l,\ldots,p_{N-1})\no \\
 & = \bm{a}(\bp) \delta[E(p_i) - E(p_{i-1})] \eql{aplusa},
\end{align}
where $\bm{a}(\bp)$ is some function of $\bp$ whose exact form need not concern us. Furthermore, from \eqsr{sidefluxd}{aplusa} we observe
\begin{align}
 \bm{A}&(p_0,\ldots,p_l,\ldots,p_{N-1}) = \bm{a}(\bp)|p_l| \delta[E(p_i) - E(p_{i-1})] + \mathcal{R}(\bp) \eql{robtain}
\end{align}
where $\mathcal{R}(\bp)$ is a residue term, arising from the position-space integral over scattering eigenstates $\bra{\psi_{s_{l-1}}}$ and $\ket{\psi_{s_l}}$ being cut short by the $h(z_l)$ term,
\begin{align}
 \int ds_l \bra{\psi_{s_{l-1}}} h(z_l) \ket{\psi_{s_l}} = a(s_l,s_{l-1})\delta[E(s_l) - E(s_{l-1})] + \mathcal{R}(s_{l-1},s_l)
\end{align}
where $a(s_l,s_{l-1})$ is some function of $s_l,s_{l-1}$, and $\mathcal{R}(s_{l-1},s_l)$ the contribution to the residue $\mathcal{R}(\bp)$ from this bra-ket.\footnote{Note that if either $s_l$ or $s_{l-1}$ are imaginary, there is no corresponding scattering eigenstate and the residue is zero \cite{alt13}.}

% \subsection{Residue cancellation}
We now consider the properties of the residue in \eqr{robtain}. 
Noting that the ellipses [in, e.g.\ \eqr{aplusa}] refer to all the intervening $p_i$ values possessing the same value in both $\ap$ terms, we obtain from \eqsr{aplusa}{robtain},
\begin{align}
 |p_l|^{-1} \mathcal{R}(p_0,\ldots,p_l,\ldots,p_{N-1}) + |\acute p_l|^{-1} \mathcal{R}(p_0,\ldots,\acute p_l,\ldots,p_{N-1}) = 0. \eql{ressym}
\end{align}
There are many ways of evaluating \eqr{aplusa}, which involve taking different paths around the hypercube. 
% \subsection{Product of deltas}
For notational simplicity, we now define
\begin{align}
 \bm{A}(j,\acute k) & = \bm{A}(p_0,\ldots,p_j,\ldots,\acute p_k,\ldots,p_{N-1}) \\
 \delta_j & = \delta[E(p_j) - E(p_{j-1})] \\
 \mathcal{R}_{\acute j} & = \mathcal{R}(\acute p_j,p_{j-1}) \mbox{ and so on.}
\end{align}
Consider
\begin{align}
 |p_j|^{-1}\bm{A}(j,k) +  |\acute p_j|^{-1}\bm{A}(\acute j, k) & = \bm{b}(j^\pm,k)\delta_j (\delta_k + \mathcal{R}_{k}) \eql{bma1}
\end{align}
where $\bm{b}(j^\pm,k)$ is some function of $\bp$ whose exact form need not concern us, and from
\begin{align}
 |p_j|^{-1}\bm{A}(j,\acute k) + |\acute p_j|^{-1}\bm{A}(\acute j, \acute k) & = \bm{b}(j^\pm,\acute k)\delta_j (\delta_k +
\mathcal{R}_{\acute k}) \eql{bma2} \\
 |p_k|^{-1}\bm{A}(j,k) + |\acute p_k|^{-1}\bm{A}(j,\acute k) & = \bm{b}(j,k^\pm)\delta_k (\delta_j + \mathcal{R}_{j}) \eql{bma3} \\
 |p_k|^{-1}\bm{A}(\acute j,k) + |\acute p_k|^{-1}\bm{A}(\acute j, \acute k) & = \bm{b}(\acute j,k^\pm)\delta_k (\delta_j +
\mathcal{R}_{\acute j}) \eql{bma4}
\end{align}
we have
\begin{align}
 |p_j|^{-1}&\bm{A}(j,k) +  |\acute p_j|^{-1}\bm{A}(\acute j, k) \no \\
 & = \frac{|p_k|}{|p_j|}\bm{b}(j,k^\pm)\delta_k
(\delta_j + \mathcal{R}_{j}) - \frac{|p_k|}{|\acute p_k|}\bm{b}(j^\pm,\acute k)\delta_j (\delta_k + \mathcal{R}_{\acute k}) +
\frac{|p_k|}{|\acute p_j|} \bm{b}(\acute j,k^\pm)\delta_k (\delta_j + \mathcal{R}_{\acute j}). \eql{bma5}
\end{align}
This procedure can be done for any $j,k$, such that for a given $j$, there will be $N-1$ independent equations,\footnote{For particular systems the equations may not be linearly dependent at isolated points in the hypercube, but the contribution from such points over the entire integral will vanish.} which must hold $\forall \bp$. The only solution to these is
\begin{align}
 |p_j|^{-1}&\bm{A}(j,k) +  |\acute p_j|^{-1}\bm{A}(\acute j, k) = \bm{a}(\bp) \prod_{i=1}^{N-1}
\delta_i, \eql{deltaprod}
\end{align}
meaning the residues in each subcube are identical, leading to \eqr{astruc}.\footnote{The product in \eqr{deltaprod} contains $N-1$ Dirac delta functions arising from the product over $N-1$ Heaviside functions in \eqr{aplusa2}, such that all $N$ energies must be equal; the $i$ value which is omitted from the product is therefore arbitrary.}

\section{Integral over residues}
\label{ap:ior}
The task is to show
\begin{align}
 \lNti \int d\bp \ \mathcal{R}(\bp) h[\bar g(\bp)] = 0,\eql{resint}
\end{align}
which is achieved by building the integral over the entire hypercube by summing contributions from the subcubes, as shown schematically in \figr{cubesum}.

Considering the $j$th step of this process, i.e.\ evaluating
\begin{align}
 \inti dp_0\ldots & \int_{-\infty}^{\infty} dp_j \ldots \int_0^{\infty} dp_{N-1} \mathcal{R}(\ldots p_j \ldots) h[\bar g(\ldots p_j \ldots)] = \no\\
 & \inti dp_0\ldots \int_{-\infty}^0 dp_j \ldots \int_0^{\infty} dp_{N-1} \mathcal{R}(\ldots p_j \ldots) h[\bar g(\ldots p_j \ldots)]  \no\\
 & + \inti dp_0\ldots \int_{0}^{\infty} dp_j \ldots \int_0^{\infty} dp_{N-1} \mathcal{R}(\ldots p_j \ldots) h[\bar g(\ldots p_j \ldots)]
\end{align}
where the ellipses between the first and second integral signs correspond to the integrals in $p_i,\ i=1,\ldots,j-1$ evaluated between $\pm \infty$, and the ellipses between the second and third integrals correspond to the integrals in $p_i,\ i=j+1,\ldots,N-2$ evaluated between $0$ and $+\infty$.

From the symmetry of the residue in \eqr{ressym} and that $p_idp_i = \acute p_i d \acute p_i$ \cite{alt13},
\begin{align}
 \inti dp_0\ldots & \int_{-\infty}^0 dp_j \ldots \inti dp_{N-1} \mathcal{R}(\ldots p_j \ldots) h[\bar g(\ldots p_j \ldots)] = \no \\
 & -\inti dp_0\ldots \int_{0}^{\infty} d p_j \ldots \inti dp_{N-1}\mathcal{R}(\ldots p_j \ldots) h[\bar g(\ldots \acute p_j \ldots)] \eql{nofthem}
\end{align}
such that
\begin{align}
 \ldots \inti & dp_j\ldots \mathcal{R}(\ldots p_j \ldots) h[\bar g(\ldots p_j \ldots)] = \no\\
 & \ldots \int_{0}^{\infty} dp_j \ldots \mathcal{R}(\bp) \left\{h[\bar g(\ldots p_j \ldots)]-  h[\bar g(\ldots \acute p_j \ldots)]\right\}.
\end{align}
The task is now evaluation of $h[\bar g(\ldots p_j \ldots)]-  h[\bar g(\ldots \acute p_j \ldots)]$. By a Taylor expansion\footnote{The range of $p$ is finite due to the presence of the Boltzmann factor removing high-momenta terms.},
\begin{align}
 h[\bar g(\ldots p_j \ldots)]-  h[\bar g(\ldots \acute p_j \ldots)] = (p_j - \acute p_j) \delta[\bar g(\ldots p_j \ldots)] \ddp{\bar g(\ldots p_j \ldots)}{p_j} \eql{hminush}
\end{align}
However, for the dividing surface to be smooth in the \largeN~limit, it must be writeable as a finite number of $K$ normal modes
\begin{align}
 P_k = \sum_{j=0}^{N-1} T_{jk} p_j,\ k=0,\ldots,K-1
\end{align}
which are defined such that $P_k$ converges in the \largeN~limit, i.e. $T_{jk} \sim 1/N$. Consequently,
\begin{align}
 \ddp{\bar g(\ldots p_j \ldots)}{p_j} & = \sum_{k=0}^{K-1} \ddp{\bar g(\ldots p_j \ldots)}{P_k} \ddp{P_k}{p_j} \\
 & =  \sum_{k=0}^{K-1} \ddp{\bar g(\ldots p_j \ldots)}{P_k} T_{jk} \\
 & \sim 1/N
\end{align}
Consequently, \eqr{hminush} vanishes as $N^{-1}$. However, there are $N$ integrals like \eqr{nofthem}, each for a different $j$, such that the overall integral over residues in \eqsr{cfsap}{resint} decays as $N^{-N}$; very rapidly. Note that the value of the residue itself is not a function of $N$ from \eqr{robtain}.

\section{Evaluation of $\Cfsbnv{t}$}
\label{ap:eoc}
From \eqr{cfsb}, we observe that the flux operator only acts on a single path-integral bead such that
\begin{align}
  \ltti \Csfbnv{t} = & \int d\bp \ e^{-\beta_N \bp^{\rm T}\bp/2m} h[\bar f(\bp)] \nonumber \\
  & \times \int d\bz \ \bk{\psi_{p_0}}{z_1}\mathcal{\hat F}[z_1]\bk{z_1}{\psi_{p_1}} \prod_{i=0,\ i\neq 1}^{N-1} \bk{\ppim}{z_i} \bk{z_i}{\ppi}.
\end{align}
Integrating out all $z_i,\ i\neq1$, and generating $N-1$ Dirac delta functions from the orthogonality of scattering eigenstates,% given in \eqr{seo},
\begin{align}
 \ltti \Csfbnv{t} = &  \int d\bp \ e^{-\beta_N \bp^{\rm T}\bp/2m} h[\bar f(\bp)] \int d z_1 \ \bk{\psi_{p_0}}{z_1}\mathcal{\hat F}[z_1]\bk{z_1}{\psi_{p_1}}   \prod_{i=0,\ i\neq 1}^{N-1} \delta(p_{i-1} -p_i).
\end{align}
Upon integrating out all $p_i,\ i\neq1$, and from the definition of a dividing surface,
\begin{align}
 \ltti \Csfbnv{t} = & \int dp_1 e^{-\beta p_1^2/2m} h(p_1) \int dz_1 \bk{\psi_{p_1}}{z_1}\mathcal{\hat F}[z_1]\bk{z_1}{\psi_{p_1}}
\end{align}
which is identical to the exact quantum rate expression of Miller et.~al.\cite{mil74,mil83} given in \eqr{exactlt}, such that
\begin{align}
  \ltti \Csfbnv{t} = & k_{\rm QM}(\beta) \Qrb 
\end{align}
as required.

%% file: uniap.tex
\chapter{Short-time limit of $C_{\rm fs}^{[\bm{\Xi}]}(t)$}

\section{Derivation of the short-time limit}
\label{ap:st}
\subsection{Co-ordinate transformation}
\label{ssec:cot}
The coordinate transform used to convert \eqr{nfs} to \eqr{simpt} is 
\begin{align}
 Q_j = & \left\{
 \begin{array}{ll}
  \tfrac{1}{2} \left( q_i + \Delta_i/2 + y_i - \zeta_i/2 \right), & j = 2i \\
  \tfrac{1}{2} \left( q_i - \Delta_i/2 + y_i + \zeta_i/2 \right), & j = 2i +1 
 \end{array}
 \right. \\
 D_j = & \left\{
 \begin{array}{ll}
  - q_i - \Delta_i/2 + y_i - \zeta_i/2, & j = 2i \\
  q_i - \Delta_i/2 - y_i - \zeta_i/2, & j =2i + 1
 \end{array}
 \right. \\
 Z_j = & \left\{
 \begin{array}{ll}
 z_i - \eta_i/2, & j = 2i \\
 z_i + \eta_i/2, & j = 2i + 1
 \end{array}
 \right.%\\
% \xi_j = &\left\{
%  \begin{array}{ll}
%  \xi_i^-, & j = 2i \\
%  \xi_i^+, & j = 2i + 1
%  \end{array}
%  \right.
\end{align}
where $j = 0, 1, \ldots, 2N-1$ and $i = 0, 1, \ldots, N-1$, and the associated Jacobian is unity. As $\fq$ is unchanged by the coordinate transformation, $f(\bQ,\bD)$ in \eqr{simpt} depends on $\bQ$ and $\bD$ through the relation
\begin{align}
q_i = & \ Q_{2i} + Q_{2i+1} + (D_{2i+1} - D_{2i})/2, \label{eq:qdef}
\end{align}
such that $f(\bQ,\bD)$ is \emph{not} a general function of $\bQ$ and $\bD$, since it remains a function of only $N$ independent variables. Similarly, $g(\bZ)$ depends on $\bZ$ through
\begin{align}
z_i = & \ (Z_{2i} + Z_{2i+1})/2. \label{eq:zdef}
\end{align}

\subsection{The \shortt~limit}
\label{ssec:stl}
The \shortt~limit of \eqr{simpt} can be obtained by using Eqs.~\eqref{eq:shorttlim}--\eqref{eq:etfop},
\begin{align}
  \lttz\Cfsun{t} = & \lttz\tphtN \int d{\bf Q} \int d {\bf P} \int d {\bf D} \ \delta[f({\bf Q,D})] S_f({\bf Q,D,P}) \nonumber \\
 \times &  h[g({\bf Q + P}t/m)]  \prod_{j=0}^{2N-1} \bra{Q_{j-1}-D_{j-1}/2} e^{-\beta \xi_j \hat H} \ket{Q_j+D_j/2} e^{i D_j P_j/\hbar}, \label{eq:stl} % Short time limit
\end{align}
where $P_j = (Z_j - Q_j)m/t$, and
\begin{align}
 S_f({\bf Q,D,P}) =  & \frac{1}{2m} \smiN \ddp{\fq}{q_i} p_i \\
= & \frac{1}{2m} \smiN \ddp{f({\bf Q,D})}{[Q_{2i} + Q_{2i+1} + (D_{2i+1} - D_{2i})/2]} \nonumber \\
  & \qquad \times \left[P_{2i}+P_{2i+1} + \frac{m}{2t}(D_{2i+1} - D_{2i}) \right]
\end{align}
with $p_i = (z_i-q_i)m/t$.

To convert \eqr{stl} to \eqr{bigeq}, we note that
\begin{align}
 \ddp{g({\bf Z})}{Z_{2i}} = \ddp{g({\bf Z})}{Z_{2i+1}},
\end{align}
[see \eqref{eq:zdef}] and hence that
\begin{align}
 \lim_{t\to0_+} g(\bQ+\bP t/m) = & g({\bf Q}) + \frac{t}{m} \sum_{i=0}^{N-1} (P_{2i} + P_{2i+1}) \ddp{g({\bf Q})}{Q_{2i}}.
\end{align}
Transforming to
\begin{align}
 P^+_i = & \ \ort (P_{2i}+P_{2i+1}) \\
 P^-_i = & \ \ort (P_{2i} - P_{2i+1})
\end{align}
where $0 \leq i \leq N-1 $ and likewise for $\bf D^+,D^-$, we obtain
\begin{align}
  \lttz\Cfsun{t} = & \lttz\tphtN \int d{\bf Q} \int d {\bf P^+} \int d {\bf P^-} \int d {\bf D^+} \int d {\bf D^-} \  \nonumber \\
  & \times \delta[f({\bf Q,D^-})] S_f({\bf Q,D^-,P^+}) h[g(\bQ+\sqrt{2}\bP^+ t/m)] \nonumber \\
  & \times \prod_{i=0}^{N-1} \Big[ e^{i D^+_i P^+_i/\hbar} e^{i D^-_i P^-_i/\hbar}  \nonumber \\
  & \qquad \times \bra{Q_{2i-1}- \otrt(D_{i-1}^+ - D^-_{i-1})} e^{-\beta \xi_{2i} \hat H} \ket{Q_{2i}+\otrt(D_i^+ + D_i^-)} \nonumber \\
  & \qquad\times \bra{Q_{2i} - \otrt(D_i^+ + D_i^-)} e^{-\beta \xi_{2i+1} \hat H} \ket{Q_{2i+1} + \otrt(D_i^+ - D_i^-)}\Big].\label{eq:bigeq}
\end{align}
We can then integrate out the $\bf P^-$ to generate $N$ Dirac delta functions in $\bf D^-$, such that $f({\bf Q,D^-})$ and $S_f({\bf Q,D^-,P^+})$ reduce to
 $\fQ$ and $S_f({\bf Q,P^+})$ respectively, and \eqr{bigeq} becomes
\begin{align}
  \lttz\Cfsun{t} = & \lttz\tphN \int d{\bf Q} \int d {\bf P^+} \int d {\bf D^+} \ \dfQ S_f({\bf Q,P^+}) h[g(\bQ+\sqrt{2}\bP^+ t/m)] \nonumber \\
  & \times   \prod_{i=0}^{N-1} \bra{Q_{2i-1}-\otrt D_{i-1}^+} e^{-\beta \xi_{2i} \hat H} \ket{Q_{2i}+\otrt D_i^+} \nonumber \\
  & \qquad \times \bra{Q_{2i}-\otrt D_i^+} e^{-\beta \xi_{2i+1} \hat H} \ket{Q_{2i+1}+\otrt D_i^+}\ e^{i D^+_i P^+_i/\hbar}.
\end{align}
Using the reasoning in section~\ref{sec:rpfsf}, this expression is non-zero only if
 $\fQ \equiv \gQ$, in which case the limit
\begin{align}
\lim_{t \to 0_+} \dfQ h[f(\bQ+\sqrt{2}\bP^+ t/m)] = &\lim_{t \to 0_+} \dfQ h[f({\bf Q}) + t S_f({\bf Q,P^+})] \nonumber \\
= & \ \dfQ h[S_f({\bf Q,P^+})]
\end{align}
 results in \eqr{nzq}.

\subsection{Normal mode transformation}
\label{ssec:intout}

To integrate out $D_{i}^+,\ i>0$ from \eqr{nzq}, we transform to the coordinates 
\begin{align}
 \tilde P_j' = & \sum_{i=0}^{N-1} P_i^+ T_{2ij}' \\
 \tilde D_j' = & \sum_{i=0}^{N-1} D_i^+ T_{2ij}'
\end{align}
where
\begin{align}
 T_{i0}' = & \frac{1}{\sqrt{B_N'}}\ddp{f({\bf Q})}{Q_{i}} \label{eq:tdef2} \\
 B_N' = & \sum_{i=0}^{N-1} \left[ \ddp{f({\bf Q})}{Q_{2i}} \right]^2 \label{eq:bnpdef}
\end{align}
such that $S_f({\bf Q,P^+}) = \tpo'\sqrt{2B_N'}$ and, from \eqr{qdef}, $T_{2i\ 0}' = T_{2i+1\ 0}'$. The other normal modes, $T_{ij}'$, $j = 1, \ldots, 2N-1$ are chosen to be orthogonal to $T_{i0}'$ and their exact form need not concern us further. Unless $\fQ$ is linear in $\bQ$ (such as the centroid), $T_{ij}'$ and $B_N$ are functions of $\bQ$. We obtain
\begin{align}
 \lttz \Cfsxv{t} = & \tphN \int d{\bf Q} \int d {\bf \tilde P'} \int d {\bf \tilde D'} \ h(\tilde P_0') \frac{\tpo'}{m} \sqrt{2B_N'} \dfQ \prod_{i=0}^{N-1} e^{i \tilde D_i' \tilde P_i'/\hbar} \nonumber \\
 \times & \prod_{j=0}^{2N-1} \bra{Q_{j-1}-\otrt \sum_{i=0}^{N-1} T_{j-1\ i}' \tilde D_{i}'} e^{-\beta \xi_j \hat H} \ket{Q_j+\sum_{i=0}^{N-1} \otrt T_{ji}' \tilde D_{i}'}
\end{align}
Integrating out $\tilde P_i', \ 1 \leq i \leq N-1$ to generate Dirac delta functions in $\tilde D_i', \ 1 \leq i \leq N-1$, which are themselves then integrated out, we obtain
\begin{align}
 \lttz \Cfsxv{t} = & \frac{1}{2\pi\hbar} \int d{\bf Q} \int d \tilde P_0' \int d \tilde D_0' \ h(\tilde P_0')\frac{\tilde P_0'}{m}  \sqrt{2B_N'} \dfQ e^{i \tilde D_0' \tilde P_0'/\hbar} \nonumber \\
 \times & \prod_{j=0}^{2N-1} \bra{Q_{j-1}- \otrt T_{j-1\ 0}' \tilde D_0'} e^{-\beta \xi_j \hat H} \ket{Q_j+  \otrt T_{j0}' \tilde D_0'}.
 \label{eq:pdp} % P, D prime
\end{align}
This transformation was made using the $N$-dimensional $\bf P^+, D^+$ coordinates. To redefine the transformation from $2N$-dimensional $\bP, \bD$ we define $\bf \tilde P\ \tilde D$ (where the absence of a prime indicates a $2N$-dimensional transformation), such that [using \eqr{qdef}]
\begin{align}
 \tilde P_0' & = \frac{\sum_{i=0}^{N-1}P_i^+\ddp{\fQ}{Q_{2i}}}{\sqrt{\sum_{i=0}^{N-1}\left(\ddp{\fQ}{Q_{2i}}\right)^2}} \\
  & =  \frac{\sum_{i=0}^{2N-1}P_i \ddp{\fQ}{Q_{i}}}{\sqrt{\sum_{i=0}^{2N-1}\left(\ddp{\fQ}{Q_{i}}\right)^2}} \\
  & = \tpo.
\end{align}
Likewise $\tilde D_0' = \tdo$. However, from \eqr{bnpdef} 
\begin{align}
 B_N' & = \frac{1}{2} \sum_{i=0}^{2N-1} \left[ \ddp{f({\bf Q})}{Q_{i}} \right]^2 \\
 & = \frac{1}{2} B_N,
\end{align}
and it follows from this result and \eqr{tdef2} that 
\begin{align}
T_{j0} = T_{j0}'/\sqrt{2}.\eql{tdeftn} %T def 2 N
\end{align}
These adjustments convert \eqr{pdp} to Eq.~\eqref{eq:likenf2}.
 
\section[Imaginary-time translational invariance]{Invariance of the dividing surface to imaginary-time translation}
\label{app:piiti}
\subsection{General case}
To show that Eq.~\eqref{eq:zerocon} is equivalent to the requirement that $\fq$ be invariant under imaginary time-translation (in the limit \largeN),
we rewrite this expression in the form
\begin{align}
\lim_{N\to\infty} \sum_{j=0}^{2N-1} T_{j0} \left(\frac{Q_{j+1}-Q_j}{\beta \hbar \xi_{j+1}}-\frac{Q_{j-1} - Q_{j}}{\beta \hbar\xi_j}\right) = 0 \label{eq:tqc}. % T Q condition
\end{align}
We then consider a shift in imaginary time by a small, positive, amount $\delta \tau$,
 represented by the operator $\mathcal{P_{+\delta \tau}}$,
\begin{equation} 
\lim_{N\to\infty} \mathcal{P_{+\delta \tau}}Q_j=Q_j+(Q_{j+1}-Q_j)\delta \tau/\xi_{j+1},
 \end{equation}
and hence\footnote{Since $f(\bQ)$ is, by construction, a continuous function which converges with $N$, and $\bQ(\tau)$ is also
smooth due to the Boltzmann operator attenuating the higher normal modes of the ring polymer.} 
\begin{equation}
 \lim_{N\to\infty} \mathcal{P_{+\delta \tau}} \fQ =  \lim_{N\to\infty}\fQ +  \sum_{j=0}^{2N-1} (Q_{j+1}-Q_j) \ddp{\fQ}{Q_j} \frac{\delta \tau}{\beta \hbar \xi_{j+1}} \label{eq:permdef}.
\end{equation}
Noting from \eqsr{tdef2}{tdeftn} that $\partial \fQ/\partial Q_j = \sqrt{B_N}T_{j0}$, we see that
 the second term on the RHS of \eqr{permdef} is proportional to the first term on the LHS of \eqr{tqc}. Using similar reasoning, we find that the second term on the  LHS of \eqr{tqc} is proportional to $-\lim_{N\to\infty} \mathcal{P_{-\delta \tau}} \fQ$, where $\mathcal{P_{-\delta \tau}}$ denotes a shift in imaginary time by a small negative amount $-\delta \tau$. \eqr{tqc} is thus equivalent to the condition
\begin{align}
\lim_{N\to\infty}\mathcal{P_{+\delta \tau}}\fQ - \mathcal{P_{-\delta \tau}} \fQ = 0,
\end{align}
namely that the dividing surface $\fQ$ is invariant to imaginary-time-translation in the \largeN\ limit.
           
\subsection{Equal imaginary time discretization}
\label{ssec:eitd}
The case of equal imaginary-time discretization considered in section~\ref{ssec:nmt} is a special case of the above, where $\xi_j = 1/N$ and invariance to imaginary-time translation reduces to invariance w.r.t.~cyclic permutation of the path-integral beads.

\section{Integrating out the ring-opening coordinate}
\label{ap:uneq}

When Eq.~(\ref{eq:zerocon}) is satisfied, the only contribution to the imaginary-time path-integral from $\tilde D_0$ in the limit \largeN\ is the term ${m \tdo^2}A({\bf Q})/{2\beta\hbar^2}$, in which
\begin{align}
A({\bf Q}) &= \lim_{N\to\infty} \sum_{j=0}^{2N-1} \frac{1}{4\xi_j} \left[T_{j-1\ 0} + T_{j0}\right]^2 \\
&= \lim_{N\to\infty}\frac{1}{B_N}\sum_{j=0}^{2N-1}\frac{1}{4\xi_j}\left[\ddp{\fQ}{Q_{j-1}} + \ddp{\fQ}{Q_{j}}\right]^2,
\end{align}
and where the last line follows from the definition of $T_{j0}$ in \eqsr{tdef2}{tdeftn}. The integral over $\tdo$ in \eqr{likenf2} is evaluated to give
\begin{align}
  \lttz\Cfsxv{t} = & \frac{1}{2\pi\hbar} \int d{\bf Q} \int d \tilde P_0 \ h(\tilde P_0) \frac{\tilde P_0}{m}  \sqrt{B_N} \dfQ \nonumber \\
  & \times \sqrt{\frac{2\pi\beta \hbar^2}{mA({\bf Q})}}e^{-\beta \tilde P_0^2/2mA({\bf Q})} \prod_{j=0}^{2N-1} \bra{Q_{j-1}} e^{-\beta\xi_j \hat H} \ket{Q_j} \label{eq:uneqd} % Unequal discretization
\end{align}
and integration over $\tpo$ gives Eq.~(\ref{eq:nop}).